\documentclass[twocolumn,tighten]{aastex61}
\usepackage{hyperref}
\usepackage{pbox}

\newcommand{\link}[1]{\href{#1}{\tt #1}}

\def\simless{\mathbin{\lower 3pt\hbox
    {$\,\rlap{\raise 5pt\hbox{$\char'074$}}\mathchar"7218\,$}}} 
\def\simgreat{\mathbin{\lower 3pt\hbox
    {$\,\rlap{\raise 5pt\hbox{$\char'076$}}\mathchar"7218\,$}}} 

\newcommand{\lya}{Ly$\alpha$}

\newcommand{\fnl}{\ensuremath{f_{\rm NL}}}

\newcommand{\Hband}{$H$-band}

\newcounter{thefigs}

\newcounter{thetabs}

\newcounter{address}

\shortauthors{Blanton {\it et al.} (2017)}
\shorttitle{Sloan Digital Sky Survey IV}

\begin{document}

\renewcommand{\arraystretch}{0.8}

\title{Sloan Digital Sky Survey IV: \\
  Mapping the Milky Way, Nearby
  Galaxies, and the Distant Universe}

\correspondingauthor{Michael R.~Blanton}
\email{michael.blanton@gmail.com}
\author[0000-0003-1641-6222]{Michael R.~Blanton}
\affiliation{Center for Cosmology and Particle Physics, Department of Physics, New York University, 4 Washington Place, New York, NY 10003, USA}
\author{Matthew A.~Bershady}
\affiliation{Department of Astronomy, University of Wisconsin-Madison, 475 N. Charter St., Madison, WI 53726, USA}
\author{Bela Abolfathi}
\affiliation{Department of Physics and Astronomy, University of California, Irvine, Irvine, CA 92697, USA}
\author{Franco D.~Albareti}
\affiliation{Instituto de F{\'i}sica Te{\'o}rica (IFT) UAM/CSIC, Universidad Aut\'onoma de Madrid, Cantoblanco, E-28049 Madrid, Spain}
\affiliation{Campus of International Excellence UAM+CSIC, Cantoblanco, E-28049 Madrid, Spain}
\affiliation{``la Caixa''-Severo Ochoa Scholar}
\author{Carlos Allende Prieto}
\affiliation{Instituto de Astrof\'isica de Canarias, E-38205 La Laguna, Tenerife, Spain}
\affiliation{Departamento de Astrof\'isica, Universidad de La Laguna (ULL), E-38206 La Laguna, Tenerife, Spain}
\author{Andres Almeida}
\affiliation{Departamento de F{\'i}sica, Facultad de Ciencias, Universidad de La Serena, Cisternas 1200, La Serena, Chile}
\author{Javier Alonso-Garc\'ia}
\affiliation{Unidad de Astronom\'ia, Fac. Cs. B\'asicas, Universidad de Antofagasta, Avda. U. de Antofagasta 02800, Antofagasta, Chile}
\affiliation{Instituto Milenio de Astrof{\'i}sica, Av. Vicu\~na Mackenna 4860, Macul, Santiago, Chile}
\author{Friedrich Anders}
\affiliation{Leibniz-Institut f\"ur Astrophysik Potsdam (AIP), An der Sternwarte 16, D-14482 Potsdam, Germany}
\author{Scott F.~Anderson}
\affiliation{Department of Astronomy, Box 351580, University of Washington, Seattle, WA 98195, USA}
\author{Brett Andrews}
\affiliation{PITT PACC, Department of Physics and Astronomy, University of Pittsburgh, Pittsburgh, PA 15260, USA}
% ...

\setcounter{address}{1}

\begin{abstract}
We describe the Sloan Digital Sky Survey IV (SDSS-IV), a project
encompassing three major spectroscopic programs. The Apache Point
Observatory Galactic Evolution Experiment 2 (APOGEE-2) is observing
hundreds of thousands of Milky Way stars at high resolution and high
signal-to-noise ratios in the near-infrared.  The Mapping Nearby
Galaxies at Apache Point Observatory (MaNGA) survey is obtaining
spatially resolved spectroscopy for thousands of nearby galaxies
(median $z\sim 0.03$). The extended Baryon Oscillation Spectroscopic
Survey (eBOSS) is mapping the galaxy, quasar, and neutral gas
distributions between $z\sim 0.6$ and $3.5$ to constrain cosmology
using baryon acoustic oscillations, redshift space distortions, and
the shape of the power spectrum. Within eBOSS, we are conducting two
major subprograms: the SPectroscopic IDentification of eROSITA Sources
(SPIDERS), investigating X-ray AGNs and galaxies in X-ray clusters,
and the Time Domain Spectroscopic Survey (TDSS), obtaining spectra of
variable sources.  All programs use the 2.5 m Sloan Foundation
Telescope at the Apache Point Observatory; observations there began in
Summer 2014. APOGEE-2 also operates a second near-infrared
spectrograph at the 2.5 m du Pont Telescope at Las Campanas
Observatory, with observations beginning in early 2017. Observations
at both facilities are scheduled to continue through 2020.  In keeping
with previous SDSS policy, SDSS-IV provides regularly scheduled public
data releases; the first one, Data Release 13, was made available in
2016 July.
\end{abstract}

\section{Introduction}
\label{sec:intro}

The Sloan Digital Sky Survey (SDSS; \citealt{york00a}) started
observations in 1998 and has completed three different phases. The
data collected includes optical imaging of most of the northern high
Galactic latitude sky as well as optical and near-infrared
spectroscopy of over 3.5 million stars, galaxies, and quasars. These
observations all used the 2.5~m Sloan Foundation Telescope at
Apache Point Observatory (APO;
\citealt{gunn06a}). This paper describes SDSS-IV, the fourth phase,
and how it builds upon and extends both the infrastructure and
scientific legacy of the previous generations of surveys.

\subsection{The SDSS-I through SDSS-III legacy}

Between 2000 April and 2005 June, as described by \citet{york00a},
SDSS-I began the SDSS Legacy Survey, imaging the sky in five
bandpasses \citep[$u$, $g$, $r$, $i$ and $z$;][]{fukugita96a} using
the SDSS imaging camera (\citealt{gunn98a}). As part of the Legacy
Survey, SDSS-I also observed spectra, mostly of galaxies and
quasars,\footnote{To refer to objects thought to have actively
accreting supermassive black holes, we use the terms ``quasar'' or
``Active Galactic Nuclei (AGN),'' sometimes interchangeably,
throughout this paper.}  using a pair of dual-channel optical fiber
spectrographs fed by 640 fibers with 3$''$ diameters
(\citealt{smee13a}). The galaxies were divided into two samples, a
flux-limited Main Sample with a median redshift of $z\sim 0.1$
(\citealt{strauss02a}) and a color-selected sample of Luminous Red
Galaxies which extended to $z\sim 0.5$ (\citealt{eisenstein01a}). The
quasar sample included both ultraviolet excess quasars out to $z\sim
2$ and a set of high-redshift quasars with redshifts beyond $z=5$
(\citealt{richards02a}).

Between 2005 July and 2008 June, SDSS-II completed the Legacy Survey
with 1.3 million spectra over 8000 deg$^2$; the area covered was a
large contiguous region in the Northern Galactic Cap (NGC) and three
long, thin stripes in the Southern Galactic Cap (SGC). SDSS-II also
executed two new programs: The Sloan Extension for Galactic
Understanding and Exploration 1 (\mbox{SEGUE-1;} \citealt{yanny09a}) obtained
around 3,000 deg$^2$ of new imaging over a larger range of Galactic
latitudes and spectra of 240,000 unique stars over a range of spectral
types to investigate Milky Way structure. The Sloan Digital Sky Survey
II Supernova Survey (\citealt{frieman08b, sako14a}) cataloged over
10,000 transient and variable sources, including 1,400 SN Type Ia,
over a 200 deg$^2$ region on the equatorial stripe in the SGC,
referred to as Stripe 82. These two surveys primarily utilized the
dark time.

Between 2008 July and 2014 June, SDSS-III conducted four surveys
(\citealt{eisenstein11a}). Stellar spectroscopy continued with
SEGUE-2, which obtained 130,000 more stars during the first year of
SDSS-III (\citealt{aihara11a}). SDSS-III continued the imaging
campaign, adding 2,350 deg$^2$ of unique area and creating a
contiguous footprint in the Southern Galactic Cap; at the end of 2009
the imaging camera was retired. In Summer 2009, for the Baryon
Oscillation Spectroscopic Survey (BOSS; \citealt{dawson13a}), SDSS-III
upgraded the optical spectrographs to cover a larger optical range and
accommodate 1000 fibers (\citealt{smee13a}). By the end of SDSS-III,
BOSS spectroscopically surveyed 10,338 deg$^{2}$, gathering 1.2
million galaxy spectra to extend the original luminous red galaxy
sample from SDSS-I and SDSS-II to $z\sim 0.7$ and to increase its
sampling density at lower redshifts. It simultaneously used the
Ly$\alpha$ forest in 140,000 spectra drawn from a sample of 180,000
observed quasars to map the fluctuations in neutral hydrogen at
redshifts $2.1<z<3.5$. Both SEGUE-2 and BOSS were conducted using the
dark time.

SDSS-III also employed the Sloan Foundation Telescope in bright
time. From Fall 2008 through 2012 July, the Multi-Object APO Radial
Velocity Exoplanet Large-area Survey (MARVELS; \citealt{ge09a})
observed 5,500 bright stars ($7.6<V<12$) with a 60-fiber
interferometric spectrograph to measure high precision radial
velocities, searching for extra-solar planets and brown
dwarfs. Starting in 2011 May through 2014 June, the APO Galactic
Evolution Experiment 1 (APOGEE-1; \citealt{majewski15a}) observed
140,000 stars with a 300-fiber, $R\sim$ 22,500, \Hband\ spectrograph.

Because the weather efficiency of BOSS exceeded expectations, it
finished its primary observations early, and during its last few
months SDSS-III conducted several special programs in dark time
(\citealt{alam15b}). The Sloan Extended QUasar, ELG and LRG Survey
(SEQUELS) observed 300 deg$^2$ using the BOSS spectrograph to obtain a
dense set of quasars, emission line galaxies (ELGs), and luminous red
galaxies (LRGs), which was used to test target selection for
SDSS-IV. The SDSS Reverberation Mapping program
(SDSS-RM; \citealt{shen15a}) observed a single field containing 849
quasars over more than 30 epochs in order to monitor quasar
variability. During dark time when the inner galaxy was visible (local
sidereal times 15--20 hr) the bulk of the time was allocated to the
APOGEE-1 program.

Data from these surveys have been publicly released.  The SDSS-I and
SDSS-II Legacy, Supernova, and SEGUE-I survey data were released in a
set of data releases beginning in 2001 and culminating in 2008 October
with Data Release 7 (DR7; \citealt{abazajian09a}). The complete
SDSS-III data set was released in 2015 January in DR12
(\citealt{alam15b}).

\subsection{SDSS-IV}

SDSS-IV has new goals that build upon the scientific results of
previous SDSS surveys in the areas of Galactic archeology, galaxy
evolution, and cosmology. In so doing, SDSS-IV observations enable the
detailed astrophysical study of stars and stellar systems, the
interstellar and intergalactic medium, and supermassive black holes;
some of the emerging science themes are described below. The primary
goals of SDSS-IV are achieved in the following three core programs,
two of which required new infrastructure.

\begin{itemize}
\item 
{\it APO Galactic Evolution Experiment 2} (APOGEE-2;
Section \ref{sec:apogee2}) aims to improve our understanding of the
history of the Milky Way and of stellar astrophysics. It expands the
APOGEE-1 probe of the Milky Way history through mapping the chemical
and dynamical patterns of the Galaxy's stars via high resolution,
near-infrared spectroscopy.  The second-generation program has
northern and southern components, APOGEE-2N and APOGEE-2S,
respectively.  APOGEE-2N continues at APO, with primary use of the
bright time. APOGEE-2S utilizes new infrastructure and a new
spectrograph now installed at the 2.5 m du Pont Telescope at Las
Campanas Observatory (LCO).  The pair of spectrographs at APO and LCO
together target a total sample of around 400,000 stars.  APOGEE-2's
near-infrared observations yield access to key regions of the Galaxy
unobservable by virtually all other existing surveys of the Milky Way,
which are predominantly conducted at optical wavelengths.

\item 
{\it Mapping Nearby Galaxies at APO} (MaNGA; \citealt{bundy15a};
Section \ref{sec:manga}) aims to better understand the evolutionary
histories of galaxies and what regulates their star formation.  It
provides a comprehensive census of the internal structure of nearby
galaxies (median redshift $z\sim 0.03$), rendered via integral field
spectroscopy (IFS) --- a new observing mode for SDSS. This census
includes the spatial distribution of both gas and stars, enabling
assessments of the dynamics, stellar populations, and chemical
abundance patterns within galaxies as a function of environment. Using
half of the dark time at APO, MaNGA relies on novel fiber bundle
technology to observe 17 galaxies simultaneously by feeding the fiber
output of independent integral field units into the optical BOSS
spectrographs.  MaNGA plans to observe 10,000 nearby galaxies spanning
all environments and the stellar mass range $10^9$--$10^{11}$
$M_\odot$. The MaNGA observations cover 3500 \AA\ to 1 $\mu$m with
about 65 km s$^{-1}$ velocity resolution and 1--2 kpc spatial
resolution.

\item
{\it extended Baryon Oscillation Spectroscopic Survey}
(eBOSS; \citealt{dawson16a}; Section \ref{sec:ets}) aims to better
understand dark matter, dark energy, the properties of neutrinos, and
inflation. It pushes large-scale structure measurements into a new
redshift regime ($0.6<z<2.2$). Using single-fiber spectroscopy, it
targets galaxies in the range $0.6<z<1.1$ and quasars at redshifts
$z>0.9$. These samples allow an investigation of the expansion of the
universe using the Baryon Acoustic Oscillation (BAO) and the growth of
structure using large-scale redshift space distortions. The
large-scale structure measurements also constrain the mass of the
neutrino and primordial non-Gaussianity. Using half of the dark time
at APO, eBOSS is to observe $\sim$ 250,000 new LRGs ($0.6<z<1.0$) and
$\sim$ 450,000 new quasars ($0.9<z<3.5$) over 7,500 deg$^{2}$. Using
300 plates to cover a portion of this footprint, it also aims to
obtain spectra of $\sim$195,000 new ELGs ($0.7<z<1.1$).

\end{itemize}

There are two major subprograms executed concurrently with eBOSS, also
described in Section \ref{sec:ets}:

\begin{itemize}
\item 
{\it SPectroscopic IDentification of ERosita Sources} (SPIDERS) 
investigates the nature of $X$-ray emitting sources, including active
galactic nuclei and galaxy clusters. It uses
$\sim$5\% of the eBOSS fibers on sources related to X-ray
emission. Most of its targets are X-ray emitting active galactic
nuclei, and a portion are galaxies associated with X-ray
clusters. Initially, SPIDERS targets X-ray sources detected mainly in
the ROSAT All Sky Survey (RASS; \citealt{voges99a}), which has
recently been reprocessed (\citealt{boller16a}). In late 2018, SPIDERS
plans to begin targeting sources from the eROSITA instrument on board
the Spectrum Roentgen Gamma satellite (\citealt{predehl10a,
merloni12a}). Together with eBOSS, SPIDERS targets a sample of 80,000
X-ray identified sources ($\sim$ 57,000 X-ray cluster galaxies and
22,000 AGNs, of which around 5,000 are already included in
eBOSS targeting).

\item 
{\it Time Domain Spectroscopic Survey} (TDSS; \citealt{morganson15a})
investigates the physical nature of time-variable sources through
spectroscopy.  It also uses $\sim$5\% of the eBOSS fibers, primarily
on sources detected to be variable in Pan-STARRS1 data
(PS1; \citealt{kaiser10a}), or between SDSS and PS1 imaging. The
targets identified in PS1 are a mix of quasars (about 60\%) and
stellar variables (about 40\%). A majority of the quasars are already
targeted by eBOSS. TDSS aims to produce a spectroscopic
characterization of a statistically complete selection of
$\sim$200,000 variables on the sky down to $i=21$. TDSS targets a
total of around 80,000 objects not otherwise included by eBOSS
targeting.

\end{itemize}

In executing these programs, we exploit several efficiencies allowed
by the SDSS observing facilities. First, there is substantial common
infrastructure and technology invested in the plate and cartridge
hardware at APO and in the associated software.  Second, the SDSS-IV
survey teams closely coordinate the observing schedule on long and
short time scales to maximize efficiency. Finally, MaNGA and APOGEE-2
are able to co-observe, which allows APOGEE-2 to observe a large
number of halo stars during dark time and for MaNGA to create a unique
optical stellar library in bright time.

In addition to these overlaps in infrastructure, there exist
substantial scientific synergies between the SDSS-IV programs. These
connections allow the surveys to explore a number of critical aspects
of baryon processing into and out of gravitational potentials from
scales of stars to galaxy clusters. We remark on two emerging themes
that we expect to grow over the course of the survey. First, the
science goals of APOGEE-2 and MaNGA are closely aligned in the context
of understanding galaxy formation and evolution. APOGEE-2 treats the
Milky Way as a detailed laboratory for asking questions about galaxy
evolution similar to those MaNGA asks using a set of more distant
galaxies observed in less detail. These vantage points are highly
complementary because APOGEE-2 has access to chemo-dynamical structure
on a star-by-star basis, while MaNGA samples all viewing angles for
both gas and stars over a wide range of galaxy masses and
environments. These disparate perspectives facilitate understanding
the kind of galaxy we live in, and by extension, the detailed
processes occurring in other galaxies.

Second, the eBOSS, TDSS, and SPIDERS programs create an
unprecedentedly large and complete sample of quasars, essentially
complete down to Seyfert luminosities out to nearly $z\sim 2$ (further
discussion of quasar science is in Section \ref{sec:quasars}).  This
sample serves as a critically important tool for understanding the
evolution and decline in accretion rates of supermassive black holes,
and in turn how active galactic nuclei impact the hosts in which they
reside.

This paper describes the facilities that make these programs possible
as well as the scientific goals, observational strategy, and
management of the project and its associated collaboration. We pay
particular attention to the new hardware developments of the program,
which are primarily related to APOGEE-2S and MaNGA. More detail on all
programs, and in particular how each survey's design addresses its
high level requirements, is or will be available in existing and
upcoming technical papers (\citealt{bundy15a, morganson15a, clerc16a,
dawson16a, dwelly17a}, and APOGEE-2 and TDSS papers in preparation).

Section 
\ref{sec:facilities} provides an overview of the APO and LCO facilities.
Section \ref{sec:imaging} describes the imaging data utilized in
SDSS-IV, which includes significant reanalysis of SDSS and {\it
Wide-field Infrared Survey Explorer (WISE)} images.
Sections \ref{sec:apogee2} through \ref{sec:ets} present the survey
programs.  Section \ref{sec:data} describes the data management and
distribution plan for the project. Section \ref{sec:epo} provides a
summary of the education and public engagement strategies employed by
the project. Section \ref{sec:management} describes the project
management and organization of the science collaboration, including
the activities associated with fostering and maintaining a healthy
climate within SDSS-IV. Section \ref{sec:summary} provides a brief
summary.

%

\section{SDSS-IV Facilities}
\label{sec:facilities}
The primary departure in SDSS-IV from previous survey generations is
the expansion of our observing facilities to include telescopes in
both hemispheres. In contrast to the requirements for extragalactic
surveys on scales where the universe is isotropic, such as MaNGA and
eBOSS, this expansion is essential for the study of the Milky Way in
APOGEE-2.  In particular, the south affords much more efficient access
to the Galactic bulge and the inner disk, even for near-infrared
surveys that can operate at high airmass; full mapping of the Milky
Way, including the disk and bulge where APOGEE's near-infrared view has
the greatest advantage, requires all-sky coverage.

Since its inception, SDSS has used the 2.5~m Sloan Foundation
Telescope at the Apache Point Observatory (APO), located in the
Sacramento Mountains of south-central New Mexico. Since the advent of
APOGEE-1 in SDSS-III, the NMSU 1~m Telescope
(\citealt{holtzman10a}) at APO has also been used with the APOGEE
spectrograph. SDSS-IV adds the 2.5~m du Pont Telescope
(\citealt{bowen73a}) located at the Las Campanas Observatory (LCO) in
the Andean foothills of Chile.  On the 2.5~m Sloan Foundation
Telescope, we continue to operate the BOSS spectrographs for the eBOSS
and MaNGA programs during dark time, and the APOGEE spectrograph
during bright time.  For the 2.5~m du Pont Telescope, a second,
nearly identical APOGEE spectrograph was constructed for the southern
component of the APOGEE-2 survey.

\subsection{Apache Point Observatory}
\label{sec:apo}
The 2.5~m Sloan Foundation Telescope at APO is a modified
two-corrector Ritchey--Chr{\'e}tien design, with a Gascoigne
astigmatic corrector, and a highly aspheric corrector designed for
spectroscopy near the focal plane.  It has a 3$^\circ$ diameter usable
field of view, and a focal ratio of $f/5$.  Commissioned during the
late 1990s, it has been acquiring survey data for the past 19 years.
It performed photometric imaging through 2009; for this purpose, there
was an alternative corrector near the focal plane designed for imaging
mode. It has performed multi-object fiber-fed spectroscopy through the
present, and is devoted to this task exclusive in SDSS-IV. The on-axis
focal plane scale is nominally 217.736 mm deg$^{-1}$.

The telescope system is maintained and operated throughout the year by
engineering and administrative staff plus a team of nine full-time
observers and two to three plate-pluggers.  On each night of
observing, two observers are on duty. The field change operation
involves the manipulation of the cartridges, which weigh 100--130
kilograms, on the telescope pier near the telescope, in dark, often
cold, and occasionally icy conditions. The presence of two observers
on site is necessary to ensure instrument and personnel safety. The
use of dedicated, full time employees as observers is necessary for
maintaining safe working conditions and contributes to the high
reliability of the system and the homogeneity of the resulting data
set.

We conduct multiplexed spectroscopic observations on the Sloan
Foundation Telescope in the following manner. Each day, the plugging
technicians prepare a set of cartridges with aluminum plates plugged
with optical fibers. Each plate corresponds to a specific field on the
sky to be observed at a specific hour angle. When the cartridge is
engaged on the telescope, the plate is bent to conform to the
telescope focal plane in the optical. Depending on the cartridge
configuration, the optical fibers feed either the BOSS optical
spectrographs (\citealt{smee13a}), the APOGEE spectrograph
(\citealt{wilson12a}), or both. The cartridges are initially staged in
a bay near the telescope and allowed to equilibrate with the outside
air temperature. During the night, the observers can swap the
cartridges efficiently so that a number of fields can be observed
throughout the night. \citet{dawson13a} provide a detailed description
of this procedure.  The APO observers submit observing reports each
morning and track time lost due to weather and technical problems on a
monthly basis. Technical issues have led to $<1\%$ time loss overall
over the past few years.

The system has seventeen cartridges used for spectroscopy. Eight have
1000 fibers that emanate at two slit heads (500 fibers each). The slit
heads directly interface with the two pairs of BOSS optical
spectrographs. Each pair consists of a red spectrograph and a blue
spectrograph that together cover the optical regime from 356 nm to
1040 nm, with $R\sim 1500$--$2500$. The fibers have 120 $\mu$m active
cores, which subtend 2'' on the sky.

The other nine cartridges contain 300 short fibers that are grouped in
sets of 30 into harnesses and terminate in US Conec MTP fiber
connectors.  The 10 fiber connectors are in turn grouped into a
precision gang connector that connects to a set of long ($\sim$40 m)
fibers extending from the telescope into the APOGEE instrument room
and terminating on the APOGEE spectrograph slit head. The APOGEE
instrument has a wavelength coverage of 1.5--1.7 $\mu$m, with $R\sim$
22,500. As in the case of BOSS, the fibers have 120 $\mu$m active
cores.  Through most of SDSS-III, there were eight APOGEE-2
cartridges; in early 2014, one BOSS cartridge was converted to an
APOGEE cartridge.

New in SDSS-IV, six of the nine APOGEE cartridges have an additional
short fiber system for MaNGA that interfaces with the BOSS
spectrographs (\citealt{drory15a}). The MaNGA fiber system consists of
17 IFUs and 12 mini-IFUs, plus 92 sky fibers, for 1423 fibers in
total. These fibers are spaced more densely on the spectrograph slit
head, which leads to a greater degree of blending between the spectra;
this blending is more tolerable in MaNGA than in BOSS because
neighboring MaNGA spectra on the spectrograph are also neighboring on
the sky, which reduces the dynamic range in flux between neighboring
spectra. As is the case for the APOGEE and BOSS systems, all of these
fibers have \mbox{120 $\mu$m} active cores; however, the cladding and
buffer on the fibers were reduced to increase the filling factor of
the IFU.  The resulting spectra have nearly the same properties of
those taken with the BOSS spectrographs. These six cartridges are
capable of simultaneous APOGEE and MaNGA observations. The first MaNGA
cartridge was commissioned in 2014 March, and the final one became
operational in 2015 January. Section \ref{sec:manga} describes the
system and its use in more detail.

In addition to the science fibers, each cartridge contains a set of 16
coherent fiber bundles that are plugged into holes centered on bright
stars and are routed to a guide camera that functions at visible
wavelengths ($\sim 5500$ \AA). The operations software uses the guide
camera feedback to control telescope position, rotator position, and
focal plane scale. During APOGEE observations, the guiding software
accounts for the chromatic differential refraction between visible
wavelengths and APOGEE wavelengths in order to best align the APOGEE
fibers with the images in the focal plane at 1.66 $\mu$m.

A special purpose fiber connection exists between the NMSU 1~m
Telescope and the APOGEE spectrograph. Seven fibers are deployed in
the NMSU 1~m focal plane in a fixed pattern; one fiber is used for
a science target and the remainder for sky measurements. This mode can
be activated when the APOGEE spectrograph is not being used by the
Sloan Foundation Telescope.

A database ({\tt apodb}) at APO tracks the status and location of all
plates and cartridges. An automatic scheduling program ({\tt
autoscheduler}) determines which plates should be plugged or observed
at any given time. The pluggers and observers use a web application
({\tt Petunia}) to interface with the database and view autoscheduler
output.  Occasionally, human intervention and re-prioritization of the
automatic schedule is required; this action is performed by {\tt
Petunia}.  The observers use a graphical user interface ({\tt STUI})
to send commands to and receive feedback from the operations software
controlling the telescope and instruments.

In SDSS-IV, APOGEE-2N, MaNGA, and eBOSS share the APO observing time
from 2014 July 1 to 2020 June 30. The observatory functions all year
except for the summer shutdown period, a roughly six-week hiatus for
engineering and maintenance in July and August, during the season with
the worst weather for observing.  Major engineering work is scheduled
for this period. The baseline plan for observations allocates the
bright time to APOGEE-2 and splits the dark time between eBOSS and
MaNGA; the exact allocations are adjusted to best achieve the overall
science goals depending on progress during the survey. We describe
here the baseline plan at the start of the survey. The overall number
of hours available in the survey is 18,826 (excluding engineering
nights, typically taken at full moon). This number (and those below)
assumes uneventful recommissioning of the telescope after each summer
shutdown.

APOGEE-2 uses the 8,424 of those hours that are deemed bright time,
because the APOGEE-2 observations are of sources typically much
brighter than the sky background. We define bright time as when the
moon is illuminated more than 35\% and is above the horizon. For
APOGEE-2, science observations occur between 8$^\circ$ twilight in the
``summer'' (roughly between the vernal and autumnal equinoxes) and
between 12$^\circ$ twilight in the ``winter,'' to avoid overworking
the observers. In the ``summer'' period, APOGEE-2 also utilizes dark
time in the morning twilight between 15$^\circ$ and 8$^\circ$, which
eBOSS and MaNGA cannot use.

eBOSS and MaNGA use the remaining hours, when the moon is below the
horizon or illuminated at less than 35\%.  For these dark time
programs, science observations occur between 15$^\circ$
twilights. Although eBOSS and MaNGA split the effective observing time
in SDSS-IV, in practice, the implementation is complicated by
observational limitations. MaNGA requires the bulk of its time to be
spent when the NGC is observable. Because MaNGA target selection is
based on the Legacy spectroscopic survey, it has available 7,500
deg$^2$ of targeting in the NGC but only 500 deg$^2$ in three isolated
stripes in the SGC \citep{abazajian09a}. Providing sufficient
targeting, and assuring that three-dimensional environmental
information is available for each target, requires MaNGA to be
NGC-focused and eBOSS to be SGC-focused. In addition, the SGC is more
difficult to observe because of Galactic dust foregrounds. Therefore,
in accounting for the time balance between eBOSS and MaNGA, 1.4 hr
of SGC dark time is effectively equivalent to 1.0 hr of NGC dark
time. As a result, eBOSS is assigned 5,497 hr and MaNGA 4,904
hr.

The inital time allocation for the three surveys as a function of
Local Sidereal Time (LST) is shown in Table~\ref{tab:timeallocation}.

\begin{table}[htp]
\centering
\caption{
\label{tab:timeallocation}
Initial allocations for SDSS-IV APO programs.}
\begin{tabular}{l c c c}
\hline\hline
LST (Hours) & \multicolumn{3}{c}{Time Allocated (Hours)}\\
  & APOGEE-2 & MaNGA & eBOSS \\ \hline
0--1 & 322.9 & 22.4 & 423.0 \cr
1--2 & 350.0 & 55.1 & 434.0 \cr
2--3 & 372.1 & 99.5 & 409.4 \cr
3--4 & 377.9 & 168.7 & 337.1 \cr
4--5 & 375.8 & 215.6 & 290.5 \cr
5--6 & 377.3 & 225.1 & 279.4 \cr
6--7 & 373.7 & 239.5 & 261.4 \cr
7--8 & 373.2 & 283.8 & 219.3 \cr
8--9 & 379.0 & 296.3 & 202.5 \cr
9--10 & 377.2 & 287.5 & 210.4 \cr
10--11 & 385.2 & 284.0 & 206.5 \cr
11--12 & 384.1 & 308.1 & 177.8 \cr
12--13 & 388.8 & 291.5 & 185.6 \cr
13--14 & 388.1 & 273.3 & 193.2 \cr
14--15 & 390.7 & 317.9 & 135.1 \cr
15--16 & 380.7 & 351.3 & 72.9 \cr
16--17 & 339.6 & 284.6 & 67.0 \cr
17--18 & 316.3 & 230.2 & 49.8 \cr
18--19 & 316.8 & 212.5 & 65.8 \cr
19--20 & 294.0 & 171.4 & 134.6 \cr
20--21 & 288.6 & 132.3 & 194.7 \cr
21--22 & 285.2 & 96.0 & 250.7 \cr
22--23 & 288.7 & 40.4 & 319.0 \cr
23--24 & 298.0 & 16.8 & 377.6 \cr
\hline
\end{tabular}
\end{table}

\subsection{Las Campanas Observatory}
\label{sec:lco}
The 2.5~m Ir\'{e}n\'{e}e du Pont telescope is a modified
Ritchey--Chr\'{e}tien optical design held in an equatorial fork mount.
With a Gascoigne corrector lens, it has a 2.1 degree diameter usable
field of view \citep{bowen73a} with a focal ratio of $f/7.5$. The
on-axis focal plane scale is nominally 329.310 mm~deg$^{-1}$. The du
Pont telescope design informed a number of features of the Sloan
Foundation telescope at APO \citep{gunn06a}.

Completed in 1977, the du Pont telescope pioneered early wide-field
fiber spectroscopy. \cite{shectman93a} describes the fiber system used
for the Las Campanas Redshift Survey \citep[LCRS,][]{shectman96a} that
formed a basis for the design of the SDSS observing systems. Since the
completion of the LCRS, the du Pont telescope has not been used for
wide-field spectroscopy.  SDSS-IV is creating the infrastructure to
return to this mode of operation with improved efficiency. The primary
system upgrades include an expanded range of motion for the corrector
lens (to optimize wide-field image quality in the \Hband), improved
servo-control of the instrument rotator, and re-design of the
secondary mirror mounting structure for increased stiffness and
enhanced collimation and focus control.  In addition, implementation
of a new flat-field system is planned to optimize observing
efficiency. The telescope drives, control electronics, and control
software have also been recently modernized.

The SDSS-IV project is designing, fabricating and installing an
optical fiber cartridge and plugging system for LCO that is similar to
that at APO. We use five interchangeable cartridges with 300 short
fibers that can be re-plugged throughout each night, with a plan to
support observations of up to ten plates per night. The short fibers
in each cartridge are precisely connected through a fiber link
(the ``telescope link'') to a set of long fibers that transmit light
to the spectrograph on the ground floor of the telescope building. The
fibers run along a long metal boom attached to the wall of the dome,
and which can rotate to lie along the wall to keep the fibers safe
during observations and to provide safe storage.

Each cartridge includes a plug plate mechanically bent to conform to
the telescope's focal surface, which at \mbox{1.6 $\mu$m} has a radius
of curvature of 8800 mm. The focal plane position parallel to the
optical axis varies around \mbox{6 mm} between the center and edge of
the field (\citealt{shectman93a}), compared to around 2 mm for the
Sloan Foundation Telescope in the optical.  To achieve this large
flexure, the outer part of the plate is held at a fixed angle with a
bending ring (as done at APO). The plate profile is verified and the
profile measurements are stored in the SDSS-IV LCO database ({\tt
lcodb}).

\begin{figure}[!t]
\centering
\includegraphics[width=0.49\textwidth,
  angle=0]{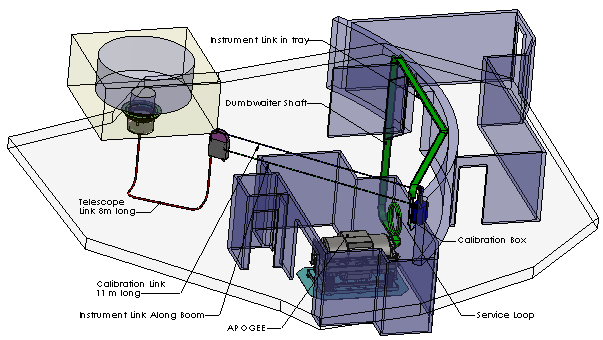}
\caption{ \label{fig:lco-layout} Model of the du Pont Telescope
configuration during APOGEE-2S observations. The yellow transparent
box indicates the bottom of the telescope, with the gray annulus
indicating the location of the primary mirror. The scaling ring
mechanism with a cartridge attached is just below the primary. A
telescope fiber link connects the cartridge to a patch panel at the end
of the boom. The instrument fibers travel down a movable boom to the
wall of the dome, and are directed to the instrument room in the level
below the telescope dome. The room on the dome level on the right side
of the diagram is used for plate plugging and mapping during the
night.}
\end{figure}

Figure \ref{fig:lco-layout} shows the configuration during
observations, in particular, the fiber run. The bottom of the du Pont
Telescope and the primary mirror are shown as the yellow box and the
inset gray annulus, respectively. The secondary focal plane is located
approximately 8 feet above the dome floor when the telescope points to
zenith. During APOGEE-2S operations, a focal plane scaling mechanism
is attached at the secondary focus. Cartridges must latch to this
scaling mechanism in order to be observed. As shown, the fibers exit
the cartridge, run along a boom to the dome wall, and travel down a
level to the instrument room.

The scaling mechanism allows real-time changes in plate position along
the optical axis. With corresponding movement of the telescope's
secondary mirror, this can be used to alter the focal plane scale to
compensate for changes introduced by differential refraction, thermal
expansion and contraction of the plate, and stellar aberration. The
scaling mechanism is controlled by the SDSS operations software as
part of the overall guiding system.

In order to implement efficient cartridge changes on the scaling
mechanism, we have constructed a stable three-rope hoist system, which
lifts the cartridges into place in the focal plane. The five
cartridges themselves are stored on custom-built dollies so they can
be maneuvered about the observing floor and plugging room.  Cartridges
are plugged in a room next to the dome, then placed in the dome to
equilibrate with the dome temperature. When a cartridge is ready to be
observed, it is rolled to the hoist, attached to the three ropes, and
lifted to the focal plane. Electrical cabinets attached to the scaling
ring house the motion control electronics, while a second electrical
cabinet at the end of the fiber boom contains an LCD touch screen
(VMI), allowing the user to control the system.  The VMI communicates
with the scaling ring electronics through a Bluetooth connection.  A set
of interlocks prevent the cartridge from being lifted in an unsafe
state (e.g., not fully attached to the hoists) or from being left
unsecured to the scaling mechanism.

The focal plane and its distortions are estimated using Zemax and an
adjusted version of the specifications from \citet{bowen73a}. From the
analysis of test images the ``best'' focal distance is 254 mm below
the rotator (993 mm from the secondary). We have directly measured the
on-axis scale and distortions at 229 mm and 279 mm below the
rotator by observing star fields using a camera positioned at various
radii in the focal plane. We have found that the specifications
in \citet{bowen73a} do not reproduce these scales well. Their Table 1
entry of the telescope focal length does not include the contribution
of the corrector.  We use a Zemax model based on the surface
specifications in their Table 2, including the corrector, with the
curvature of the primary and secondary adjusted to be consistent with
our observed scales. The resulting nominal scale and distortion is
modeled with a quintic function $s = s_0 \theta + s_3\theta^3 +
s_5\theta^5$. Our best current estimates yield, in the $H$-band, $s_0
= 329.342$ mm deg$^{-1}$, $s_3 = 2.109$ mm deg$^{-3}$, and $s_5 =
0.033$ mm deg$^{-5}$, and at the guider camera wavelength of 7600 \AA,
$s_0 = 329.297$ mm deg$^{-1}$, $s_3 = 2.168$ mm deg$^{-3}$, and $s_5 =
0.021$ mm deg$^{-5}$.  These estimates may be further refined in the
course of commissioning the system.

The cartridges contain guide systems similar to those used on the
telescope at APO. Because the system is being solely designed for use
with APOGEE-2S, we have designed a camera with effective wavelength
around 7600 \AA, which should increase its ability to use guide stars
in the more reddened part of the Milky Way.  The camera is an Andor
iKon-M 394 with a 1024 $\times$ 1024 pixel CCD, with 13 $\mu$m
pixels. This configuration is similar to that currently used at APO
(\citealt{smee13a}).  The effective wavelength is defined by an
Astrodon Photometrics Gen 2 Sloan $i$ filter. The filter is mounted in
the parallel beam between the two Nikon $f/1.4$ 35~mm lenses that
comprise the transfer optics from the output fiber block to the CCD.
The guide fibers and transfer optics preserve the telescope focal
plane scale.  Each 13 $\mu$m guider pixel subtends 0.142'' on the
sky.  The camera is operated binned $2\times2$ for guiding; thus each binned
pixel subtends 0.284'' on the sky.

The plug plates for APOGEE-2S are nearly identical to those used at
APO.  On the du Pont telescope, we use a 1.9$^\circ$ diameter field of
view, which is similar in physical size to the Sloan Foundation
Telescope. As at APO, the fibers have 120 $\mu$m diameter cores to
preserve the instrumental resolution of the spectrograph.  The fiber
core size corresponds to 1.3'' on the sky. The smaller angular size at
LCO relative to APO is appropriate for the better median seeing at LCO
($\sim$0.7'' FWHM in the
\Hband). Relative to APO, this configuration does place stricter
constraints on telescope pointing and focus (despite the slower beam of
the du Pont).

The fibers feed the APOGEE-South spectrograph, a near-clone of
the APOGEE spectrograph at APO. Changes in the new spectrograph are
described in more detail in Section \ref{sec:apogee2}.

APOGEE-2S uses approximately the equivalent of 75 nights per year on
the du Pont telescope starting in 2017 and continuing through 2020
June. In addition, up to 25 nights per year are available to guest
observers through Carnegie Observatories and the Chilean Time
Allocation Committee. All observations are conducted in $\sim$10 night
observing runs throughout the year. The southern \mbox{APOGEE-2}
program has led to a developing partnership between SDSS-IV and
astronomers at seven Chilean universities that have joined the SDSS-IV
project in a collaboration on the design, construction, engineering,
and execution of the survey. This Chilean Participation Group is an
unprecedentedly broad collaboration among Chilean universities in
astronomy and dovetails with the interest of the Chilean government in
developing astronomical engineering as a national strategy in
technology transfer and development of science.

\subsection{Plate Drilling}
\label{sec:plates}
The plates used at APO and LCO are produced for SDSS-IV using the same
systems used in previous SDSS programs.  The plates themselves are 3.2
mm thick aluminum plates, 80 cm in diameter, with a 65.2 cm diameter
region in which holes can be drilled to place fibers. Each fiber or
IFU is housed in a metal ferrule whose tip ranges in size from 2.154
mm to 3.25 mm in diameter. The larger diameter ferrules are employed
in the MaNGA and LCO systems; all others use a 2.154 mm diameter (see
Sections \ref{sec:apogee2:observations}, \ref{sec:manga:observations},
and \ref{sec:eboss:observations} for details). The ferrules have a
larger base that rests on the back side of the plate to keep the fiber
tip position fixed in focus.

Each survey plans potential observations several months in advance and
determines the sky coordinates and optimal Local Sidereal Times (LSTs)
for a set of plates. Based on the target selection results, the
potential targets in each field are assigned fibers. The fiber
placements have some physical constraints, most significantly with
regard to the minimum separation of fibers. Other constraints on the
fiber assignment based on target type and brightness can be applied.
These constraints are described below for APOGEE-2, MaNGA, and eBOSS.

Given a desired observation at a given celestial location and LST, the
target coordinates are translated into observed altitude and azimuth
given atmospheric refraction and the observatory location.  These
coordinates are translated into the physical focal plane location of
each target image, based on telescope scale and distortions. Finally,
the focal plane location is translated into a drilling location taking
into account the relative bending of the plate and the thermal
expansion of the plate due to the difference between the drill shop
temperature and the estimated observing temperature.

A large format vertical milling machine (a Dah Lih MCV-2100) at the
University of Washington drills each plate
(\citealt{siegmund98a}). During drilling, the APO plates are bent on a
mandrel such that the fiber angle will be aligned with the chief ray
at that position on the focal plane. The LCO plates are fixed to a
flat fixture, since, for the du Pont Telescope, the chief ray is
normal to the focal plane.

When observed at APO, the plates are bent to match the focal plane
curvature at around 5400 \AA. The \Hband\ focal plane has a slightly
smaller radius of curvature. In order for APOGEE fibers to remain near
the \Hband\ focus in the outer parts of the plate, a shallow
``counterbore'' is drilled on the back side of the plate, so that when
the base of the ferrule rests inside this counterbore, the fiber tip
extends beyond the plate surface slightly in order to reach the
\Hband\ focal plane.  When observed at LCO, the plates are bent to
match the focal plane in the \Hband, so no counterboring is
necessary.

At both observatories, the bending is achieved using a center post
with a 4.87 mm radius. We insert a further 1.1 mm buffer between the
post and the outer diameter of any ferrule, restricting the placement
of targets very near the centers of plates.

A Coordinate Measuring Machine measures a subset of holes on each
plate for quality assurance purposes. The typical errors measured in
hole position are 10 $\mu$m. This error has increased somewhat over
time from 7 $\mu$m since the system was first installed in
1996. However, this contribution to the total fiber position error is
subdominant.  As plugged, the median fiber position offset is 13
$\mu$m; 90\% of fibers do better than 22 $\mu$m.  The most important
error contribution arises from the slight ``clearance'' tilts induced
when each fiber is plugged, because the holes are by necessity
slightly larger than the ferrules.

\section{SDSS-IV Imaging Data}
\label{sec:imaging}
For the purposes of the SDSS-IV survey targeting, we have undertaken
the reanalysis of a variety of existing imaging data sets.  We will refer
to these data sets in subsequent sections describing the survey
programs.

We have applied a photometric recalibration to the SDSS imaging data
set. Using the PS1 photometric calibrations
of \citet{schlafly12a}, \citet{finkbeiner16a} have rederived the $g$,
$r$, $i$, and $z$ band zero points and the flat fields in all five SDSS
bands (including $u$).  The residual systematics are reduced to 0.9,
0.7, 0.7, and 0.8\% in the $griz$ bands, respectively; several
uncertain calibrations of specific imaging scans are also now much
better constrained. The resulting recalibrated images and imaging
catalogs are the basis for the eBOSS and MaNGA targeting.  They are
now included as the default imaging data set in SDSS-IV public data
releases, starting in DR13. 

All the targeting based on SDSS imaging in SDSS-IV uses the DR9
astrometric calibration \citep{pier03a,ahn12a} for both targets and
for guide stars. The SDSS-III BOSS survey used the previous DR8
astrometric calibration, which has known systematic errors. Because
the systematic errors were fairly coherent over the SDSS
field-of-view, the fiber flux losses due to these errors were
relatively minor.

For the purposes of the MaNGA target selection, we are using the
NASA-Sloan Atlas (NSA; \citealt{blanton11a}), a reanalysis of the SDSS
photometric data using sky subtraction and deblending better tuned for
large galaxies. Relative to the originally distributed version of that
catalog, we have used the new calibrations mentioned above, increased
the redshift range to $z=0.15$, and have added an elliptical aperture
Petrosian measurement of flux, which MaNGA targeting is based upon.

For the purposes of eBOSS target selection, \citet{lang16a} reanalyzed
data from {\it WISE} (\citealt{wright10a}). Using positions and galaxy
profile measurements from SDSS photometry as input structural models,
they constrained {\it WISE} band fluxes using the {\it WISE}
imaging. These results agree with the standard {\it WISE} photometry
to within 0.03 mag for high signal-to-noise ratio, isolated point
sources in {\it WISE}.  However, the new reductions also provide flux
measurements for low signal-to-noise ratio ($<5\sigma$) objects
detected in the SDSS but not in {\it WISE} (over 200 million
objects). Despite the fact that the objects are undetected, their flux
measurements are nevertheless informative to target selection, in
particular, for distinguishing stars from quasars.  These results have
been used for eBOSS targeting and have been released in DR13.

Several additional imaging analyses have been performed for targeting
SDSS-IV data; these extra sources of imaging will not necessarily be
incorporated into the SDSS public data releases, although some of them
have been released separately. 
\begin{itemize}
\item Variability analysis of Palomar Transient Factory
(PTF; \citealt{law09a}) catalogs to detect quasars 
(\citealt{palanquedelabrouille16a};
Section \ref{sec:eboss:targeting}). 
\item Selection of variable sources from PS1 (\citealt{morganson15a};
Section \ref{sec:tdss:targeting}). 
\item Intermediate-band imaging in Washington $M$, $T_2$ and 
DDO~51 filters for APOGEE-2 (\citealt{majewski00a,zasowski13a};
Section \ref{sec:apogee2:targeting}).
\item Selection of emission-line galaxies from the Dark Energy Camera
Legacy Survey (DECaLS), a $g$, $r$ and $z$ band photometric survey
being performed in preparation for the Dark Energy Spectroscopic
Instrument (DESI; \citealt{levi13a}) project.
\end{itemize}

For the purposes of eBOSS and MaNGA targeting, we correct magnitudes
for Galactic extinction using the \citet{schlegel98a} models of dust
absorption.  Galactic extinction coefficients have been updated as
recommended by \citet{schlafly11a}.  The extinction coefficients
$R_u$, $R_g$, $R_r$, $R_i$, and $R_z$ are changed from the values used
in BOSS (5.155, 3.793, 2.751, 2.086, and 1.479) to updated values
(4.239, 3.303, 2.285, 1.698, and 1.263).  We set $R_{W1}=0.184$ for
the {\it WISE} 3.4 $\mu$m band and $R_{W2}=0.113$ for the 4.6 $\mu$m
band \citep{fitzpatrick99a}.

\section{APOGEE-2}
\label{sec:apogee2}
\subsection{APOGEE-2 Motivation} 

APOGEE-2 is conducting high-resolution, high signal-to-noise ratio
spectroscopy in the near infrared for a large sample of Milky Way
stars.  A key challenge in astrophysics is the characterization of the
archeological record, chemical evolution, dynamics, and flows of mass
and energy within galaxies. The Milky Way provides a unique
opportunity to examine these processes in detail, star-by-star. Large
spectroscopic samples are critical for mapping the Galaxy's numerous
spatial, chemical, and kinematic Galactic sub-populations.

APOGEE-2 is creating a Galactic archeology sample designed to
understand the history of all components of the Milky Way, including
the dust-obscured ones (Fig.~\ref{fig:apogee2:overview}), and to
better understand the stellar astrophysics necessary to uncover that
history. APOGEE-2 is accomplishing this goal by continuing the overall
strategy of APOGEE-1 (\citealt{zasowski13a, majewski15a}), increasing
to 400,000 the number of stars sampled, and expanding to cover the
inner Galaxy from the Southern Hemisphere. The primary sample is a set
of red giant branch stars that trace Galactic structure and
evolution. Several smaller sets of targets explore more specific
aspects of Galactic and stellar astrophysics.  These spectra yield
precise radial velocities, stellar parameters, and abundances of at
least 15 elements. The Sloan Foundation Telescope at APO and the du
Pont Telescope at LCO are mapping both hemispheres of the Milky Way.

APOGEE-2 is distinguished from all other Galactic archeology
experiments planned or in progress by its combination of high spectral
resolution, near infrared wavelength coverage, high signal-to-noise
ratio, homogeneity, dual-hemisphere capability, and large statistical
sample.  It improves upon other Milky Way spectroscopic surveys that
lack the combined high resolution and $S/N$ needed by current
methodology for the determination of accurate stellar parameters and
chemical abundances (RAVE, \citealt{steinmetz06a, kordopatis13a};
BRAVA, \citealt{howard08a}; SEGUE-1 and SEGUE-2, \citealt{yanny09a};
ARGOS, \citealt{freeman13a}; and LAMOST, \citealt{cui12a,
  zhao12a}). APOGEE-2 complements existing or future wide-angle,
high-resolution stellar spectroscopic surveys or instruments that are
single-hemisphere and are optical, experiencing heavy dust extinction
at low Galactic latitudes and in the inner Galaxy (GALAH,
\citealt{zucker12a, desilva15a}; {\it Gaia}-ESO,
\citealt{gilmore12a}; WEAVE, \citealt{dalton14a}; 4MOST,
\citealt{dejong14a}). MOONS (\citealt{cirasuolo14a}) is the closest
analog and is complementary in ambition; it is a near-infrared
instrument under construction for the Very Large Telescope in the
Southern Hemisphere, with a larger number of fibers (1024) and
telescope aperture size (8.2~m), but twenty times smaller
field of view (500 arcmin$^2$).

Like other high resolution surveys and instruments, APOGEE-2
complements the optical {\it Gaia} satellite measurements of parallax,
proper motion, and spectroscopy of a much larger number of stars
(\citealt{prusti16a}). APOGEE-2 will benefit from the accurate
measurements of distance and proper motion from {\it Gaia} for its
stars. Our understanding of the Galactic chemical and dynamical
structure will be strengthened using the APOGEE-2 information
available for these stars: more precise radial velocities, more
precise stellar atmospheric parameters, and more precise abundances
for a larger set of elements.

\subsection{APOGEE-2 Science}

The combined APOGEE-1 and APOGEE-2 data sets yield multi-element
chemical abundances and kinematic information for stars from the inner
bulge out to the more distant halo in all longitudinal directions and
include both Galactic satellites and star clusters.  To effectively
exploit these data, APOGEE-2 is collecting additional observations on
fundamental aspects of stellar physics necessary to promote the
overall understanding of the formation of the Galaxy.

Near-infrared spectra are excellent for studies of stars in the
Galactic disk and bulge. The bulk of these regions suffer high
extinction from foreground dust in the visible, with regions in the
Galactic plane frequently yielding $A_{V} > 10$ \citep{nidever12a}.
With $A_{H}/A_{V} \sim 0.16$, NIR observations can peer through the
dust far more efficiently than optical data. The \Hband\ is rich in
stellar atomic (e.g., Fe, Ti, Si, Mg, and Ca) and molecular (e.g., CO, OH,
CN) absorption lines that can be used to determine stellar properties
and elemental abundances \citep{meszaros13a, holtzman15a,
  shetrone15a}. In particular, lines in the \Hband\ are sensitive to
the most common metals in the universe, C, N, and O, which are
difficult to measure in the optical.  The luminous red giant branch
(RGB) population dominates useful source catalogs like 2MASS, and
selecting targets by \Hband\ flux and red $J-K_s$ color
yields a population relatively unbiased in age and metallicity.

As shown in APOGEE-1 \citep{holtzman15a} typical APOGEE-2 spectra
enable measurements of at least 15 separate chemical abundances with
0.1 dex precision and high precision radial velocities (better than
100~m~s$^{-1}$).  The final spectra are the result of coadding several
observations spaced up to a month or more apart; these time series
data can identify radial velocity variables and detect interesting
binaries and substellar companions.

APOGEE-2's magnitude and color selection criteria result in a main
survey sample dominated by distant red giant, subgiant, and red clump
stars, but with some contribution from nearby late-type dwarf stars.
Through the inclusion of supplementary science programs, the final
APOGEE-2 program also includes observations of RR Lyrae stars,
high-mass and early main-sequence objects, as well as
pre-main-sequence stars. Combined, these programs will address a
number of topics in Galactic and stellar astrophysics.

\begin{itemize}
\item Mapping of the thick and thin disk at all Galactic
  longitudes, including the inner disk regions, and at the full range
  of Galactic radii, with substantial samples at least 6 kpc from the
  sun and with a significant subsample having reliably determined
  ages. These maps expand upon APOGEE-1 results
  (\citealt{anders14a, nidever14a, hayden15a}), and further test
  scenarios of inside-out growth, radial migration, and the origin of
  the $\alpha$-enriched population (\citealt{chiappini15a, martig15a,
    bovy16c}).

\item 
  Accurate stellar ages and masses from the combination of APOGEE data
  with asteroseismology (e.g., \citealt{epstein14a, chiappini15a,
    martig15a}), establishing critical benchmarks in the analysis of
  Galactic chemistry and dynamics in numerous directions sampled by
  {\it Kepler} and its subsequent K2 mission.

\item Dynamics of the disk and the Galactic rotation curve, including
  non-axisymmetric influences of the bar and spiral arms (e.g.,
  \citealt{bovy12a, bovy15a}).

\item Three-dimensional mapping of the Galactic bulge and bar, measuring
  dynamics of the bar, bulge, and nuclear disk (\citealt{nidever12a,
    schonrich15a, ness16a}) and their chemistry
  (\citealt{garciaperez13a, ness15a}).  Southern Hemisphere operations
  as well as the inclusion of standard candles such as red clump and
  RR Lyrae stars will make this mapping more complete and precise than
  APOGEE-1.

\item Chemistry and dynamics in the inner and outer halo across all
  Galactic longitudes, including a large area of the NGC, and sampling
  known halo substructure and stars reaching to at least 25 kpc.

\item Stellar populations, chemistry and dynamics of nascent star
  clusters, open clusters, globular clusters at various evolutionary
  stages, dwarf spheroidals, the Magellanic Clouds, and other
  important components of the Milky Way system (e.g.,
  \citealt{frinchaboy13a, majewski13a, meszaros13a, cottaar14a,
    cottaar15a, foster15a, garciahernandez15a, meszaros15a, bovy16b}).

\item Exoplanet host observations in {\it Kepler} fields to
  characterize host versus non-host properties and assess false
  positive rates (\citealt{fleming15a}).

\item 
  Detection of stellar companions of stellar, brown dwarf, and
  planetary mass across the Galaxy (e.g., \citealt{troup16a}).

\item Mapping the interstellar medium using Diffuse Interstellar Bands
  (\citealt{zasowski15b, zasowski15a}), or dust reddening effects
  (\citealt{schultheis14a}).

\end{itemize}

APOGEE-2 is also pursuing ancillary science programs with a small
fraction of the available fibers to utilize more targeted and
exploratory uses of the APOGEE instruments.

\subsection{APOGEE-2 Hardware}

APOGEE-2 utilizes one existing spectrograph at APO
(\citealt{eisenstein11a, wilson12a, majewski15a}) and a second
instrument at LCO. Each spectrograph is fed with 300 fibers with 120
$\mu$m cores; both yield nearly complete spectral coverage between
1.51 $\mu$m $< \lambda < $ 1.70 $\mu$m, high spectral resolution
($R\sim$ 22,500, as measured for the first spectrograph) and high S/N
($> 100 $ pixel$^{-1}$) for most targets \citep{majewski15a}. The
APOGEE spectrographs each utilize a large mosaic volume-phase
holographic (VPH) grating.  At APO, the first spectrograph's VPH
grating consisted of three aligned panels on the same substrate. The
spectrograph cameras consist of four monocrystalline silicon lenses
and two fused silica lenses. The spectra are dispersed onto three
Teledyne H2RG array detectors with 18 $\mu$m pixels, sampling three
adjacent spectral ranges; all elements of each array are sampled
``up-the-ramp'' at 10.7 second intervals within each exposure.  This
procedure yields an effective detector read-noise of $\sim$10 e$^-$
per pixel.  The geometric demagnification of the camera and collimator
optics delivers slightly over 2 pixels sampling of the fiber diameter
in the spatial dimension, but the spectra are slightly undersampled in
the blue part of the spectrum. To fully sample the spectra, the three
detectors are dithered by a half pixel in the spectral dimension
between exposures, which therefore are routinely taken in pairs. The
measured throughput of the APOGEE-1 instrument is $20\pm2$\%
\citep{majewski15a}.

At APO, the spectrograph is fed by long fibers extending from the
Sloan Foundation Telescope and the NMSU 1-meter Telescope, as
described in Section \ref{sec:apo}.  The NMSU 1-meter Telescope is
used to observe bright stars, such as previously well-characterized
spectral standards and HIPPARCOS targets (\citealt{feuillet16a}), when
the spectrograph is not otherwise in use with the Sloan Telescope.

The APOGEE-South spectrograph at LCO is a near-clone of the APOGEE
spectrograph with some slight differences. First, the mosaic VPH
grating uses two panels instead of three, a simplification with
negligible impact on the net instrument throughput. Nevertheless, the
pair of panel exposures were not perfectly aligned; therefore, an
optical wedge is added to compensate for this misalignment to optimize
spectral resolution.  Second, the spectrograph optical bench is
mounted within the instrument cryostat with greater consideration of
seismic events, given its location in Chile.  Other more minor
modifications in the optical bench and cryostat configuration have
been adopted as well.

We anticipate that the data from the second spectrograph will be, in
most respects, quite similar to those from the original.  The fibers
will typically have lower sky backgrounds because they subtend a
smaller angular size. In addition, the du Pont optical correctors have
less loss in the \Hband, which is $\sim 40\%$ on the Sloan Foundation
Telescope.

The APOGEE-South spectrograph was installed at the du Pont Telescope
in 2017 February and survey operations are planned to start soon
thereafter.

\begin{figure}[t!]
\centering
\includegraphics[width=0.49\textwidth,
  angle=0]{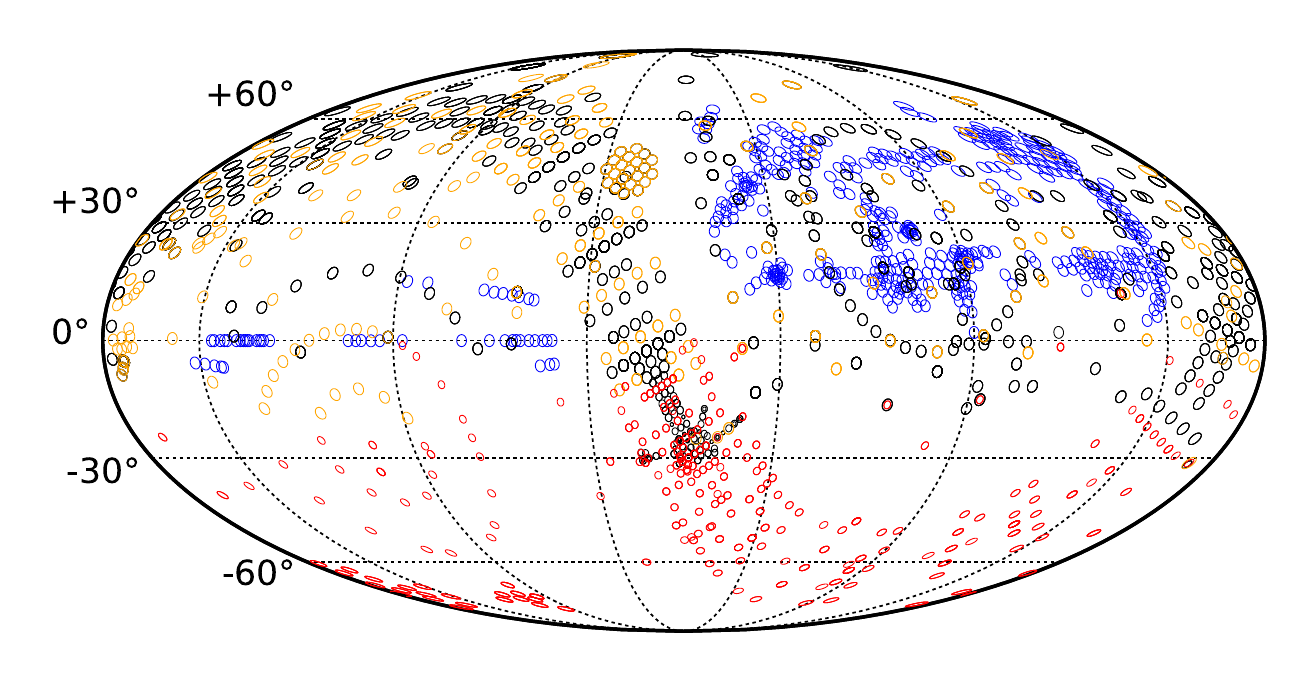}
\caption{ \label{fig:apogee2:overview} APOGEE-1 and planned APOGEE-2
  spectroscopic footprint in equatorial coordinates, centered at
  $\alpha_{\rm J2000}=270^\circ$, with East to the left.  Black shows
  APOGEE-1 data, orange indicates APOGEE-2N, and red is APOGEE-2S.
  Blue shows projected MaNGA coverage for which APOGEE-2 can
  potentially have observations of stars (see also Figure
  \ref{fig:manga-footprint}). Because of logistical constraints and
  potential changes in the MaNGA plans, the final coverage of the halo
  may differ somewhat from this figure.}
\end{figure}

\subsection{APOGEE-2 Targeting and Observing Strategy}
\label{sec:apogee2:targeting}

APOGEE-2 continues much of the observational strategy for APOGEE-1
(\citealt{zasowski13a}).  Its standard targeting uses the 2MASS
survey, selecting stars based on dereddened $J-K_s$.  Additional
information from the Optical Gravitational Lensing Experiment (e.g.,
OGLE-III and OGLE-IV; \citealt{udalski08a, udalski15a}), Vista
Variables in the Via Lactea (VVV: \citealt{minniti10a, saito12a,
  hempel14a}), and the VVV Extended ESO Public Survey (VVVX) surveys
are incorporated for certain subsamples. Dereddened magnitude limits
range from $H=12.2$ to $13.8$ mag (depending on cohort, as explained
below) for the bright-time observations, and are $H=11.5$ during
co-observing with MaNGA.

To estimate extinction in the disk and bulge, APOGEE-2 supplements
2MASS imaging with the Spitzer-IRAC Galactic Legacy Infrared Mid-Plane
Survey Extraordinaire and extensions (GLIMPSE; \citealt{benjamin03a,
  churchwell09a}).  Where GLIMPSE data are not available, APOGEE-2
uses data from the all-sky {\it WISE}
mission (\citealt{wright10a}).  The reddening estimates employ the
Rayleigh-Jeans Color Excess method (\citealt{zasowski09a,
  majewski11a}).

To efficiently separate dwarfs and giants in the stellar halo,
APOGEE-2 obtained Washington $M$ and $T_2$ and DDO~51 stellar
photometry using the Array Camera on the 1.3~m telescope of the
U.S. Naval Observatory in Flagstaff, with additional data anticipated
for the Magellanic Cloud targeting in the Southern survey
component. In the ($M-T_2$) versus ($M-{\rm DDO~51}$) color plane, dwarfs
and giants lie in distinct locations, which allows relatively clean
separation of these stellar classes (\citealt{geisler84b, munoz05a,
  zasowski13a}).

To collect sufficient signals on fainter stars while still acquiring
data on large numbers of brighter stars, APOGEE-1 and APOGEE-2 employ
a system of ``cohorts,'' groups of stars observed together for the
same length of time.  The 3-visit cohorts correspond to the brighter
magnitude limits ($H=12.2$) and the longer cohorts correspond to
deeper magnitude limits (down to $H=13.8$). Each 3$^\circ$ diameter
field on the sky is observed with one or more plate designs, each of
which consists of a combination of cohorts.  Stars are predominantly
divided into cohorts according to brightness, and observed
(``visited'') long enough to obtain the required S/N goals: typically
S/N $\sim$ 100 per half-resolution element for the core programs
sampling Milky Way giant stars; S/N $\sim$ 70 for some exceptional
target classes such as luminous stars in Local Group dSph and the
Magellanic clouds; and S/N $\sim$ 10 for RR Lyrae in the bulge. For
example, in a 12-visit field, ``short'' cohort stars are observed on 3
visits, ``medium'' cohort stars are observed on 6 visits, and ``long''
cohort stars are observed on all 12 visits.  \citet{zasowski13a}
provide additional examples. Each visit corresponds to 67 minutes of
exposure time in nominal conditions (see
\S\ref{sec:apogee2:observations} for further visit details), with
fields visited anywhere from 3 to 24 times.

Visits per field have cadences between 3 and 25 days. This strategy is
adopted to yield detections of spectroscopic variability, most
commonly velocity shifts due to binary companions with a typical
radial velocity precision of $\sim$100--200 m s$^{-1}$.  For stars
observed more than the nominal three visits, it is possible to detect
brown dwarf and planet mass companions (\citealt{fleming15a,
  troup16a}).

The APOGEE-2 observations are divided into northern and southern
components, and each of these are sub-divided into different target
classes identifying different Galactic regions or special target
classes. The sky coverage is summarized in Figure
\ref{fig:apogee2:overview}. The target categories summarized in
Table~\ref{tab:apogeetargeting}, providing the number of plates,
visits, and stars observed in each class from respective hemispheres
(N or S). All targeted stars will have observations yielding radial
velocities and stellar atmospheric parameters, but, depending on the
target faintness (e.g., giants in the Magellanic clouds) or type
(e.g., RR Lyrae), abundance information may only be partial or
unavailable, as noted.

\begin{figure}[t!]
\centering
\includegraphics[width=0.49\textwidth, angle=0]{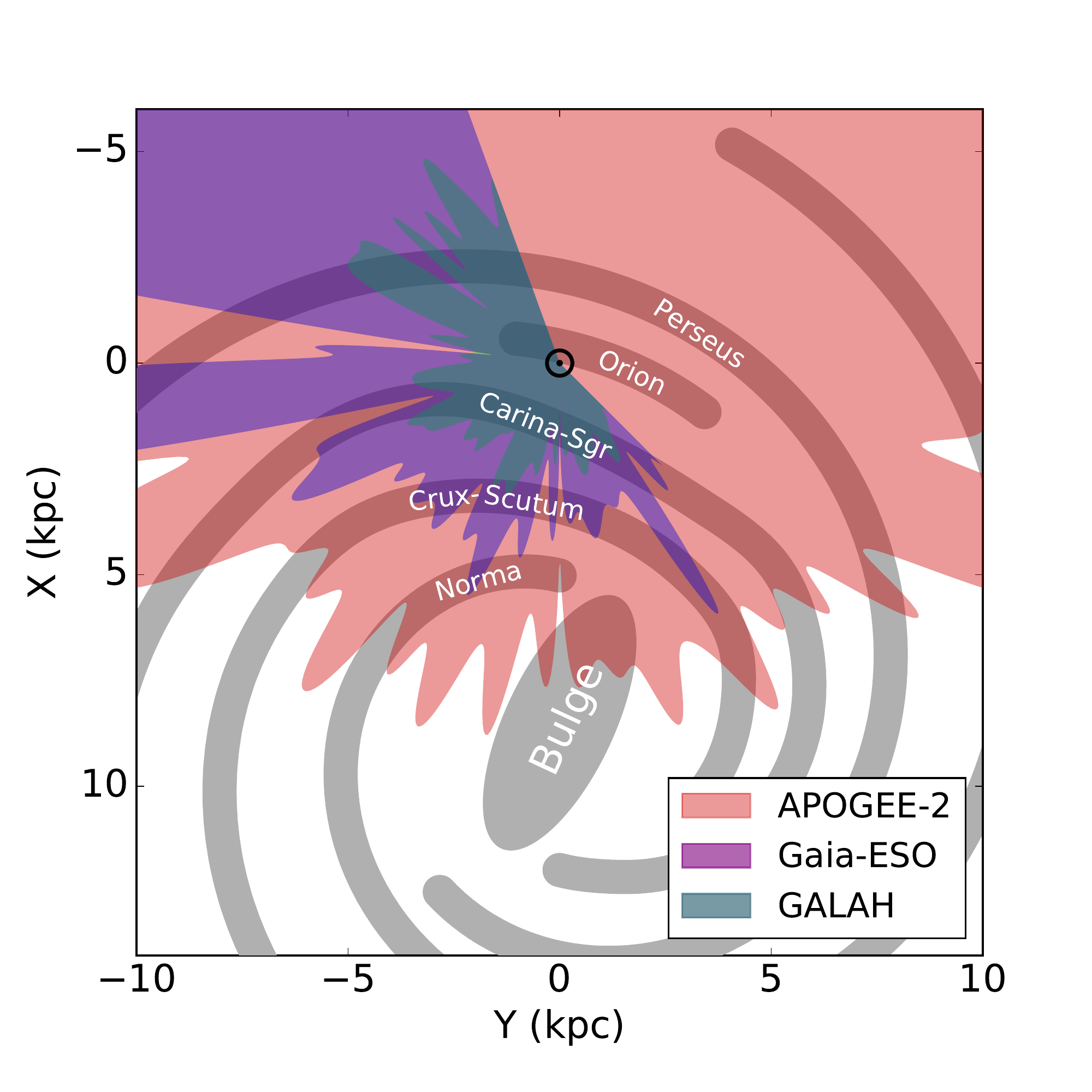}
\caption{ \label{fig:apogee2-galaxy} Map of APOGEE-1 and APOGEE-2
  distance limits at $b=0^\circ$ within the Galactic plane, compared
  to other Galactic plane spectroscopic surveys. These limits assume
  observations of stars at the tip of the red giant branch (for
  solar metallicity and 2 Gyr of age) using isochrones from
  \citet{bressan12a}. To calculate the distance limit, we use the dust
  extinction prescription of \citet{bovy16a} and limits of $H=12.2$
  for APOGEE-2, $V=14$ for GALAH, and $V=19$ for {\it Gaia}-ESO (their
  faintest limit across all fields). Longer cohorts in APOGEE-2 
  extend correspondingly further.}
\end{figure}

APOGEE-2N continues observations of red giant branch (RGB) and red
clump (RC) stars in the inner and outer Galactic disk, and of the
stellar halo in the NGC and in the SGC. The distance limits for this
sample in the Galactic plane ($b=0^\circ$) are shown in Figure
\ref{fig:apogee2-galaxy}. Some halo fields specifically target areas
with known tidal streams; these samples are anticipated to total
$\sim$58,000 stars.  Additional Galactic evolution programs target
dwarf spheroidals, as well as open and globular clusters. Because it
shares cartridges with MaNGA, APOGEE-2N is co-observing with MaNGA
during dark time. Due to the MaNGA observing strategy, these exposures
are typically three hours of integration.  However, MaNGA's dithers
mean a lower overall throughput (see Section
\ref{sec:apogee2:observations}) and therefore the magnitude limit in
these fields is $H=11.5$.  We anticipate an additional 120,000 stars,
primarily selected as red giants, in ``halo'' (i.e., high latitude)
fields. These locations are displayed in blue in Figure
\ref{fig:apogee2:overview}. These co-observed stars represent a
substantial increase in numbers of halo stars over what was possible
in APOGEE-1.

APOGEE-2 expands an ancillary APOGEE-1 program in the {\it Kepler}
satellite Cygnus field into a main survey objective including the
fields observed with the K2 mission.  Two main goals focus on
asteroseismology and gyrochronology targets and observations relating to
Kepler exoplanets. The APOKASC collaboration combines the resources of
APOGEE and the {\it Kepler} Asteroseismology Science Consortium (KASC)
to determine precise age and mass constraints on stars of a range of
stellar types (\citealt{pinsonneault14a}).  The {\it Kepler} Object of
Interest (KOI) program provides multi-epoch observations on five of
the modules in the original {\it Kepler} Cygnus field, targeting KOIs
to characterize planet-host versus control star properties as well as
to improve our understanding of the frequency of false positives
within the KOI sample.  In addition to these observations of the
primary Kepler field, APOGEE-2N is conducting a campaign of {\it
  Kepler} K2 fields, using the combined space
asteroseismology/gyrochronology plus APOGEE spectral data to determine
high-quality ages for stars in a wide range of Galactic directions.

The samples listed in Table~\ref{tab:apogeetargeting} complete
APOGEE-2's homogeneous sampling of all Galactic regions with the RGB
and RC survey. We are also targeting fainter stars from the upper RGB of
the LMC, SMC, and several dSphs, and probing the chemistry of open and
globular clusters. A new program observes RR Lyrae stars in the
bulge from OGLE-IV and VVV to measure the detailed structure and
kinematics of the ancient bulge.

Both the northern and the southern components also contain ancillary
program targets with a diverse range of science goals. These programs
include using low extinction windows to examine the far disk at
distances of over 15 kpc in the plane, measuring Cepheid metallicities
across the disk, characterizing young moving groups, determining the
detailed and precision abundance trends in clusters, and studying
massive AGB stars. APOGEE is also conducting an extensive
cross-calibration program between APOGEE, SEGUE, GALAH, and {\it
  Gaia}-ESO, and between the APOGEE and APOGEE-South spectrographs.

\begin{table}[htp]
\centering
\caption{
\label{tab:apogeetargeting}
APOGEE-2 Targeting Description}
\begin{tabular}{l c r r r l}
\hline\hline
Target      & N or S & $N_{\rm plate}$ & $N_{\rm visit}$ & $N_{\rm
  star}$ & Abundances \\ \hline
Clusters    & N &  31 &   63 &  2340 & complete \\
            & S &  63 &  158 &  8715 & complete \\
\multicolumn{6}{c}{} \\ [-0.1in]
Bulge       & N &   1 &   18 &   230 & complete \\
            & S & 213 &  321 & 38310 & complete \\
\multicolumn{6}{c}{} \\ [-0.1in]
Inner Disk  & N & 116 &  348 & 20010 & complete \\
Outer Disk  & N &  93 &  279 & 21390 & complete \\
Disk        & S & 179 &  537 & 30470 & complete \\
dSph        & N &  12 &   72 &   780 & partial \\
            & S &  12 &   72 &   780 & partial \\
\multicolumn{6}{c}{} \\ [-0.1in]
Halo-NGC    & N &  84 &  504 &  5460 & complete \\
            & S &   4 &   48 &   480 & complete \\
\multicolumn{6}{c}{} \\ [-0.1in]
Halo-SGC    & N &  28 &   87 &  6670 & complete \\
            & S &  24 &   72 &  5520 & complete \\
\multicolumn{6}{c}{} \\ [-0.1in]
Streams-NGC & N &  48 &  288 &  3840 & partial \\
Streams-SGC & N &   9 &   39 &  1410 & partial \\
            & S &   2 &   12 &   345 & partial \\
\multicolumn{6}{c}{} \\ [-0.1in]
APOKASC     & N &  56 &   56 & 12880 & complete \\
KOI         & N &   5 &   90 &  1150 & complete \\
Halo Co-obs & N & 600 &  600 &120000 & complete \\
LMC         & S &  51 &  153 &  4930 & partial \\
SMC         & S &  24 &   78 &  1920 & partial \\
SGR         & S &   4 &   30 &  1405 & complete \\
RRLyrae     & S &  31 &   31 &  4000 & ---\\
\multicolumn{6}{c}{} \\ [-0.1in]
TOTALS      & N &1084 & 2444 &196160 & \\
            & S & 607 & 1512 & 96875 & \\ 
\hline
\end{tabular}
\end{table}

\clearpage
\subsection{APOGEE-2 Observations}
\label{sec:apogee2:observations}

APOGEE-2N utilizes the bright time at APO. Details of the division of
observations across the SDSS-IV surveys at APO are given in Section
\ref{sec:apo}. APOGEE-2S primarily utilizes the bright time at
LCO, and conducts observations 75 nights each year. Section
\ref{sec:lco} describes the operational model; otherwise, APOGEE-2S
largely employs the same observing strategies as APOGEE-2N. 

Each APOGEE-2N fiber is encased in a metal ferrule whose tip is
relatively narrow at 2.154 mm and is inserted fully into the plate
hole, but whose base is around 3.722 in mm diameter and sits flat on
the back of the plate.  A buffer of 0.3 mm around each ferrule is
maintained to prevent plugging difficulty. Given the plate scale on
the Sloan Foundation Telescope, on the same plate no two APOGEE-2N
fibers can be separate by less than 72$''$ on the sky. As described in
Section \ref{sec:plates}, the APOGEE-2N holes are counterbored so that
the fiber tips lie on the \Hband\ focal plane.

Each APOGEE-2S fiber has a larger 3.25 mm tip and a 4.76 mm base. No
buffer is used around each ferrule. Given the plate scale of the du
Pont Telescope, on the same plate no two APOGEE-2S fibers can be
separated by less than 52$''$ on the sky. Because at LCO the plate is
curved to match the \Hband\ focal plane, there is no counterboring of
the APOGEE-2S plates.

Each plate is designed for a specific hour angle of observation.  The
observability window is designed such that no image falls more than
0.3$''$ from the fiber center during guiding.  These limits on the
LST of observation are slightly larger than for eBOSS because APOGEE-2
operates in the near-infrared where the refraction effects are
smaller. In addition, for APOGEE-2N, we add 30 minutes on either side
to ease scheduling constraints.


An APOGEE-2 visit typically consists of eight 500~s exposures taken
in two ABBA sequences (a total of 66.7 minutes), where A and B are two
detector dither positions in the spectral dimension described above to
ensure critical sampling. Each exposure consists of 47 non-destructive
detector reads spaced every 10.7~s.  Each visit requires 20 minutes
overhead in cartridge changes, calibrations, and field acquisition.
Whereas in APOGEE-1 and the beginning of APOGEE-2, we had a fixed
number of exposures per visit, starting in 2016 we have adapted the
number of exposures based on the accumulated signal-to-noise ratio
relative to the requirement, as eBOSS and MaNGA do. This change allows
more efficient use of resources; initial estimates from the first few
months indicate that the net increase in the survey completion rate is
significant (roughly $15\%$).

During MaNGA time, APOGEE fibers are placed on APOGEE-2 targets.  The
MaNGA observations are dithered on the sky and their schedule
constrains the APOGEE exposures to have 10\% shorter exposure times
than the standard APOGEE exposures. Both of these effects lead to a
net throughput reduction per exposure of almost a factor of two; a
reduction of about 40\% due to the offset under typical seeing, and
about 10\% more due to the shorter exposure times. In some cases, the
MaNGA-led observing yields more than the standard number of APOGEE
exposures per field, but this is generally insufficient to compensate
for the reduced throughput per exposure. As a result, the faint limit
for targets on the MaNGA-led co-observing plates is chosen to be
$\sim$0.7 mag brighter than it is for standard APOGEE plates ($H<11.5$
instead of $H<12.2$), so that the standard APOGEE signal-to-noise
ratio requirement is met for targets in the MaNGA fields.

\begin{figure*}[t!]
\centering
~\\
~\\
\includegraphics[width=0.99\textwidth, angle=0]{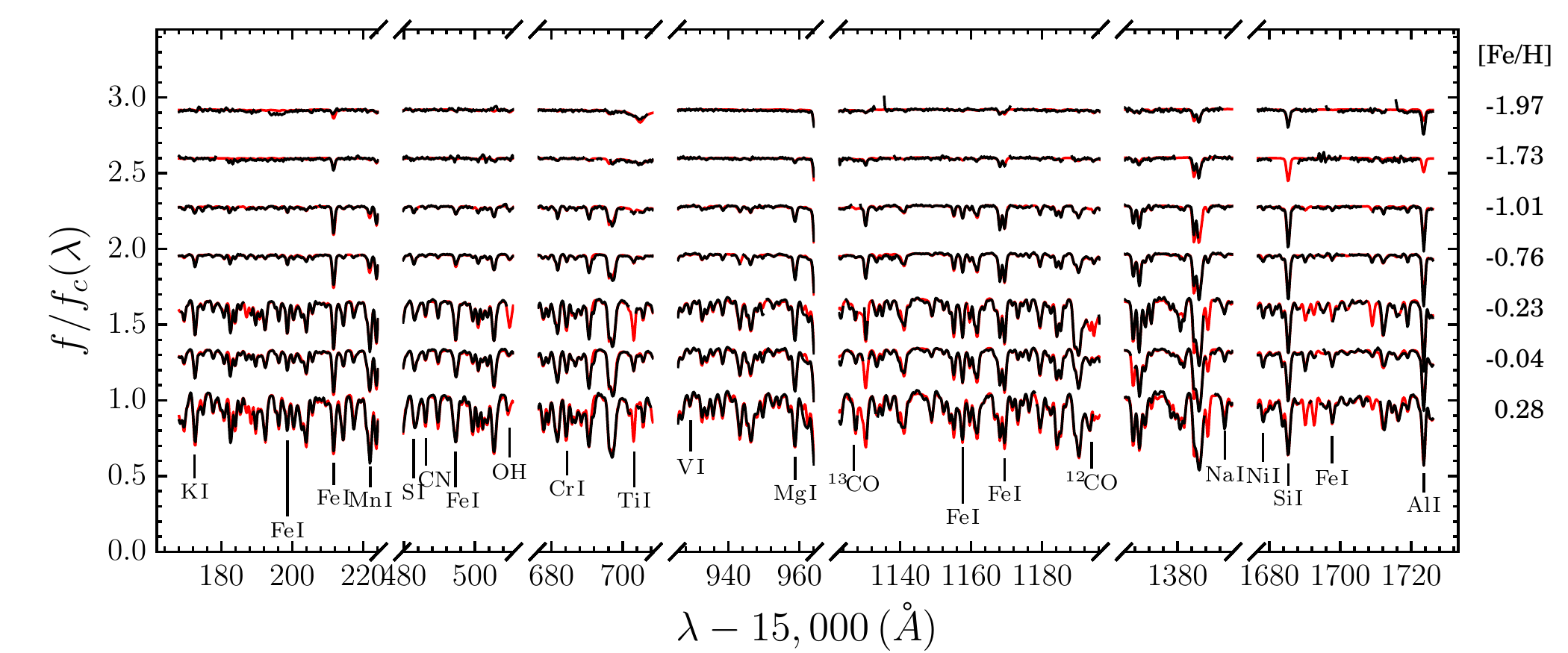}
~\\
\includegraphics[width=0.99\textwidth,
  angle=0]{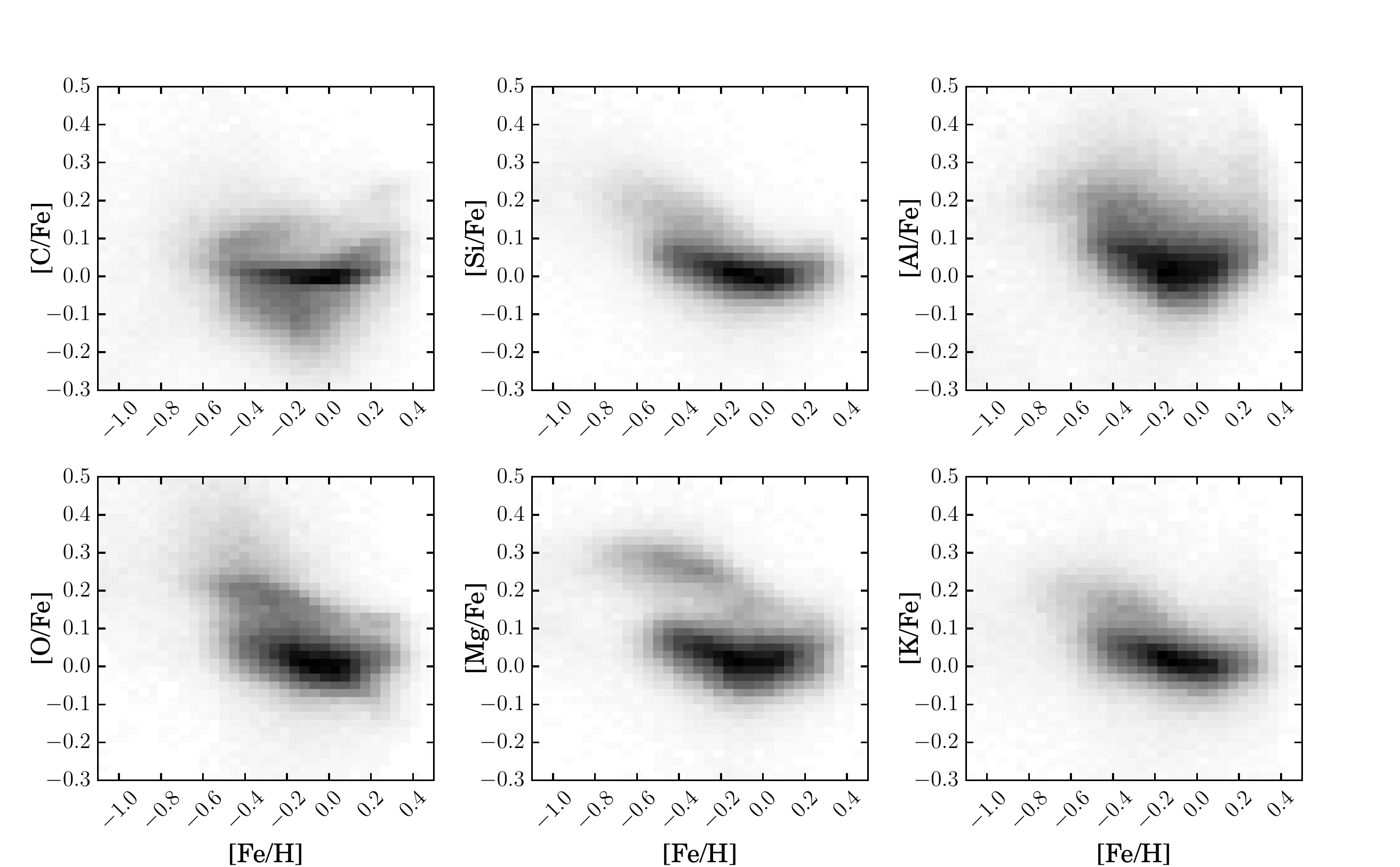}
\caption{ \label{fig:apogee2-data} Top panel: Several subregions of
  the full APOGEE spectra for seven stars of a range of metallicities,
  as labeled on the right (plotted using the software described in
  \citealt{bovy16b}). The black lines are the data; the red lines are
  the best-fit {\tt ASPCAP} model; the areas where the data are
  missing are masked due to sky contamination or other issues. Both
  data and model have been normalized to the pseudo-continuum
  $f_c(\lambda)$ (\citealt{holtzman15a}). Clean, strong lines
  identified by \citet{smith13a} are labeled.  Bottom panels:
  Elemental abundances relative to Fe for several of the species whose
  lines exist in the top panel, as a function of [Fe/H], for the
  APOGEE DR13 sample of 164,562 stars. APOGEE-2 can examine the major
  patterns as a function of Galactic location (e.g.,
  \citealt{nidever14a}, \citealt{hayden15a}).  \\ ~\\ ~\\ }
\end{figure*}

\subsection{APOGEE-2 Data}

The APOGEE-2 spectroscopic data consist of $R\sim 22,500$ spectra in
the $H$ band (1.51 $\mu$m $< \lambda < $ 1.70 $\mu$m), at high
signal-to-noise ratio ($> 100 $ pixel$^{-1}$) for most targets
\citep{majewski15a}. From these data, we determine radial velocities,
stellar parameters, and abundances. \citet{garciaperez16a},
\citet{holtzman15a}, and \citet{nidever15a} describe the APOGEE data
processing pipelines. The fundamentals remain unchanged for APOGEE-2,
and are summarized below.

The APOGEE Quicklook pipeline ({\tt apogeeql}) analyzes the
observations during each exposure to estimate the signal-to-noise
ratio and make decisions about continuing to subsequent exposures. The
observers use these data but they are not used for scientific
analysis.

Each morning, the APOGEE Reduction Pipeline ({\tt APRED}) produces
spectra for each new visit for the observed plates, extracting
individual spectra (\citealt{horne86a}).  Multiple exposures taken on
the same night are combined into ``visit'' spectra. In most cases,
multiple visits are made to each star, sometimes with the same plate
and sometimes with multiple plates. APOGEE-2 measures radial velocities from
each visit spectrum, aligns the spectra in their rest frame, and
creates a combined spectrum. 

The APOGEE Stellar Parameters and Chemical Abundance Pipeline ({\tt
  ASPCAP}) analyzes the combined spectrum.  This pipeline divides each
spectrum by a pseudo-continuum, and then performs two analyses.
First, {\tt ASPCAP} determines the key stellar parameters influencing
the spectrum --- effective temperature ($T_{\rm eff}$), surface
gravity ($\log g$), overall scaled-solar metal abundance [M/H],
$\alpha$-element abundance [$\alpha$/M], carbon abundance [C/M], and
nitrogen abundance [N/M] --- via optimization against a set of large,
multidimensional libraries of synthetic spectra (\citealt{zamora15a}).
{\tt ASPCAP} uses the FERRE\footnote{\tt
  http://github.com/callendeprieto/ferre} code to minimize $\chi^2$
differences between the pseudo-continuum-normalized spectrum and
synthesized stellar spectra interpolated from a precomputed grid
(\citealt{allendeprieto06a}). The synthetic spectra used in ASPCAP are
computed using the model atmospheres described by \citet{meszaros12a}
based on the ATLAS9\footnote{\tt
  http://www.iac.es/proyecto/ATLAS-APOGEE/} (\citealt{kurucz79a}) or
MARCS\footnote{\tt http://marcs.astro.uu.se} (\citealt{gustafsson08a})
model atmospheres. These models consider variations in carbon and the
$\alpha$ elements of $\pm 1$ dex from the solar abundance ratios. In
DR13 and DR14, the radiative transfer calculations are performed with
the code Turbospectrum (\citealt{alvarez98a, plez12a}). This code
differs from the code ASS$\epsilon$T (\citealt{koesterke09a}) used in
DR12, and includes an upgrade of the $H$-band atomic and molecular
line lists presented by \citet{shetrone15a}. In the fitting, we
usually tie the micro-turbulence ($v_{\rm micro}$) to the surface
gravity.  In the models, oxygen abundance is taken to scale with
$\alpha$.

Second, {\tt ASPCAP} performs a detailed chemical abundance
determination, conducting a series of one-dimensional parameter
searches for a set of 15 elements (C, N, O, Na, Mg, Al, Si, S, K, Ca,
Ti, V, Mn, Fe, and Ni).  For each element, a set of weighted regions
of the pseudo-continuum-normalized spectrum is compared to the models
(\citealt{garciaperez16a}).  The same underlying stellar parameter
grid is used for these searches as for the stellar parameter
determination. In each case $T_{\rm eff}$, $\log g$, and $v_{\rm
  micro}$ are fixed; only one metallicity parameter is varied. For C
and N, the [C/M] and [N/M] dimensions are varied, respectively; for O,
Mg, Si, S, Ca, and Ti, the [$\alpha$/M] dimension is varied; for Na,
Al, K, V, Mn, Fe, and Ni, the [M/H] dimension is varied. The spectroscopic
windows defined by \citet{garciaperez16a} are designed such that the
procedure in each case is sensitive primarily to the variation in the
desired element; the precise windows have changed since
DR12. Additional elemental abundances can be estimated from the
spectra and {\tt ASPCAP} is being developed over time to incorporate
these.

The {\tt ASPCAP} pipeline abundances are calibrated in several ways to
minimize systematic errors both internally and with respect to other
abundance scales. An internal temperature-dependent calibration of the
raw abundances returned by {\tt ASPCAP} is derived using the
assumption that abundances within open clusters and first-generation
stars in globular clusters (apart from C and N in giants) are
homogeneous (\citealt{desilva06a, desilva07a}). Some elements show
temperature-dependent abundance trends that are removed by this
calibration.  To improve the external accuracy, APOGEE-2 applies an
external correction that sets the median abundances of solar
metallicity stars ($-0.1<$[M/H]$<0.1$) near the solar circle to have
solar abundance ratios; this differs from DR12, where no external
correction was applied to quantities other than [M/H].  After this
calibration, most abundances have a typical precision near 0.05 dex,
though uncertainties for some elements with just a few weak lines
can be considerably larger; in detail, the precision is a function of
effective temperature, metallicity, and signal-to-noise.

The top panel of Figure \ref{fig:apogee2-data} displays several
spectra of varying metallicities from APOGEE-2 along with the best-fit
{\tt ASPCAP} model. The bottom panel presents the distribution of
several abundance ratios within the sample.

The first SDSS-IV data release (DR13; 2016 July) contains a
rereduction of APOGEE-1 data through the latest version of the
pipeline. In DR14 (summer 2017) the first two years of APOGEE-2 data
will be released.

\section{MaNGA}
\label{sec:manga}
\subsection{MaNGA Motivation} 

MaNGA is gathering two-dimensional optical spectroscopic maps
(integral field spectroscopy) over a broad wavelength range for a
sample of 10,000 nearby galaxies. In contrast, the original SDSS
Legacy survey of the nearby galaxy population, and all similar efforts
of similar scope to it, obtained single fiber spectroscopy. Single
fiber spectroscopy constrains the ionized gas content, stellar
populations, and kinematics of each galaxy, but only averaged over one
specific (typically central) region. These surveys revealed in broad
terms how the properties of galaxies, including their stellar mass,
photometric structure, dynamics, and environment, relate to their
star-formation activity and its bimodal distribution.  However, to
fully understand how galaxy growth proceeds, how star-formation ends,
and how the assembly process shapes the final observed galaxy
properties, detailed mapping of gas and stellar structure across the
entire volume of each galaxy is required. MaNGA's integral field
spectroscopic data allows study and characterization of the spatial
distribution of stars and gas as well as of the detailed dynamical
structure, including rotation, non-circular motions, and spatial maps
of higher moments of the velocity distribution function.

MaNGA is the latest and most comprehensive of a series of integral
field spectroscopic galaxy surveys of ever-increasing size. The
Spectrographic Areal Unit for Research on Optical Nebulae (SAURON;
\citealt{dezeeuw02a}), DiskMass (\citealt{bershady10a}), ATLAS$^{\rm
  3D}$ (\citealt{cappellari11a}), and the Calar Alto Legacy Integral
Field Area Survey (CALIFA; \citealt{sanchez11a}) have created a total
sample of around $1000$ well-resolved galaxies. The Sydney-AAO
Multi-object Integral field spectrograph (SAMI; \citealt{croom12a})
survey is now operating at the Anglo-Australian Observatory and plans
to observe 3400 galaxies.

MaNGA's distinguishing characteristics in this context are as
follows. First, it is the largest planned survey. Relative to CALIFA
and ATLAS$^{\rm 3D}$, the larger sample sizes of both MaNGA and SAMI
are made possible through multiplexing; by having multiple,
independently positionable IFUs across the telescope field of view,
both surveys are able to observe more than one galaxy at once, and
hence dramatically increase survey speed.  A consequence of requiring
all targets to be contained within the telescope field of view is that
both MaNGA and SAMI target more distant objects than SAURON or CALIFA,
and achieve lower physical resolution. Second, MaNGA uses the BOSS
spectrograph, which has broader wavelength coverage than SAMI, CALIFA,
or previous surveys. MaNGA is the only large integral field survey
with spectroscopic coverage out to 1 $\mu$m to allow coverage of the
calcium triplet and iron hydride features informative of stellar
populations, and [S III] emission lines from ionized gas. Third, MaNGA
covers the radial scale of galaxies in a uniform manner regardless of
mass or other characteristics; one-third of MaNGA galaxies have
coverage to at least 2.5$R_e$ and two-thirds have coverage to at least
1.5$R_e$ ($R_e$ is equivalent to the half-light radius for any profile
shape).  Finally, MaNGA has statistically well-defined selection
criteria across galaxy mass, color, environment, and redshift.

\subsection{MaNGA Science}

The primary science goal of MaNGA is to investigate the evolution of
galaxy growth. It is designed to supply critical information for
addressing four questions. (1) How are galaxy disks growing at the
present day and what is the source of the gas supplying this growth?
(2) What are the relative contributions of stellar accretion, major
mergers, and secular evolution processes to the present-day growth of
galactic bulges and ellipticals? (3) How is the shutdown of star
formation regulated by internal processes within galaxies and
externally driven processes that may depend on environment? (4) How is
mass and angular momentum distributed among different components and
how has their assembly affected the components through time?

MaNGA's resolved spectroscopy provides critical observations to
address these questions. The stellar continuum of the galaxies reveals
the star-formation history and stellar chemistry (e.g.,
\citealt{thomas03a}). Nebular emission characterizes active galactic
nuclei, star formation, and other processes
(e.g. \citealt{osterbrock06a}).  When star formation dominates the
emission, line fluxes and flux ratios indicate the rate of star
formation and the metallicity of the ionized gas around the stars
(e.g. \citealt{tremonti04a}). Both nebular emission and stellar light
provide key dynamical information related to the mass and mass profile
of the galaxies (e.g. \citealt{cappellari08a, li16a}).

The MaNGA hardware and survey are designed with the aim to constrain
the distribution of physical properties of galaxies by gathering a
sample large enough to probe the natural variation of these properties
in the three dimensions of environment, mass, and galaxy
star-formation rate. The sample size (10,000 galaxies) is justified by
the desire to resolve the variation of galaxy properties in six bins
in each of these three dimensions with about 50 galaxies in each
bin. This number of galaxies per bin is sufficient such that
differences between bins can be determined accurately.

The major areas of study for MaNGA follow from and map into the four
science questions above.
\begin{itemize}
\item Growth of galaxy disks, through the determination of
  star-formation rate surface densities and gas metallicity gradients.
\item Quenching of star formation, through star-formation rates and
  star-formation history gradients.
\item Assembly of bulges and spheroids, through star-formation
  histories and metallicity and abundance gradients.
\item The distribution and transfer of angular momentum in the stellar
  and gas components.
\item Weighing galaxy subcomponents, using the dynamically determined
  masses (from both gas and star kinematics) and the stellar masses.
\end{itemize}

The MaNGA exposure times are designed to achieve sufficient
signal-to-noise ratio spectra to address these questions.  The driving
requirements on exposure time are the precision requirements at
1.5$R_e$ on star-formation rates (0.2 dex per spatial resolution
element), stellar population ages, metallicities, and
$\alpha$-abundances (0.12 dex when averaged over an annular ring), and
dynamical mass determinations (10\%). When these goals are achieved,
other precision requirements on ionized gas and stellar population
properties necessary to study the above questions are typically
satisfied. For the majority of galaxies in the MaNGA sample, these
requirements are met by achieving the signal-to-noise ratio criteria
described below (Section \ref{sec:manga:observations}).

\subsection{MaNGA Hardware}
\label{sec:manga:hardware}

\citet{drory15a} describe the MaNGA fiber bundle technology in
detail. This technology allows precise hex-packed bundles of optical
fibers to be fed to the BOSS spectrograph. As described in Section
\ref{sec:apo}, for each of six cartridges there are 17 fiber bundles,
12 7-fiber minibundles used for standard stars, and 92 single fibers
for sky. The 17 large bundles are normally used to target galaxies and
have a range of sizes tuned to the MaNGA target galaxy distribution;
there are 2 19-fiber bundles, 4 37-fiber bundles, 4 61-fiber bundles, 2
91-fiber bundles, and 5 127-fiber bundles. Each fiber has a 120 $\mu$m
active core (2$''$ on the sky); in addition, there are 6 $\mu$m of
cladding and 9 $\mu$m of buffer, for a total diameter of 150 $\mu$m,
which defines the hexagonal spacing. When deployed, the fiber system
has high throughput (97\% $\pm$ 0.5\% in lab throughput tests).
Each fiber has a focal ratio degradation that is small and is
equivalent to the BOSS single fiber system. The overall throughput is
improved slightly relative to BOSS through the use of antireflective
coatings.

Each fiber bundle has associated sky fibers. Minibundles have
a single sky fiber, 19-fiber and 37-fiber bundles have two, 61-fiber
bundles have four, 91-fiber bundles have six, and 127-fiber bundles
have eight.  These sky fibers are constrained physically to be placed
in holes within 14$'$ of their associated IFU. This
configuration leads to sky fibers always being available close to the
science fibers both on the focal plane and on the BOSS slit head (see
\citealt{law16a}).

\begin{figure}[t!]
\centering
\includegraphics[width=0.49\textwidth,
  angle=0]{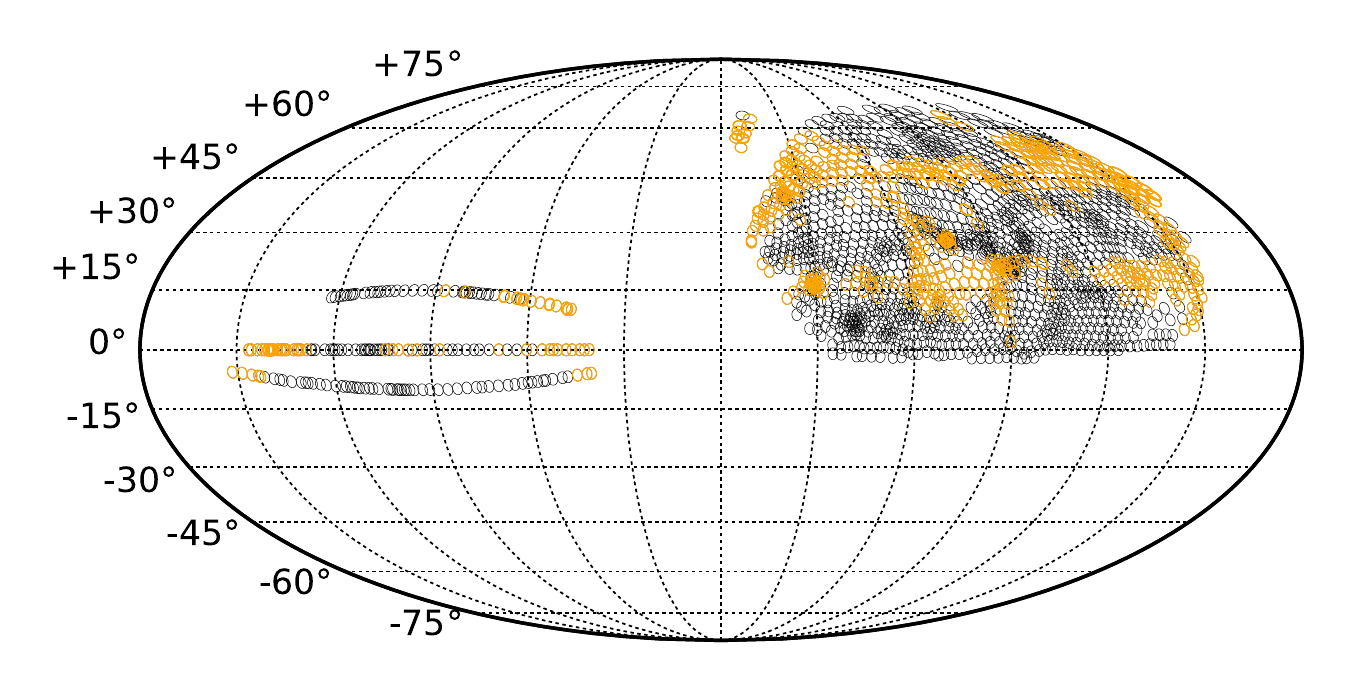}
\caption{ \label{fig:manga-footprint} Planned MaNGA spectroscopic
  footprint in equatorial coordinates, centered at $\alpha_{\rm
    J2000}=270^\circ$, with East to the left.  Black shows the
  available MaNGA tiles; orange indicates example coverage for a
  simulated SDSS-IV MaNGA survey.}
\end{figure}

\subsection{MaNGA Target Selection}

Wake et al.~(submitted) describe the galaxy targeting
strategy. The primary goals are to obtain a statistically
representative sample of 10,000 galaxies with uniform spatial
coverage, an approximately flat distribution in $\log M_\ast$, and the
maximum spatial resolution and signal-to-noise ratio with these
constraints. To ensure that the sample definition is simple and fully
reproducible, selection functions are defined in redshift, rest-frame
$r$-band absolute magnitude, rest-frame $g-r$ color, and (for the
color-enhanced sample) rest-frame NUV$-i$ color only.

MaNGA selects galaxies from the NASA-Sloan Atlas (NSA;
\citealt{blanton11a}), which is based on the Main Galaxy Sample of
\citet{strauss02a} but includes a number of nearby galaxies without
SDSS spectroscopy and incorporates better photometric analysis than
the standard SDSS pipeline.  The version of NSA used ({\tt v1\_0\_1})
is limited to galaxies with $z<0.15$.  For selection and targeting
purposes, $R_e$ is defined in the MaNGA survey as the major-axis
elliptical Petrosian radius in the $r$ band.  Galaxies are matched to
IFUs of different size based on this $R_e$ value and the effective
size of the IFU.

MaNGA target selection is limited to the redshift range
$0.01<z<0.15$. We seek an approximately flat stellar mass
distribution, and to cover most galaxies out to a roughly uniform
radius in terms of $R_e$. Achieving these goals requires targeting
more luminous, and consequently intrinsically larger, galaxies at
larger redshifts.  MaNGA defines three major samples across the
footprint of the Main Sample of galaxies from the SDSS-II Legacy
Survey; about one-third of this full sample is targeted for
observation. The observed sample is to include the following.
\begin{itemize}
\item 5000 Primary galaxies: selected in a narrow band of rest-frame
  $i$-band luminosity and redshift such that 80\% have coverage out to
  $1.5 R_e$.
\item 1700 Color-enhanced galaxies: selected according to $i$-band
  luminosity and redshift as for Primary, but with a well-defined
  upweighting as a function of NUV$-i$ color to better sample the
  rarer colors. The Primary and the Color-enhanced sample together
  are referred to as the Primary+ sample.
\item 3300 Secondary galaxies: selected in a band of rest-frame
  $i$-band luminosity and redshift, somewhat higher redshift relative
  to Primary, such that 80\% have coverage out to $2.5R_e$.
\end{itemize}
The Primary sample has a median redshift of $\langle
z\rangle\sim 0.03$, whereas the Secondary sample is at a larger median
redshift $\langle z\rangle\sim 0.05$. 

These targets are defined over most of the 7800 deg$^{2}$ area of the
SDSS Main Galaxy Sample, which is a large contiguous region in the NGC
and three 2.5$^\circ$ stripes in the SGC. Since the density of MaNGA
target galaxies varies substantially over the sky, Wake et
al.~(submitted) have designed the potential field locations to adjust
to cover the dense regions more densely, using a version of the
algorithm described by \citet{blanton03a}.  Figure
\ref{fig:manga-footprint} shows these potential locations as black
circles (each 1.5$^\circ$ in radius). As in eBOSS, each pointing is
referred to as a tile, which is typically associated with a single
physical plate.  MaNGA will be able to observe about one-third of the
available tiles during its six years of operations. Figure
\ref{fig:manga-footprint} shows a simulated projection of this
coverage (depending on weather patterns).

For each plate, minibundles are associated with standard stars, which
are F stars selected similarly to those in eBOSS and are used for
spectrophotometric calibration (\citealt{yan16a}). The sky fibers
associated with each bundle are assigned to locations that are empty
in SDSS imaging.

In addition, MaNGA is targeting a set of ancillary targets observed in
fields for which the above samples do not use all the bundles. These
ancillary samples are described in the data release papers
(e.g. for DR13 in \citealt{albareti16a}).

\begin{figure*}[t!]
\centering
\includegraphics[width=0.98\textwidth,
  angle=0]{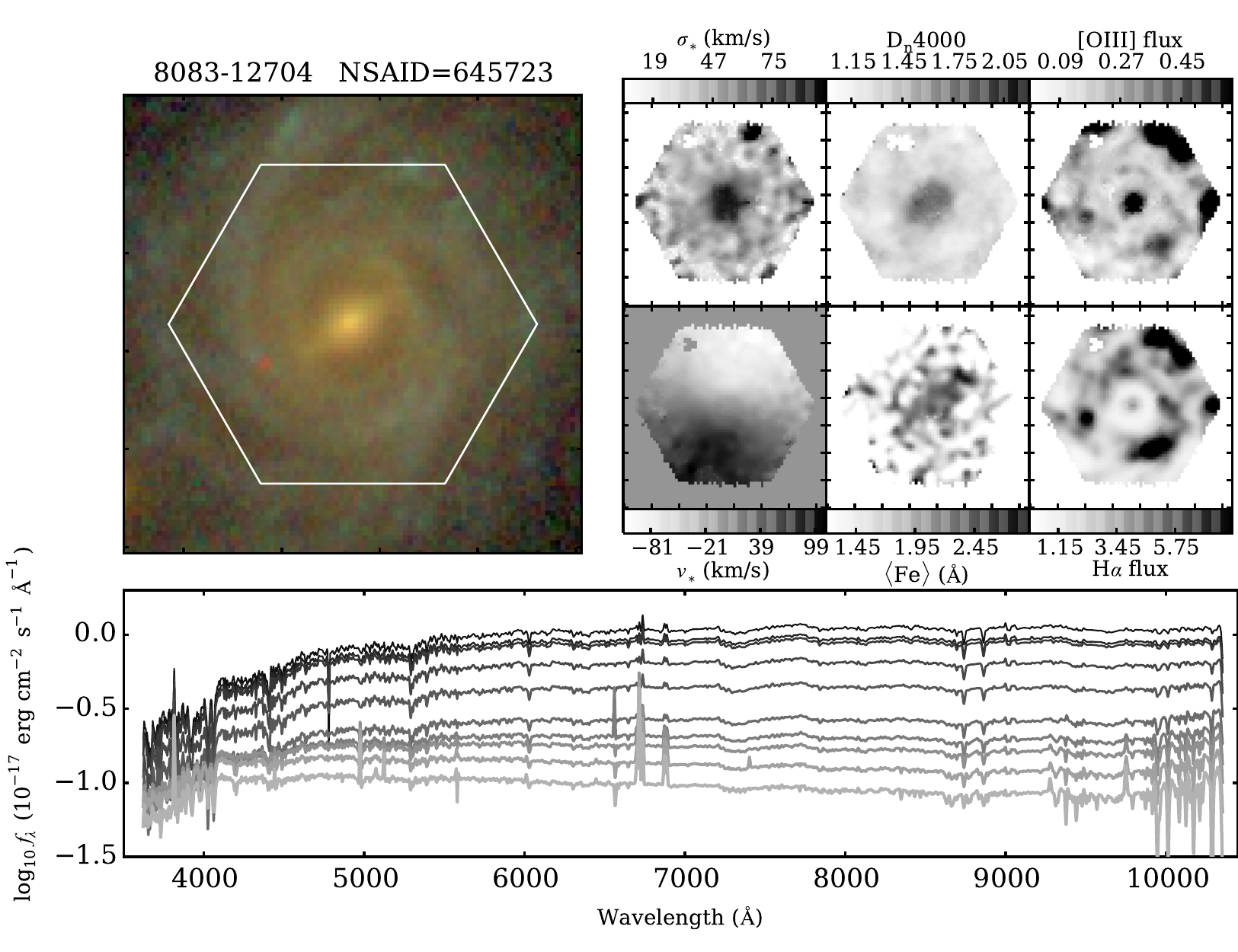}
\caption{Top left: image of a MaNGA target (UGC 02705) from SDSS, with
  MaNGA 127-fiber bundle footprint overlaid (37$''$ $\times$ 37$''$).
  Top right: maps of derived quantities from the DAP pipeline: stellar
  velocity dispersion $\sigma_\ast$, stellar mean velocity $v_\ast$,
  the stellar population age indicator $D_n$4000, the metallicity
  indicator $\langle$Fe$\rangle = 0.5($Fe5270$+$Fe5335$)$, the [OIII]
  $\lambda$5007 flux in $10^{-17}$ erg cm$^{-2}$ s$^{-1}$, and the
  H$\alpha$ flux in the same units.  Bottom: sum of MaNGA spectra in
  elliptical annuli of increasing radii.
\label{fig:manga-data}}
\end{figure*}

\subsection{MaNGA Observations}
\label{sec:manga:observations}

MaNGA utilizes approximately 50\% of the dark time at APO. Details of
the division of observations across the SDSS-IV surveys are given in
Section \ref{sec:apo}.

Each MaNGA fiber bundle is encased in a small metal ferrule 20 mm in
length, which protects the bundle and contains a pin for keeping the
ferrule in constant alignment on the plate. The resulting ferrule is 7
mm in diameter, larger than that for individual eBOSS or APOGEE-2
fibers. This constraint prevents two fiber bundles on the same plate
from being closer than about 116$''$.

The fiber bundles do not optimally sample the typical atmospheric and
telescope point spread function. To provide better sampling, each
plate is observed in a set of three successive 15 minute exposures
offset from each other by 1.44$''$ in a triangular pattern on the sky
(\citealt{law15a}). Typically, these dithered exposures are all taken 
in succession to make sure a full set exists for each plate and night.

Each plate is designed for a specific hour angle of observation and is
observable over a certain visibility window, as described in
\citet{law15a}. The window is defined according to how quickly the
position of the IFU shifts in sky coordinates due to differential
refraction across the field (accounting for the telescope's guiding
adjustments). The condition is that the maximum shift at any
wavelength for an IFU at any location on the plate over an hour
duration is 0.5$''$ or less.  If a dither set is begun within any part
of the observing window, all subsequent dithers must be taken at
similar hour angles in order to be combined, such that they are all
within an hour of each other (the data do not have to be taken on the
same night).

MaNGA requires a signal-to-noise ratio of 5 \AA$^{-1}$ fiber$^{-1}$ in
the $r$-band continuum at a Galactic extinction corrected $r$-band
surface brightness of 23 mag arcsec$^{-2}$ \citep[AB
  magnitude;][]{oke83a}. This goal is achieved by setting a threshold
for determining whether the plate is complete as follows for the blue
and red BOSS spectrograph data. We do so using the $(S/N)^2$ per
spectroscopic pixel summed across exposures. A plate is deemed
complete when this $(S/N)^2$ exceeds a threshold at a fiducial $g_{\rm
  fiber2}$ and $i_{\rm fiber2}$ (these are magnitudes from SDSS DR13
imaging \citep{ahn12a} within a 2$''$ diameter aperture convolved with
2$''$ FWHM seeing).  For Galactic extinction corrected $g_{\rm
  fiber2}=22$, the threshold is $(S/N)^2>20$ in the blue spectrograph.
For Galactic extinction corrected $i_{\rm fiber2}=21$, the threshold
is $(S/N)^2>36$ in the red spectrograph.  Typically three sets of
dithers (nine total exposures) are required for completion; in regions
of greater Galactic extinction more than three sets are required.
Usually, only two sets can be taken in succession while still
satisfying the hour-angle criteria described above. Observations of
the same plate are therefore typically split across nights.

For some sets, if the observing conditions are changing rapidly, some
dithers are good quality but others are not. The good-quality dithers
in this situation are considered ``orphan'' exposures since they
cannot be easily combined with exposures in other sets. These good
exposures are processed but are not included in the reconstructed data
cubes because they would lead to non-uniform images. Major changes in
the reduction procedure might allow a more efficient use of these
otherwise good-quality observations. Doing so is not in the pipeline
development plans; nevertheless, the fully calibrated row-stacked
spectra are available for such analysis.

In the mean, each plate requires 3.3 sets of 3 exposures, or about 2.5
hr of open shutter time. Each set requires 20 minutes overhead in
cartridge changes, calibrations, and field acquisition. The orphaned
exposures produce an additional 10\% loss in efficiency.

In addition to the galaxy survey, MaNGA uses their IFUs for the
development of a new optical stellar library (the MaNGA Stellar
Library, or MaSTAR). Because MaNGA IFUs share cartridges with APOGEE
fibers, during APOGEE-2N time the MaNGA IFUs are placed on MaSTAR
targets. These observations are not dithered. The MaSTAR library
provides several advantages over existing libraries. Totaling around
6,000 stars, MaSTAR is several times larger than previous efforts,
including those few that span a comparable spectral range, e.g.,
STELIB (\citealt{leborgne03a}) or INDO-US (\citealt{valdes04a}).  Its
target selection utilizes stellar parameter estimates from
\mbox{APOGEE-1} (\citealt{garciaperez16a}), SEGUE
(\citealt{allendeprieto08a}), and LAMOST (\citealt{lee15b}) to better
cover underrepresented ranges of parameter space of effective
temperature, surface gravity, metallicity, and abundance.  While the
Milky Way imposes certain practical limits, say, on the available
dynamic range in age and abundance, there are known significant gaps
in parameter coverage, e.g., at low temperatures for both dwarfs and
giants, and at low metallicity, that MaSTAR is be able to fill.  While
SEGUE (\citealt{yanny09a}) sampled a large number of stars over a
range of spectral types and surface gravities, their goal of broadly
studying the kinematics and stellar populations of our Galaxy did not
lead to an adequate sampling of some of these regions of parameter
space where stars in the Milky Way are rare in the magnitude ranges
probed. MaSTAR is the first stellar library of significant size with
wavelength coverage from 3600 \AA\ to beyond 1 $\mu$m. Finally, for
the purposes of stellar population synthesis of MaNGA galaxies, using
an empirical library with the same instrument minimizes systematics in
resolution mismatch and offers significant improvements and
consistency in spectrophotometry.

\subsection{MaNGA Data}
\label{sec:manga:data}

MaNGA spectroscopic data consists of $R\sim 2,000$ spectra in the
optical (approximately 3600 \AA $< \lambda <$ 10,350 \AA), at
signal-to-noise ratios of at least 5 per pixel, spatially resolved
across galaxies at $\sim 2.5''$ resolution FWHM, from which we create
maps of velocities, velocity dispersion, line emission, and stellar
population indicators. MaNGA data are processed using a pipeline
derived from and similar to that used for eBOSS, and utilizing similar
infrastructure.

MaNGA data are processed through a quicklook pipeline (Daughter Of
Spectro; DOS) during each observation to estimate the signal-to-noise
ratio in real time and make decisions about continuing to subsequent
exposures. Quality assurance plots are studied each day to identify
unexpected failures of the observing system or pipelines.

A Data Reduction Pipeline (DRP; \citealt{law16a}) reduces the single
fibers in each exposure into individual spectra using optimal
extraction. This pipeline is similar to and shares a code base
with the pipeline that processes BOSS spectrograph data
(\citealt{bolton12a}).  There is a subtle difference in the sky
estimation. As in BOSS and eBOSS, all fibers are used to define the
model sky spectrum; however, this model spectrum can be scaled in the DRP to
match the local sky background near each IFU.  A second and more
fundamental difference is the spectrophotometric calibration
procedure.  An important factor in the single fiber eBOSS
spectrophotometric calibration is the wavelength-dependent loss due to
atmospheric differential refraction (ADR; for a detailed discussion,
see \citealt{margala16a}).  However, for MaNGA, this effect is better
interpreted as a variation with wavelength of the effective location
of the fiber center on the sky; i.e., the blue light samples a slightly
different part of the galaxy than the red light. Loosely speaking,
light is no longer ``lost'' from a given fiber due to ADR, but instead
shifted toward a neighboring fiber.  Thus, the spectrophotometric
correction should not include ADR losses. As \citet{yan16a} describe,
the correction is performed using standard stars observed through 7-fiber
minibundles, which allow for the geometric effects to be disentangled
from the effective throughput of the system. 
The DRP produces a set of wavelength and flux calibrated ``row
stacked spectra'' for each exposure.

In the second stage of processing, the DRP associates each fiber in a
given exposure with its effective on-sky location using the
as-measured fiber bundle metrology in combination with the known
dither offsets and a model for the ADR and guider corrections.  This
astrometry is further refined on a per-exposure basis by comparing the
fiber fluxes to reference broadband imaging in order to correct small
rotations and/or offsets in the fiber bundle location from the
intended position.  The DRP then uses a flux-conserving variation of
Shepard's method \citep{sanchez12} to interpolate the row-stacked spectra onto a
three-dimensional data cube with regularly spaced dimensions, one in
wavelength and two Cartesian spatial dimensions.  Details on the DRP
can be found in \citet{law16a}.

Based on the row-stacked spectra and data cubes, a Data Analysis
Pipeline (DAP) calculates maps of derived quantities such as Lick
indices \citep[e.g.,][]{worthey94}, emission-line fluxes, and
kinematic quantities such as gas velocity, stellar velocity, and
stellar velocity dispersion. The list of calculated quantities remains
under development.  Future plans for DAP include deriving high-level
quantities such as stellar mass and abundance maps, metallicity maps,
and kinematic models.

Figure \ref{fig:manga-data} shows some typical MaNGA data for
UGC~02705, for which observations through a 127-fiber bundle finished
on 2014 October 26.

The first SDSS-IV data release (DR13; 2016 July) contains MaNGA
results data taken through 2015 July.  In DR14, the MaNGA data through
2016 May will be released.

\section{eBOSS, TDSS \& SPIDERS}
\label{sec:ets}
eBOSS, TDSS, and SPIDERS are three surveys conducted simultaneously at
APO on the 2.5~m telescope during dark time using the 1000
single-fiber configuration with the BOSS spectrograph. The overall
survey strategy is driven by eBOSS, which is the largest program. TDSS
and SPIDERS each use approximately 5\% of the fibers on each eBOSS
plate.  Table \ref{table:eboss_tdss_spiders} summarizes the three
programs.

\begin{table}[htp]
\caption{
\label{table:eboss_tdss_spiders} Target classes in eBOSS, TDSS,
and SPIDERS }
\begin{tabular}{lccr}
\hline\hline
Program & Target Class & Area (deg$^2$) & Spectra \\
\hline
eBOSS & LRG & 7500 & 266,000 \\
eBOSS & New Quasar tracers & 7500 & 400,000 \\
eBOSS & Total Quasar tracers & 7500 & 500,000 \\
eBOSS & New Ly$\alpha$ quasars & 7500 & 60,000 \\
eBOSS & Repeat Ly$\alpha$ quasars & 7500 & 60,000 \\
eBOSS & ELG & 1000--1500 & 200,000 \\
eBOSS & ``Contaminants''\tablenotemark{a} & 7500 & 320,000 \\ 
TDSS & PS1/SDSS Variables (total) & 7500 & 200,000 \\
TDSS & Few-epoch spectra & 7500 & 10,000 \\
TDSS & Repeat quasar spectra & 1000--1500 & 16,000 \\
SPIDERS & Point sources (total) & 7500 & 22,000 \\
SPIDERS & Cluster galaxies (total) & 7500 & 60,000 \\
\hline
\end{tabular}
\tablenotetext{a}{High-quality redshifts outside the range of
interest.}
\end{table}

\subsection{eBOSS}
\label{sec:eboss}
\subsubsection{eBOSS Motivation}
\label{sec:eboss:motivation}

eBOSS is conducting cosmological measurements of dark matter, dark
energy, and the gravitational growth of structure.  Current data from
other large-scale structure measurements, Supernovae Type Ia, and the
cosmic microwave background are consistent with a spatially flat cold
dark matter model and a cosmological constant
($\Lambda$CDM; \citealt{weinberg13a, aubourg15a}). The cosmological
constant or some other mechanism is required due to the observed
late-time acceleration in the cosmic
expansion \citep[e.g.,][]{riess98, perlmutter99a}.

The cosmological constant can be generated through a nonzero, but very
small, vacuum energy density; however, the particle physics mechanism
to generate this level of vacuum energy is unknown. The acceleration
could also be caused by some more general fluid with negative
pressure, referred to typically as ``dark energy;'' the equation of
state of this fluid is constrained to be fairly similar to that of the
vacuum energy. Alternatively, the acceleration may be caused due to
modifications of general relativity that affect gravity at large
scales (e.g. \citealt{randall99a, dvali00a, sahni03a, sotiriou10a,
battye12a}). Many of these explanations of the acceleration are
theoretically plausible, and the challenge is to observationally bound
the possibilities. One critical constraint arises from precisely
measuring the rate of expansion and gravitational growth of structure
throughout all cosmic epochs.

eBOSS is creating the largest volume map of the universe usable for
large-scale structure to date. This data set will allow exploration of
dark energy and other phenomena in epochs where no precision
cosmological measurements currently exist, pursuing four key goals:
BAO measurements of the Hubble parameter and distance as a function of
redshift, redshift space distortion measurements of the gravitational
growth of structure, constraints on and possible detection of the
neutrino mass sum, and constraints on inflation through measurements
of non-Gaussianity.

Among currently operating experiments, only the Hobby-Eberly Telescope
Dark Energy Experiment (HETDEX; \citealt{hill08a}) and the Dark Energy
Survey (DES; \citealt{abbott16a}) will measure the universe's
expansion history at comparable precision and accuracy. HETDEX is a
wide-field integral field spectrograph survey that will map \lya\
emitting objects at $z\sim 2$--$3$. DES is an imaging survey that will
measure BAO as a function of redshift using angular clustering and
photometric redshifts.  Future spectroscopic experiments are planned
that will exceed the precision in measuring expansion of any current
program. These experiments include DESI (\citealt{levi13a}) and the
Prime Focus Spectrograph at {\it Subaru} (PFS; \citealt{takada14a}). eBOSS's
large-scale structure results precede the beginning of either of these
experiments and is poised to deliver the first accurate measurements
of expansion in the redshift range $1<z<2$.

\subsubsection{eBOSS Science}
\label{sec:eboss:science}

The primary cosmological constraints from eBOSS are BAO measurements
of the angular diameter distance $D_A(z)$ relative to that of the CMB,
and the Hubble parameter $H(z)$ as a function of
redshift. \citet{weinberg13a} includes a recent review of this
technique. The LRG, ELG, and low-redshift quasar samples are used as
tracers to measure BAO in large-scale structure; the high-redshift
quasar sample is used for \lya\ forest measurements of BAO in the
neutral gas clustering. These measurements in real and redshift space
yield constraints on the Hubble parameter $H(z)$ and the angular
diameter distance $D_A(z)$, which can be combined into a constraint on
a combined distance $R(z)$. Full details on the definition of these
quantities, and projections regarding the precision on BAO from eBOSS
can be found in \citet{dawson16a} and \citet{zhao16a}.
Table \ref{table:eboss_samples} summarizes the expected precision from
the LRG, ELG, quasar, and \lya\ samples.  In terms of the Dark Energy
Task Force (DETF) Figure of Merit (FoM; \citealt{albrecht06a}), the
eBOSS sample improves the FoM over the existing constraints to date by
a factor of three. These projections assume only measurements of the
BAO feature itself. Addition of the broadband power spectrum, redshift
space distortions, and geometric distortions is expected to produce a
further increase in the FoM \citep{mcdonald09a}, though with greater
theoretical systematics.

\begin{table}[t!]
\caption{
\label{table:eboss_samples} Cosmological precision in eBOSS}
\begin{tabular}{lccccc}
\hline\hline
Target Class & $z$ & $\sigma_H/H$ & $\sigma_{D_A}/D_A$
& $\sigma_R/R$ & $\sigma_{f\sigma_8}/f\sigma_8$\tablenotemark{a}\\
\hline
LRG\tablenotemark{b} & 0.71 & 0.025 & 0.016 & 0.010 & 0.025 \\
ELG\tablenotemark{c} & 0.86 & 0.050 & 0.035 & 0.022 & 0.034 \\
Quasar & 1.37 & 0.033 & 0.025 & 0.016 & 0.028 \\
Ly-$\alpha$ & 2.54 & 0.014 & 0.017 & --- & --- \\
\hline
\end{tabular}
\tablecomments{
Results derived from \citet{zhao16a}. 
}
\tablenotetext{a}{$f\sigma_8$ forecasts use assumptions similar to the
model-independent constraints cited in
Section \ref{sec:eboss:science}, holding other cosmological parameters
fixed.}
\tablenotetext{b}{Includes LRGs observed in SDSS-III within the overlapping
redshift range.}
\tablenotetext{c}{Numbers correspond to the ``high density'' ELG sample
in \citet{zhao16a}, which is close to the current
plan. }
\end{table}

Redshift space surveys, as opposed to imaging surveys, yield a unique
additional constraint on cosmology; since galaxy motions reflect the
gravitational growth of structure, measuring the anisotropic
distortion they produce in clustering yields constraints on cosmological
parameters and general relativity (GR) (\citealt{weinberg13a}). In the
context of cosmic acceleration, clustering measurements can
distinguish between models for acceleration that rely on dark energy
and those that require modified gravity (\citealt{huterer15a}).  This
measurement yields $f\sigma_8$, where $f$ measures the growth rate and
$\sigma_8$ measures the amplitude of matter fluctuations. Currently
the most robust constraints on $f\sigma_8$ are from BOSS, with
large-scale model-independent constraints of $\sim 6$\% (9\% when
marginalizing over other parameters;
\citealt{beutler14a, samushia14a, alam15a}) and model-dependent
constraints on smaller scales of 2.5\% (\citealt{reid14a}). These
critical tests distinguishing dark energy and modified gravity models
are possible only with a spectroscopic redshift program such as eBOSS.

The fundamental properties of neutrinos are imprinted in the
distribution of galaxies. eBOSS's large volume permits tight new
constraints on, and perhaps finally allows for a measure of, the
neutrino mass.  Flavor oscillation measurements place lower limits on
the neutrino masses of $0.05$--$0.10$~eV depending on the model
(\citealt{fogli12a}).  Cosmological observations place upper limits on
the sum of neutrino flavor masses, due to the suppression of power by the
neutrino component in fluctuations at scales smaller than 100 Mpc. The
best existing cosmological constraint is that $\sum m_\nu < 0.23$ eV
(95\% confidence, when assuming zero curvature; \citealt{planckXVI14}),
from CMB measurements and BAO.  Adding eBOSS constraints from the LRG,
ELG, and $z<2.2$ quasars improves this limit to $\sum m_{\nu}
<0.108$ eV, close to the minimum allowed neutrino mass in conventional
particle physics theories.  eBOSS clustering data therefore have a
significant chance of measuring the neutrino mass sum, which would be
a major breakthrough in fundamental physics.

eBOSS pioneers tests of cosmic inflation through the measurement of
very-large-scale fluctuations.  Departures from the standard
inflationary scenario commonly yield small deviations from Gaussian
fluctuations, quantifiable by \fnl ($=0$ for Gaussian). A natural form
of non-Gaussianity (the ``local'' form; \citealt{wands10a}) can be
tested using two-point statistics at $>200$ Mpc
(\citealt{dalal08a}). eBOSS yields the only constraints ($\sigma_{\rm
fnl}=12$) comparable in precision to (but completely independent of)
current Planck limits (local
$\fnl=2.5\pm5.7$; \citealt{ade16a}). Furthermore, galaxy bispectrum
measurements have the potential to improve eBOSS constraints
dramatically.  Future improvements will likely be best achieved with
redshift surveys such as eBOSS.

eBOSS yields the largest existing statistical sample available for a
broad array of other science topics.
\begin{itemize}
\item Galaxy formation and evolution through interpretation of the
small-scale correlation functions (\citealt{zheng07a, leauthaud12a,
guo13a}).
\item Evolution of the most luminous galaxies out to \mbox{$z\sim 1$}
(e.g., \citealt{maraston13a, bundy15b, monterodorta16a}).
\item Nature of the circumgalactic medium through statistical absorption
studies (\citealt{steidel10a, zhu14a, zhu15a}).
\item Calibration of photometric redshifts through cross-correlation;
eBOSS provides this calibration for DES and validates this method
for use in future surveys such as LSST (\citealt{newman15a}).
\item Nature of the intergalactic medium in the range
$2<z<3.5$, using the damped Lyman$alpha$ systems, Lyman limit systems
, and the Lyman-$\alpha$ and Ly$\beta$ forests and their
cross-correlations with other tracers of structure.
(e.g. \citealt{becker13a, pieri14a, lee15a}). These techniques can
reveal signatures of He II reionization, the clustering of ionizing
sources, and can potentially detect Ly$\alpha$ emission.
\end{itemize}

We will discuss the quasar science in more detail in
Section \ref{sec:quasars}.

\subsubsection{eBOSS Targeting Strategy}
\label{sec:eboss:targeting}

\begin{figure}[t!]
\centering
\includegraphics[width=0.49\textwidth,
  angle=0]{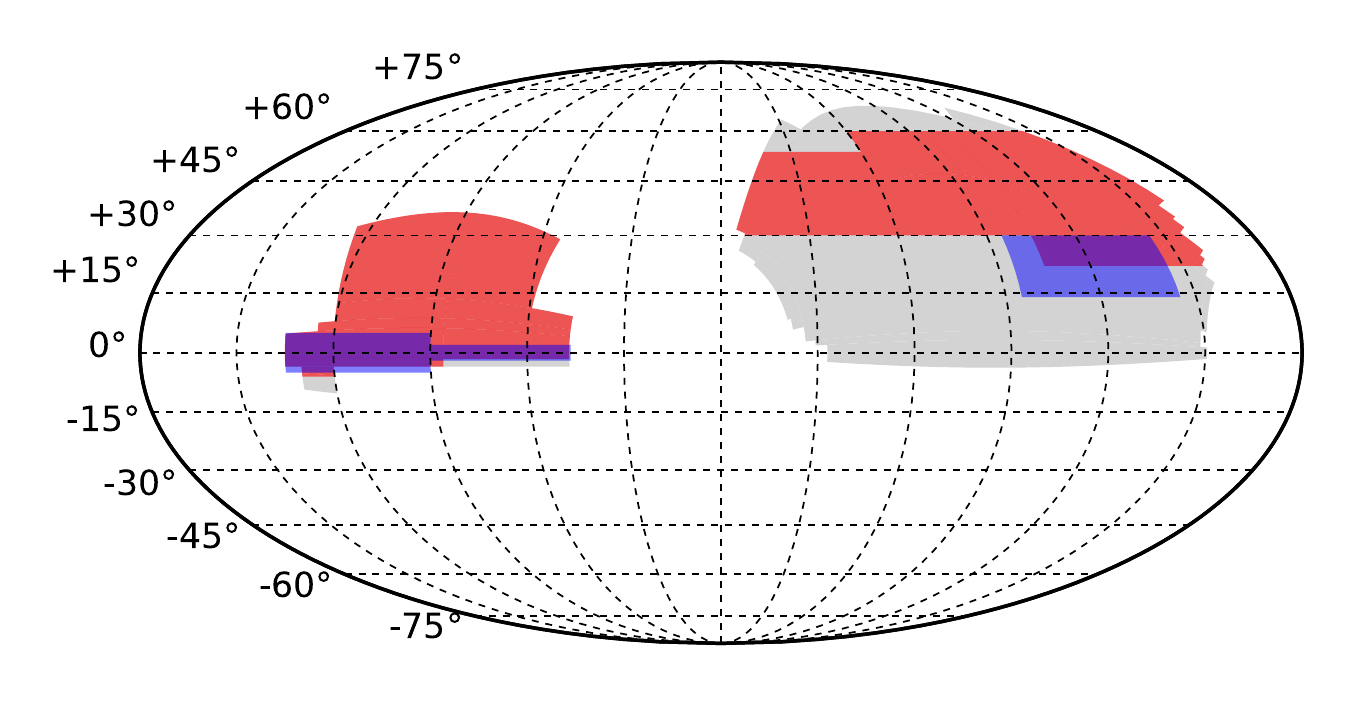}
\caption{ \label{fig:eboss_footprint}
Planned eBOSS spectroscopic footprint in equatorial coordinates,
centered at $\alpha_{\rm J2000} = 270^\circ$, with East to the left.
Grey areas are the BOSS spectroscopic footprint, and for eBOSS red
represents the planned LRG and quasar sample footprint, and blue shows
the planned ELG footprint.}
\end{figure}

\citet{dawson16a} presents an overview of the eBOSS targeting strategy, 
which aims primarily at surveying a large volume of the universe. The
eBOSS footprint covers 7500 deg$^{2}$, with approximately 4500
deg$^{2}$ in the North Galactic Cap (NGC) and 3000 deg$^{2}$ in the
South Galactic Cap (SGC). Luminous red galaxies (LRGs) and quasars
are targeted over the full eBOSS footprint.  An emission-line
galaxy (ELG) sample is targeted over 1000--1500 square degrees
starting in Fall 2016.  A 466 deg$^2$ pilot program was conducted in
SDSS-III and early SDSS-IV, designated the Sloan Extended Quasar, ELG,
and LRG Survey (SEQUELS; \citealt{dawson16a, alam15b}). SEQUELS tested
these target selection techniques. Figure \ref{fig:eboss_footprint}
shows the the currently planned eBOSS footprint, and
Table \ref{table:eboss_samples} summarizes the planned eBOSS samples
and the resulting cosmological constraints.

The targeting strategy is driven by a desire to fill the existing gap
in cosmological large-scale structure measurements between $z\sim 0.6$
and $z\sim 2.5$, which is the transition from cosmic deceleration to
acceleration. With existing facilities, this range cannot be covered
over wide fields using a single tracer. Thus, we adopt a multi-tracer
strategy: extend the BOSS LRG sample to $z\sim 0.8$, introduce an
emission-line galaxy sample that can be selected and successfully
observed to $z\sim 1.1$, conduct a dense survey of quasars to $z\sim
2.2$, and enhance the BOSS quasar sample at $z>2.2$.
 
The full quasar sample is designed to cover $0.9<z<3.5$. The quasars
at redshifts $z<2.2$ are utilized as tracers of large-scale structure
themselves.  The quasars at $z>2.1$ are utilized as backlights for
\lya\ absorption, which measures the density of neutral gas
along the line of sight at those redshifts.  The core quasar target
selection is described by \citet{myers15a}, utilizing a redshift-binned
version of the Extreme Deconvolution (XD) algorithm applied to quasars
(XDQSOz; \citet{bovy11a, bovy12b}).  In the SDSS-IV case, we apply
XD on the SDSS photometry and its associated uncertainties to select
quasars, and then consult {\it WISE} photometry to veto sources likely to be
stars.  We do not observe quasars at $z<2.1$ that were
spectroscopically classified in prior SDSS surveys (which have a
density $\sim 13$ deg$^{-2}$), but these are included in
clustering analyses. eBOSS re-observes the fainter quasars at $z>2.1$ to
improve the signal-to-noise ratio in the \lya\ forest by a factor of
1.4.

The LRG sample is designed to cover $0.6<z<1.0$, with a median $z\sim
0.71$. eBOSS achieves this selection using a combination of SDSS $r$,
$i$, and $z$ photometry and {\it WISE} 3.4 $\mu$m photometry, as
described by
\citet{prakash15a}. The sample is limited at $z<19.95$ (using
Galactic extinction corrected SDSS model magnitudes).

The ELG sample is designed to cover $0.7<z<1.1$, with a median $z\sim
0.86$ \citep{comparat16a, jouvel16a}. The selection uses the deep $g$,
$r$, and $z$ band imaging from the Dark Energy Camera
(DECam; \citealt{flaugher12a}). The imaging is primarily drawn from a
combination of DES imaging and of the DECam Legacy Survey
(DECaLS\footnote{{\tt http://legacysurvey.org}}), a wide footprint
extragalactic imaging survey being conducted in preparation for DESI.
The ELG targets are observed at a high density ($>180$ deg$^{-2}$)
over 1000--1500 deg$^{2}$ split about equally between the SGC and
NGC. Because of the available imaging depth, the target density in the
SGC is high ($\sim 240$ deg$^{-2}$) and the efficiency of selecting
ELGs in the desired redshift range is around 80\%, whereas the density
($\sim 190$ deg$^{-2}$) and efficiency ($75\%$) are lower in the
NGC. In both regions, the median redshift is similar. These targets are
observed on separate plates from the LRG and quasar cosmological
surveys. These plates do not contain SPIDERS targets, but, as
described in Section \ref{sec:tdss}, they do include Repeat Quasar
Spectroscopy targets. ELG observations began in Fall 2016. A future
paper will describe the exact selection function, its redshift
distribution, as well as systematic weights to be applied for
large-scale structure analysis.

The eBOSS team also considered the use of other imaging data sets.  In
SEQUELS, \citet{comparat15a} drew ELG targets from the South Galactic
Cap U-band Sky Survey (SCUSS; \citealt{zou15a}) and SDSS. In the last
round of tests before the ELG program was
finalized, \citet{comparat16a} and \citet{raichoor16a} combined {\it WISE}
(\citealt{wright10a}), SCUSS, and SDSS to select ELG targets. The
final selection functions are nearly as efficient as the DECaLS
targeting but yielded a lower effective redshift.

For the LRG, ELG, and quasar clustering samples, eBOSS aims to create
uniform target selection with a maximum absolute variation (peak to
peak) of 15\% in the expected target number density. The expected
target number density is defined with respect to its estimated
dependence on imaging survey sensitivity, calibration errors, stellar
density, and Galactic extinction (\citealt{myers15a, prakash15a,
dawson16a}).

The targets are assigned to plates using a descendant of the tiling
algorithm adopted in the Legacy and BOSS surveys
(\citealt{blanton03a}). The eBOSS pointings are designed to cover
large contiguous areas in the NGC and SGC. Each pointing is referred
to as a tile, which typically (but not always) is associated with a
single physical plate.  Of the 1000 available fibers, 80 are assigned
to estimate the sky and 20 are assigned to bright $F$ stars used as
standard sources.  The TDSS and SPIDERS programs are included in the
tiling assignments and observed on the same plates as the eBOSS
targets.

eBOSS adopted a tiered-priority system for assigning survey targets to
plates, which leads to an efficient assignment of fibers and a
satisfactory level of completeness. All non-LRG targets receive
maximal priority and the tiling solution must achieve 100\% tiling
completeness for a set of all non-LRG targets that do not collide with
each other (a ``decollided'' set; see
\citealt{blanton03a}). For LRGs, eBOSS does not require full
decollided completeness. Rather, the density of LRG targets
intentionally oversubscribes the remaining fiber budget.  The average
density of LRGs assigned to fibers spectra is about 50 deg$^{-2}$.  In
areas of lower density in non-LRG targets, the LRGs can be observed up
to a density of about 60 deg$^{-2}$.  In areas of higher density in
non-LRG targets, the LRGs can be incomplete; however, eBOSS does require
that the total completeness of the decollided LRG targets be greater
than 95\%.  This layered tiling scheme allows 8\% more area to be
covered than otherwise would, at the cost of the variable completeness
of LRGs.

In the first round of fiber assignments --- the non-LRG targets --- 
eBOSS specifies the priority for fiber assignments when fiber collisions
occur.  Because the quasar targets have significantly higher density
than TDSS and SPIDERS targets, quasar-TDSS/SPIDERS collisions are
fractionally more common for TDSS/SPIDERS target classes.  Collisions
are resolved in the following order (highest to lowest priority):
SPIDERS, TDSS, reobservation of known quasars, clustering quasars, and
variability-selected quasars.  Quasars found in the FIRST survey
\citep{becker95a} and white dwarf stars that can be used as possible
calibration standards are given the lowest priorities for resolving
fiber collisions.

\citet{dawson16a} summarizes the overall expected numbers of
spectra. Nominal weather performance provides completion of $\sim$1800
plates, which would yield 1.62 million object spectra including about
180,000 unique TDSS and SPIDERS targets. Table
\ref{table:eboss_tdss_spiders} lists the numbers of confirmed
quasars at $z<2.1$, new and repeated BOSS quasars at $z>2.2$,
confirmed LRGs, and confirmed ELGs, assuming our estimated
efficiencies and redshift success rates.  The spectra that are
contaminants to the eBOSS cosmological sample are primarily blue stars
for quasar targeting and M stars for LRG targeting.

\begin{figure}[t!]
\centering
\includegraphics[width=0.49\textwidth,
  angle=0]{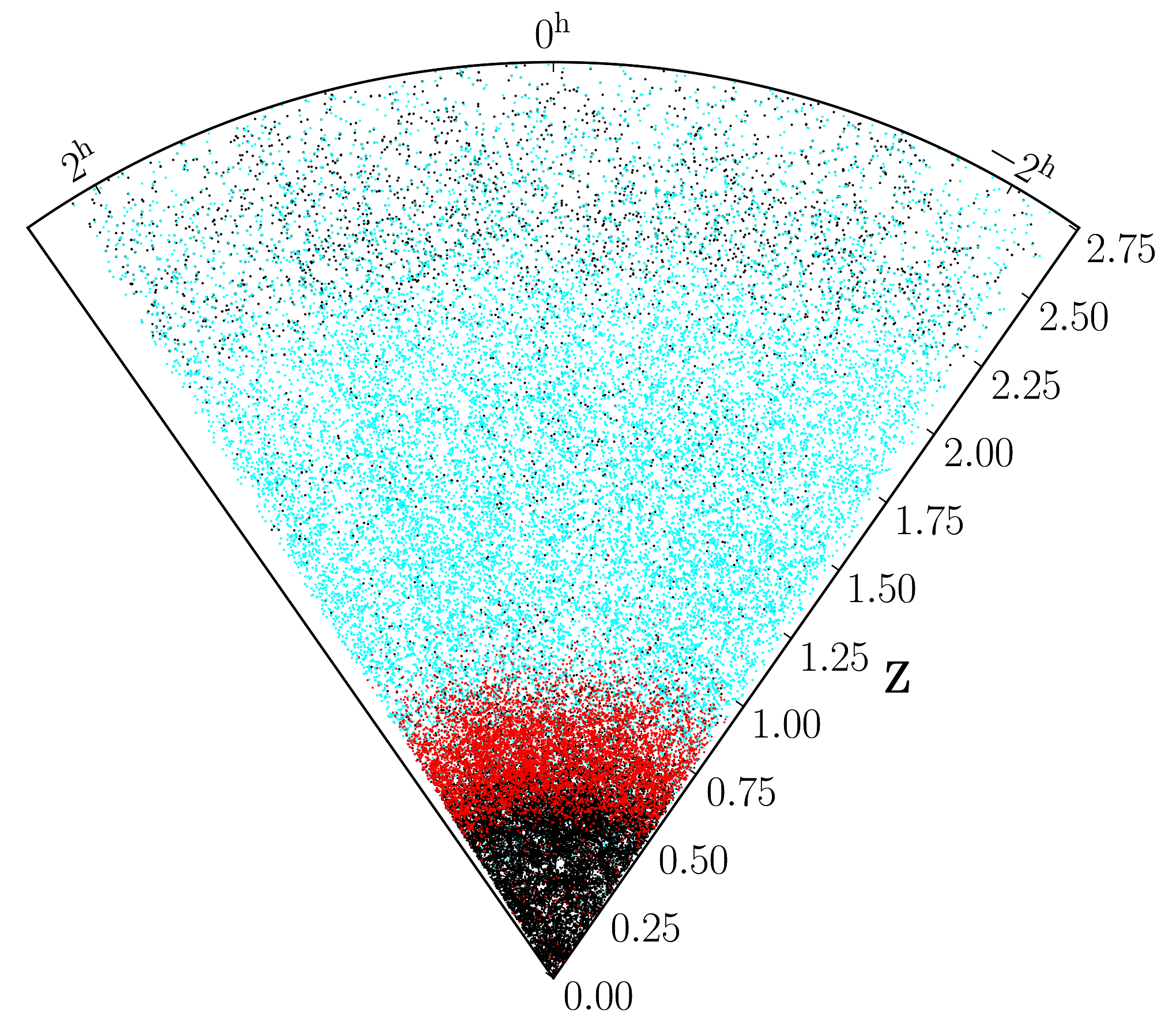}
\caption{ \label{fig:eboss_pie} Slice along right ascension through
the eBOSS redshift sample, 5$^\circ$ wide in declination and centered
at $\delta = +22^\circ$. Black points indicate previously known
redshifts from SDSS-I through SDSS-III. Cyan points show eBOSS quasars
and red points represent eBOSS LRGs, each category selected as
described in Section \ref{sec:eboss:targeting}.  }
\end{figure}

\subsubsection{eBOSS Observations}
\label{sec:eboss:observations}

eBOSS utilizes approximately 50\% of the dark time at APO. Details of the
division of observations across the SDSS-IV surveys are given in
Section \ref{sec:apo}.

Each BOSS fiber is encased in a metal ferrule whose tip is relatively
narrow (2.154~mm) and is inserted fully into the plate hole, but whose
base is 3.722~mm in diameter and sits flat on the back of the plate.
Two fibers on the same plate therefore cannot be placed more closely
than 62$''$ from each other on the sky.  Thus, except where two tiles
overlap, only one of such a pair can be observed; these fiber
collisions affect both the small- and large-scale clustering signal
from the sample and must be accounted for in the analysis
(e.g. \citealt{guo12a}).

Each plate is designed for a specific hour angle of observation.  The
observability window is designed such that no image falls more than
0.3$''$ from the fiber center during guiding. This restriction limits
the range of LSTs in which a plate is observable.

eBOSS is designed for LRGs, ELGs, and quasars with $z<1.5$ to have a
redshift accuracy $<300$\,km s$^{-1}$ (root mean squared) at all
redshifts. Larger redshift errors have the potential to damp the BAO
feature in the radial direction, thus diluting the precision
achievable on $H(z)$. We require catastrophic errors (defined as
redshift errors exceeding $1000$\,km s$^{-1}$ that are not flagged) to
be $<1\%$. At higher redshifts, we aim for quasars to have a redshift
measurement accuracy $< 300 + 400(z - 1.5)$\,km s$^{-1}$. The increase
at higher redshift reflects the expected rising difficulty of accurate
redshift measurement. A small number of repeat spectra are obtained
where fibers are available, which allow an estimate of the
uncertainties in the redshifts.

To achieve these goals, eBOSS observations are designed to obtain
median $i$-band (S/N)$^2>22$ per pixel at a fiducial target magnitude
$i_{\rm fiber2}=21$ and median $g$-band (S/N)$^2>10$ per pixel at a
fiducial target magnitude $g_{\rm fiber2}=22$.  The dispersion of the
BOSS spectrographs delivers roughly 1~\AA\ per pixel.  Plates are
exposed until they satisfy this signal-to-noise ratio requirement.
First year data indicate that plates require 4.7 15-minute exposures
to exceed these requirements; during the first year, we slightly
exceeded the requirements and averaged 5.3 exposures per plate. The
mean overhead per completed plate is around 22 minutes (this time
averages over cases where a plate was observed on multiple nights).
These thresholds are designed to satisfy the above requirements on
redshift accuracy.  The observing depths are also established to
achieve a reliable classification of all targets, whereby catastrophic
errors are required to occur at a rate of less than 1\% for all target
classes.

\begin{figure*}[t!]
\centering
\includegraphics[width=0.99\textwidth,
  angle=0]{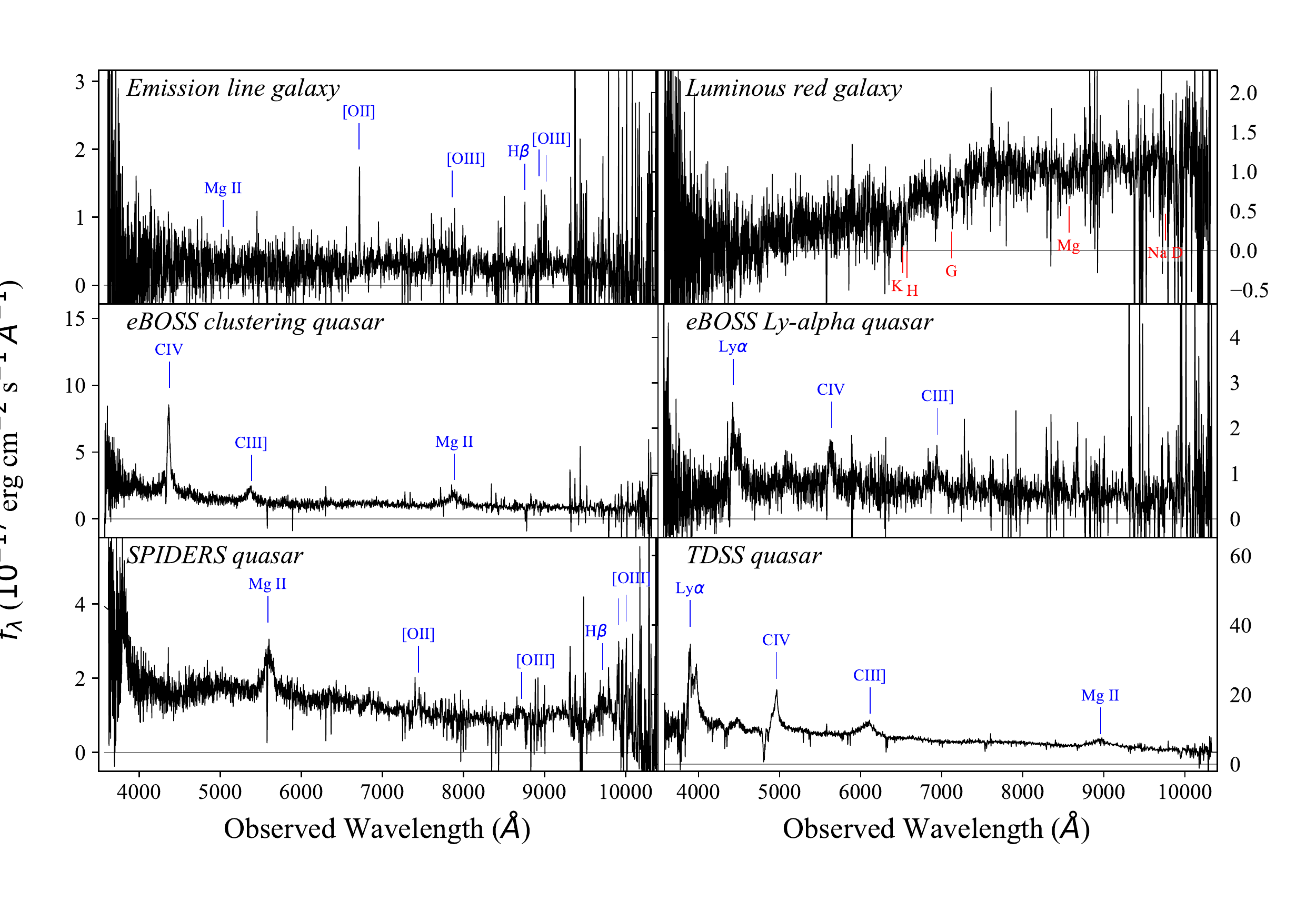}
\caption{ \label{fig:eboss-spectra}
Six representative eBOSS spectra, showing an emission line galaxy, a
luminous red galaxy, a quasar from the core ``cosmological'' sample, a
quasar selected at $z>2.2$ for Ly-$\alpha$ forest studies, an X-ray
emitting quasar selected by SPIDERS, and a TDSS-selected variable
broad absorption line quasar (listed left-to-right, and
top-to-bottom). The locations of emission lines are labeled in blue,
and for the luminous red galaxy, those of absorption features are
labeled in red.}
\end{figure*}

\subsubsection{eBOSS Data}
\label{sec:eboss:data}

eBOSS spectroscopic data consists of single-fiber $R\sim 2,000$
spectra in the optical (approximately 3600 \AA $< \lambda <$
10,350 \AA), at signal-to-noise ratios of $\sim$ 2--4 per pixel for
most targets, from which we determine redshifts and
classifications. The eBOSS pipeline is a slightly modified version of
the BOSS pipeline described by \citet{bolton12a}.
Figure \ref{fig:eboss-spectra} displays six example spectra from the
first year of eBOSS, processed through a preliminary version of the
eBOSS pipeline.

eBOSS data are processed through a quicklook pipeline (Son Of Spectro,
SOS) during each observation to estimate the signal-to-noise ratio in
real time and inform decisions about continuing to subsequent
exposures.  Quality assurance plots are examined each day to identify
unexpected failures of the observing system or pipelines.

Each morning following a night of eBOSS observations the data are
processed by the pipeline and made available for the collaboration.
The pipeline extracts the individual spectra using optimal extraction
(\citealt{horne86a}), and builds a spatially dependent model of the
sky spectrum from the 80 sky fibers and subtracts that model from each
object fiber. It determines the spectrophotometric calibration, which
includes the telluric line correction, using a set of 20 calibrator
standard stars observed on each plate, selected to have colors similar
to F stars and in the magnitude range $16<r_{\rm fiber2}<18$.
Redshifts are determined using a set of templates, with separate sets
for stars, galaxies, and quasars. For stars, the templates consist of
individual archetypes; for galaxies and quasars, the templates consist
of Principal Component Analysis (PCA) basis sets that are linearly
combined to fit the data at each potential redshift. The best redshift
and classification (star, galaxy, or quasar) is determined based on
the $\chi^2$ differences between the models and the data. For
galaxies, the pipeline also fits the velocity dispersion of the
galaxy, by comparing the spectra with linear combinations of a set of
high-resolution stellar templates. The pipeline conducts emission-line
flux and equivalent width measurements as well for a number of major
emission lines.

The pipeline undergoes continuous improvement as problems are
identified and repaired. Future versions will benefit from ongoing
efforts to improve sky subtraction and spectrophotometric
calibration. A new procedure and set of templates for fitting
redshifts is being developed to handle better the lower
signal-to-noise ratio of the fainter eBOSS targets. Specifically,
quasars and galaxies will use a large number of fixed archetypes
rather than a PCA basis set (\citealt{hutchinson16a}).

The eBOSS pipeline has been applied to all SDSS-III BOSS data as well,
which were taken with the same instrument. We do not have plans to
reanalyze the previous SDSS-I and SDSS-II data from the SDSS
spectrographs.

The first SDSS-IV data release (DR13; 2016 July) contains a
rereduction of BOSS data through the latest version of the pipeline
and includes plates from SDSS-IV completing the SEQUELS sample. In
DR14, the first two years of eBOSS data will be released.

The quasar science team within eBOSS plans to continue to maintain the
SDSS quasar catalog, the latest version of which is DR12Q
(\citealt{paris14a}). This catalog includes visually vetted redshifts
and classifications and has greater reliability than the standard
pipeline results. In DR12Q, all quasar spectra were inspected visually
by at least two people. However, in eBOSS a greater amount of
automatic vetting reduces the number of quasars that need to be
inspected visually.

\subsection{SPIDERS}
\label{sec:spiders}
\subsubsection{SPIDERS Motivation}

Within the main eBOSS program of quasars and LRGs, an average of 50
fibers per plate are allocated to sources associated with X-ray
emission, primarily AGNs and cluster galaxies. The goal of these
observations are twofold: first, to obtain a statistically complete
sample of X-ray emitting accreting black holes to better understand
quasar evolution and physics; second, to obtain redshifts and velocity
dispersions for a large sample of X-ray clusters.  The samples are
defined using the ROSAT All-Sky Survey
(RASS; \citealt{voges99a,boller16a}), the XMM Slew Survey
(XMMSL; \citealt{warwick12a}), and the upcoming eROSITA instrument
(\citealt{merloni12a}). In total, 22,000 spectra of X-ray emitting
AGNs
will be acquired, about 25\% of which will be targets in common with
the eBOSS cosmological program, and redshifts of about 58,000 galaxies
in 5,000 galaxy clusters.

SPIDERS uses this X-ray census of AGNs to better understand the
relationships among the growth of galaxies, the growth of their
central black holes, and the growth of their dark matter halos;
Section \ref{sec:quasars} describes these goals in more detail. The
SPIDERS cluster sample better establishes cluster scaling relations
and their evolution, and to use them to constrain cosmological
parameters through the evolution of the cluster mass function
(\citealt{allen11a, weinberg13a}).  For all of these science goals,
the existing statistically complete X-ray selected samples are too
small; they consist primarily of the sample of RASS sources observed
in SDSS-I and -II (\citealt{anderson03a}) and of much narrower
field of view and deeper observations in, for example, COSMOS
(\citealt{cappelluti09a, civano16a}), AEGIS (\citealt{laird09a,
nandra15a}), CDFS (\citealt{luo08a, xue11a}), and XBo{\"o}tes
(\citealt{kenter05a, murray05a}). Systematic, moderate resolution
spectroscopic follow-up of large area X-ray surveys, which sample
massive galaxy clusters and the bright end of the AGN luminosity
function, are currently lacking, and can yield important insights into
demographics, evolution, and physical characteristics of galaxies in
the densest large-scale structure environments, and of AGNs, including
the obscured populations.

\subsubsection{SPIDERS Target Selection}

eROSITA's planned launch is in early 2018 and data will become
available in Fall 2018. The satellite will observe the whole sky every
six months, and over four years will produce a series of eight
successively deeper eROSITA All Sky X-ray Survey catalogs (eRASS:1
through eRASS:8). Given this timeline, the targeting strategy for
SPIDERS is divided into several tiers depending on the available data
at the time of observation.
\begin{itemize}
\item {\it Tier 0}: Prior to the availability of eRASS data, SPIDERS 
targets RASS and XMMSL targets. 
\item {\it Tier 1}: SPIDERS will begin targeting eROSITA data with 
eRASS:1, which will be a factor of four to five times deeper than RASS (for
point sources). eRASS:1 data is planned to be available in Fall 2018
and SDSS-IV observations can begin in early 2019.
\item {\it Tier 2}: eRASS:3 is planned to be available mid-2019, and 
SPIDERS will target it beginning late 2019.
\end{itemize}

SDSS-IV does not observe eRASS sources over the entire sky. The survey
only has access to sources in the half of the sky defined in
Galactic coordinates ($180^\circ<l<360^\circ$). This hemisphere is
accessible to the eROSITA-DE consortium, with which SDSS-IV has a data
sharing agreement. Under current plans, the other half of the sky
is accessible only to the Russian eROSITA consortium.

For Tier 0 point sources, RASS identifies on average 3 deg$^{-2}$, of
which about 0.8 deg$^{-2}$ are not previously observed
spectroscopically and not too bright to observe within an eBOSS
exposure (which means, typically, $r>17$).  The uncertainty in the
coordinates of each point source is about 20$''$--30$''$, making the
identification of optical counterparts challenging. The match to the
optical counterpart is performed in two steps: (1) the {\it WISE}
counterparts are found using a Bayesian method based on that
of \citet{budavari09a}, taking into account priors in color-magnitude
space; (2) counterparts in the SDSS DR9 imaging data are determined
with a simple positional match to the {\it WISE} coordinates.  XMMSL covers
about 50\% of the eBOSS area and provides an additional 0.2 deg$^{-2}$
new point sources on average. The selection of the RASS and XMMSL
point sources is limited at $r=22$ (Galactic extinction
corrected). Details of the targeting scheme for Tier 0 AGN will be
described in \citet{dwelly17a}.

For Tier 0 extended sources, the Constrain Dark Energy with X-ray
Clusters (CODEX) team has identified photon overdensities in RASS that
correspond to galaxy clusters (\citealt{finoguenov12a}). These
clusters, plus Planck-detected clusters, have been matched to likely
cluster members using SDSS DR9 imaging, specifically using the
red-sequence Matched-filter Probabilistic Percolation method
(redMaPPer; \citealt{rykoff14a}). There are about 5,000 such clusters
within the eBOSS footprint. In addition, $\sim 300$ clusters are
identified serendipitously by XMM and also matched to DR9
(XCLASS; \citealt{clerc12a, sadibekova14a}). SPIDERS targets cluster
galaxies down to $i_{\rm fiber}=21$ (Galactic extinction
corrected). From these cluster samples, there is a target density of
up to 20 deg$^{-2}$ on average; because these targets are concentrated
in dense clusters and are subject to fiber collisions, only 7--8
deg$^{-2}$ are assigned fibers. When including previous SDSS legacy
spectroscopic observations, SPIDERS reaches a median of approximately
10 galaxies per cluster with spectroscopic redshifts. Details of the
clusters targeting algorithms and of the analysis steps are presented
in Clerc et al.~(2016).

For Tiers 1 and 2 point sources (AGN), eRASS:1 and eRASS:3 will be
matched to SDSS DR9 imaging. We will target AGNs with $17<r<22$. In the
eROSITA-DE sky area, this procedure will yield about 4,000 targets in
eRASS:1 and 7,000 in eRASS:3 that are not already targeted by
eBOSS. Including both eBOSS and SPIDERS, there will be $\sim 15,000$
eROSITA-detected AGNs with optical spectra from SDSS-IV.

For Tiers 1 and 2 extended sources (clusters), member galaxies will be
identified using the same methods as for CODEX and XCLASS, but the
improved spatial resolution and depth of eRASS relative to RASS will
allow the targeting of intrinsically less massive and/or more distant
clusters. The number of galaxies assigned fibers per cluster range
from 1 to 10 depending on distance and cluster richness. Based on
estimated cluster counts in eROSITA simulations, SPIDERS expects
target densities of 7 deg$^{-2}$ in eRASS:1 and 10 deg$^{-2}$ in
eRASS:3.

SPIDERS data are processed through the same pipeline that
processes eBOSS data. Figure \ref{fig:eboss-spectra} shows an example
spectrum from the first year of SPIDERS: an AGN selected as an X-ray
emitter in RASS.

\subsection{TDSS}
\label{sec:tdss}
\subsubsection{TDSS Motivation} 

The variable sky is the focus of many recent and upcoming large-scale
photometric surveys. For example, the SDSS Supernova program included
100 epochs of $ugriz$ imaging on a $2.5^\circ$ wide region on the
Celestial Equator in the SGC (Stripe 82; \citealt{sesar07a}). Recently
concluded and ongoing surveys include Pan-STARRS1
(PS1; \citealt{kaiser10a}), the Catalina Real-Team Transient Survey
(CRTS; \citealt{drake09a}), and the Palomar Transient Factory
(PTF; \citealt{law09a}), to be followed by the Zwicky Transient
Factory (ZTF; \citealt{bellm14a, smith14a}). In the 2020s, the Large
Synoptic Survey Telescope (LSST; \citealt{lsst09a}) will provide an
unprecedented number of transients and variable stars and quasars. The
study of variable sources will improve our understanding of
fundamental processes regarding the evolution of astrophysical
objects.  Accreting supermassive black holes, manifesting themselves
as active galactic nuclei, quasars, and blazars, often vary by tens of
percent or more in the optical on month- to year-long time
scales. Stellar variability reveals magnetic activity on stellar
surfaces, interactions between members of binaries, and pulsations.

To physically characterize the variable objects in these surveys, a
number of targeted programs have conducted spectroscopy on selected
variable types such as quasars, RR Lyrae stars, subdwarfs, white
dwarfs, and binaries (e.g., \citealt{geier11a, palanquedelabrouille11a,
rebassamansergas11a, badenes13a, drake13a}). The aim of TDSS
is to conduct a large-scale, statistically complete survey of all
variable types, without an imposed bias to either color or specific
light-curve character. This survey provides critical information
necessary to map photometric variability properties onto physical
classifications for currently ongoing projects, and future endeavors
such as LSST.

TDSS is creating a sample of single-epoch spectroscopy of 200,000
variable sources selected from PS1 over the 7,500 deg$^2$ of eBOSS;
about 140,000 of these are selected already for eBOSS or have had
spectra in SDSS-I/II/III. For a subset of selected objects ($\sim
10,000$) TDSS is conducting few-epoch spectroscopy (two to three
visits over the duration of SDSS-IV) to use spectroscopic variability
to characterize the objects.

\subsubsection{TDSS Target Selection} 
\label{sec:tdss:targeting}

\citet{morganson15a} describes the target selection for TDSS single-epoch
spectroscopy, and \citet{ruan16a} and describes early spectroscopic
results.  In brief, $griz$ imaging is used to select targets from SDSS
DR9 and PS1. SDSS data were taken between 1998 and 2009, with
typically only one epoch per observation. The PS1 3$\pi$ survey
acquired 10--15 epochs of imaging between 2010 and 2013.  TDSS uses
the SDSS-PS1 comparison as a measure of long-term variability, and the
variation among PS1 epochs as a measure of short-term
variability. Adopting the Stripe 82 database as a
testbed, \citet{morganson15a} developed an estimator $E$ related to
the probability of a specific source being variable based on the
short- and long-term variability, and the apparent magnitude. This
estimate is applied to a set of isolated point sources with $17<i<22$
and defined a threshold $E$ above which to select objects as likely
variables. Across most of the sky (80\%) TDSS randomly selects 10 targets
per deg$^2$ that pass this threshold and are not already eBOSS quasar
targets. In the remaining sky (20\%) there are fewer than 10 unique
targets that pass the threshold, and TDSS selects some targets at lower
$E$.

About 10\% of the fibers devoted to TDSS are dedicated to repeat
spectroscopy of previously known objects already having at least one
extant SDSS spectrum in the archive, and which are anticipated to
reveal astrophysically interesting spectral variablity with an
additional epoch or two of further spectroscopy. This few-epoch
spectroscopy was initially conducted in eight planned programs. The
subjects of these programs are: radial velocities of dwarf carbon
stars; M-dwarf/white dwarf binaries; active ultracool dwarfs; highly
variable ($>0.2$ mag) stars; broad absorption line quasars
(\citealt{grier16a}); Balmer-line variability in bright quasars
(\citealt{runnoe16a}); double-peaked broad emission-line quasars; and
Mg II velocity variability in quasars.

TDSS data is processed through the same pipeline that processes eBOSS
observations. Figure \ref{fig:eboss-spectra} displays an example
spectrum from the first year of TDSS: a variable broad absorption line
quasar selected for few-epoch spectroscopy.

\subsection{Quasar Science with eBOSS, SPIDERS, and TDSS}
\label{sec:quasars}
\begin{figure*}[t!]
\centering
\includegraphics[width=0.9\textwidth,
  angle=0]{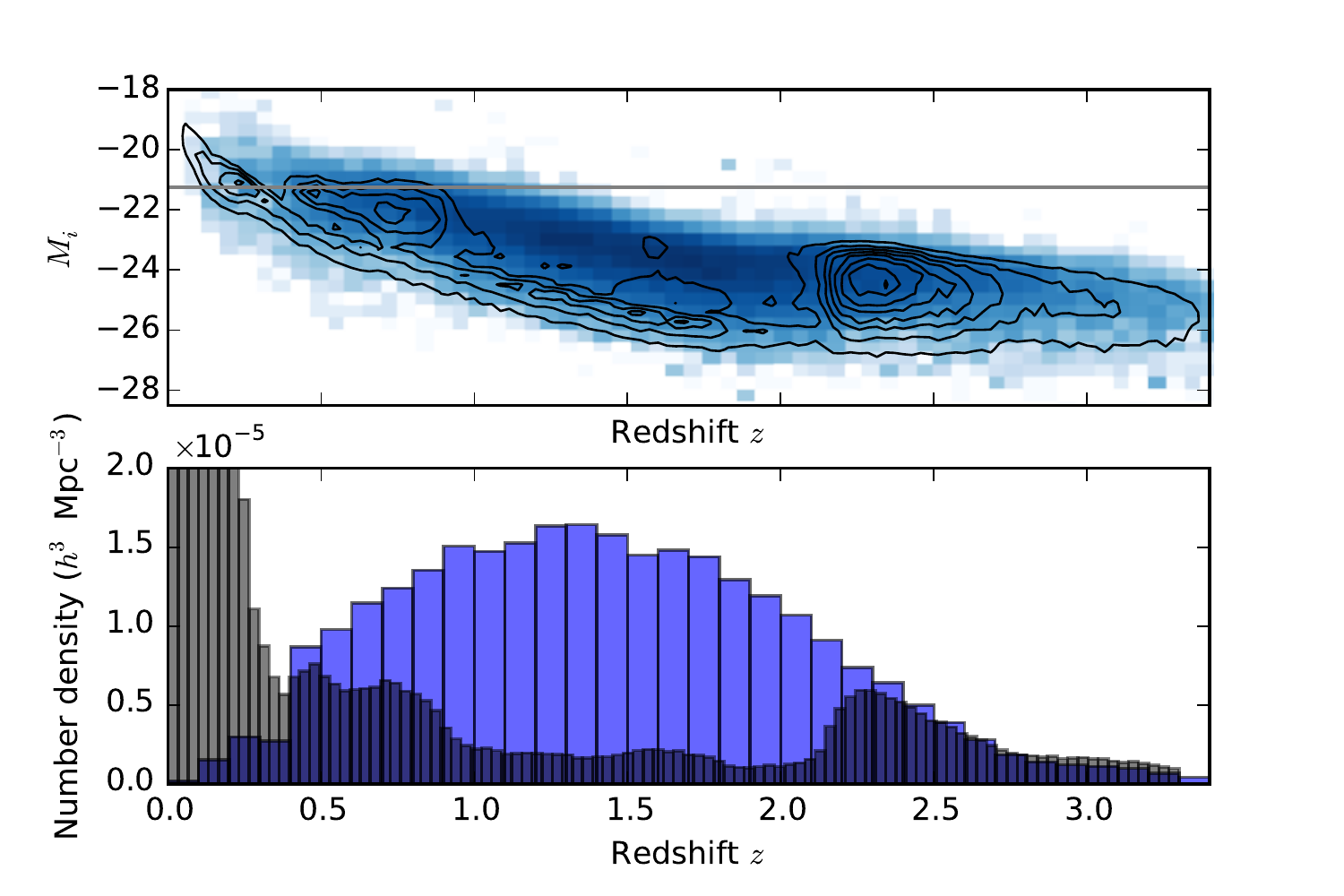}
\caption{ \label{fig:nbar-qso} Distribution of quasars in redshift and
rest-frame $i$-band absolute magnitude. Top panel: contours
show the density of Legacy and BOSS quasars in this plane from SDSS-I
through SDSS-III.  The grayscale represents the density of eBOSS,
TDSS, and SPIDERS quasars from SDSS-IV from the first year results. In
the range $1<z<2$ the SDSS-IV quasars probe much lower luminosities
than previous SDSS samples. The gray horizontal line corresponds to
$M_\ast$ for galaxies (\citealt{blanton04b}); the SDSS-IV quasars out
to $z\sim 2$ approach the faintness of Seyfert galaxies in optical
luminosity. Bottom panel: each histogram shows the density of
quasars as a function of redshift. The gray histogram is for Legacy
and BOSS quasars from SDSS-I through SDSS-III. The blue histogram
shows the estimated density of eBOSS quasars from the first year
results. In the range $1<z<2$ the eBOSS sample represents an increase
in density by factors of 5--10.}
\end{figure*}

eBOSS, TDSS, and SPIDERS together select more than half a million
quasar targets.  This enormous quasar catalog (tripling the world's
number of quasar spectra) includes objects targeted by optical and
mid-IR ({\it WISE}) colors, variability (TDSS), radio (FIRST), and
X-ray emission (SPIDERS).  Combined with previous SDSS and BOSS
observations, the catalog spans a factor of more than $\sim1000$ in
accretion luminosity from $z=0$ to $z=5$.  Whereas previous surveys
have sampled different quasar luminosity classes at different
redshifts, the SDSS-IV sample enables an understanding of individual
classes of quasars across epochs and better trace the full history of
active BH growth since $z\approx3$.  Figure \ref{fig:nbar-qso} shows
the increased density of quasars in SDSS-IV relative to previous SDSS
surveys, as well as its extension to fainter luminosities in the range
$1<z<2$.

The best measurements of the Type I quasar luminosity function at
$z<2$ from optical survey data come from 10,000 quasars compiled by
the 2dF-SDSS LRG and QSO (2SLAQ) survey \citep{croom09a}; using deeper
data, previous SDSS programs have extended to higher redshifts but
have not probed these lower redshifts as densely
(\citealt{palanquedelabrouille13a}). This survey targeted quasars to a
similar depth as eBOSS (though the eBOSS limit of $r<22$ reaches
many more quasars than the 2SLAQ limit of $g<21.85$), but over an area
$\sim40$ times smaller.  The statistical power provided by the large
--- and highly complete --- eBOSS sample provides a powerful new
probe of the evolution of the faint-end slope of the luminosity
function over the interval from $z=1$ to $z=2$, strongly constraining
feedback models for black hole growth \citep[e.g.,][]{hopkins06b}.

Combining measurements of the faint end of the luminosity function
with precision probes of quasar clustering constrains models for
quasar lifetimes, the typical halos hosting quasars, the co-evolution
of quasars and spheroidal galaxies, and the evolution in black hole
mass of active quasars (using virial mass estimators). Within the
redshift range $1<z<2$, the mass of black holes powering quasars is
expected to decrease with increasing redshift by an order of
magnitude, perhaps symptomatic of the characteristic fueling mechanism
shifting from major mergers to secular processes
(\citealt{hopkinshernquist06}). This prediction can be robustly tested
with eBOSS's measurements of the luminosity dependence of quasar
clustering. Finally, cross-correlation analyses of eBOSS galaxies and
quasars at redshifts where samples overlap provides unique insight
into the connection between quasars and galaxies (both quenched and
star-forming).

Selecting quasars using several different techniques within eBOSS,
TDSS, and SPIDERS allows SDSS-IV to account for the selection biases
that affect any individual quasar selection technique.  For example,
the SDSS-iV data enables the comparison of high-redshift quasars
with lower luminosity, X-ray selected AGNs at low redshift that may
represent their descendants.  A further advantage provided by SDSS-IV
is the ability to tie together the faint quasar population at optical
(eBOSS) and X-ray (SPIDERS) wavelengths within the same
survey. Reaching the optically fainter quasar population provides
access to a much larger number of significantly reddened quasars,
yielding a more complete census of narrow-line and reddened broad-line
AGNs.

Large quasar samples are useful not only for demographic studies, but
also for yielding rare phenomena. Repeat spectroscopy of known quasars
through TDSS captures changes in the absorption profiles of clouds
along the line of sight to quasar nuclear
regions \citep[e.g.,][]{ak13a}, rare state changes when the nuclear
emission effectively vanishes \citep[so-called ``changing-look''
quasars;][]{lamassa15a,runnoe16a}, and a variety of other
time-dependent phenomena traced by multi-epoch quasar
spectroscopy. The unprecedented density of quasar targeting
within SDSS-IV, particularly when considering that most known quasars
will not be re-targeted and thus can have nearby objects targeted
within the fiber collision radius, probes the environments of
quasars through small-scale clustering with far greater numbers and
more uniformity than achieved even by dedicated surveys of quasar
pairs \citep[e.g.,][]{hennawi06a}.  Combining small-scale quasar pairs
with the large-scale clustering sample from eBOSS constrains halo
occupation models of quasars over a wide range of both luminosity and
spatial scales and permit detailed examination of the relationship
between quasar triggering and environment.

There are three quasar programs that SDSS-IV is executing to
enhance quasar science: a complete sample of AGNs on Stripe 82, a
continuation of the SDSS-RM program, and a program for repeat quasar
spectroscopy.

First, ``Stripe82X'' provides a focused effort to build
a complete sample of AGNs with SDSS-IV spectroscopy, with a set of six
spectroscopic plates dedicated to AGN targets.  The plates span a
footprint of $\sim35~{\rm deg}^2$ within the SDSS Stripe 82 region,
bounding the area defined by the Stripe 82 X-ray survey of
\citet{lamassa2016} between $\alpha_{\rm J2000}=14^\circ$ and 
$\alpha_{\rm J2000}=28^\circ$. X-ray sources drawn
from \citet{lamassa2016} with optical counterparts having
$r<22.5$ provide the primary target class for the Stripe82X survey,
totaling nearly 900 objects. The remaining fibers on each plate are
primarily assigned to {\it WISE}-selected AGN \citep[using the R75 color
criteria of][]{assef13a} and variability-selected quasars
\citep{peters15a, palanquedelabrouille16a}. A small number of high-redshift 
quasar candidates and repeat observations of ``changing look" and
related quasar candidates using TDSS selection criteria are also
included. The tiling includes roughly 5,000 AGN targets. The primary
goals of the Stripe82X program are: (1) to better characterize AGN
bolometric corrections by combining the spectroscopy with the
extensive multiwavelength photometry available on Stripe 82; (2) to
explore and compare the diverse classes of AGN selected by different
wavelength regimes; and (3) to construct a bolometric AGN luminosity
function from a highly complete, faint AGN sample.

Second, during dark time, SDSS-IV is continuing the SDSS-RM program
(\citealt{shen15a}) initiated during the last observing semester of
SDSS-III in 2014 (\citealt{alam15b}). SDSS-RM monitors a sample of 849
quasars within a single 7 deg$^2$ field with BOSS spectroscopy and
accompanying photometry to measure quasar broad-line time lags with
the reverberation mapping technique (e.g., \citealt{blandford82a,
peterson93a}).  In eBOSS, the SDSS-RM spectroscopy has a cadence of 2
epochs (similar depth to eBOSS) per month (12 epochs/year) since 2015,
and provides an extended temporal baseline to detect broad-line lags
on multi-year timescales in high-redshift quasars when combined with
earlier SDSS-RM data.

Third, a Repeat Quasar Spectroscopy (RQS) program emphasizing known
quasars is being observed in the eBOSS ELG region discussed in
Section \ref{sec:eboss:targeting}, supplementing the TDSS few-epoch
spectroscopy. In this $\sim10^3$~deg$^2$ region, TDSS is also
obtaining a new epoch of spectroscopy for previously-known SDSS
quasars. In this region, we include quasars with $17<i<21$ (also
including morphologically extended AGNs) from the DR7 or DR12 quasar
catalogs, or SDSS-IV objects with spectro-pipeline class ``QSO'' that
have been vetted as quasars/AGNs by our own visual inspection of the
spectra. As part of the ELG plates, TDSS observes a total of
$\sim10^4$ known quasars/AGNs for an additional epoch of spectroscopy,
including the bulk of all known SDSS quasars in this region to
$i<19.1$, as well as filling additional available fibers for RQS with
either: known SDSS quasars extending to $i<20.5$ already having more
than one extant epoch of on-hand spectroscopy; and/or additional of
the most highly variable known SDSS quasars in the ELG region, as
determined from a reduced chi-squared measure of their photometric
variability in SDSS and PS1 imaging. Details of RQS target selection
will be reported in a future publication (MacLeod et al. 2017, in
preparation).

SDSS-IV maintains the tradition established by the previous
incarnations of the survey to publicly release quasar catalogs
\citep[e.g.,][]{schneider10a,paris14a} associated with each release of
new spectroscopic data.  In SDSS-III, starting from the output of the
the SDSS pipeline \citep{bolton12a}, the spectrum of each quasar
target was visually inspected to confirm both its identification and
redshift.  This procedure ensured the high purity of the catalog
content and contributed to improvements in the SDSS pipeline.  The
quasar target density of SDSS-IV is approximately three times larger
than in SDSS-III. This increase combined with the amount of time
required to perform a systematic visual inspection of all quasar
targets forces us to adapt our strategy to construct quasar catalogs.
Hence, we developed a semi-automated scheme: starting from the output
of the SDSS pipeline, we identify spectra for which the identification
and/or redshift produced by the automated pipeline are
questionable. The spectra of these objects ($\sim$7\% of the targets)
are then visually inspected. This automated strategy was tested
against a fully visually inspected sample drawn from the SDSS-IV pilot
survey performed at the end of SDSS-III and its design delivers a
quasar catalog with a purity larger than 99\% and a loss of less than
1\% of actual quasars \citep[see ][ for more details]{dawson16a}.

The content of the SDSS-IV quasar catalog is similar to the previous
ones. Multiwavelength information is provided when available along
with spectroscopic properties such as emission-line fitting, presence
of broad absorption lines and improved redshift estimates.  At the
conclusion of SDSS-IV, the photometric and spectroscopic properties of
about a million quasars will be released.

\section{Data Management}
\label{sec:data}
\begin{table*}[t!]
\caption{
\label{table:drs} SDSS-IV Data Releases}
\begin{tabular}{ccccccc}
\hline\hline
Name & Release Date & Data Through & eBOSS & MaNGA & APOGEE-2N &
APOGEE-2S \\
\hline
DR13 & 2016 Jul & 2015 Jul & SEQUELS\tablenotemark{a}
& New data and products\tablenotemark{b} & New products & --- \cr
DR14 & 2017 Jul & 2016 Jul & New data & New data & 
 New data & --- \cr
DR15 & 2018 Jul & 2017 Jul & --- & New data and
products\tablenotemark{c}
& --- & --- \cr
{\it DR16} & {\it 2019 Jul}  & {\it 2018 Jul } & {\it New data} &
{\it New data} & {\it New data} & {\it New data} \cr
{\it DR17} & {\it 2020 Dec} & {\it 2019 Jul } & {\it New data} &
{\it New data} & {\it New data} & {\it New data} \cr
\hline
\end{tabular}
\tablecomments{
The timing of the last two data releases will be based on available
funding.  ``New data'' means that new data are being released. ``New
products'' means that new types of data analysis are being
released.
\tablenotetext{a}{DR13 contains the remainder of the SEQUELS program, begun
in SDSS-III and completed in SDSS-IV, and new reductions for BOSS
data, but no new eBOSS data.}
\tablenotetext{b}{DR13 and DR14 contain MaNGA Data Release Pipeline results;
these are calibrated spectral data cubes.}
\tablenotetext{c}{DR15 contains MaNGA Data Analysis Pipeline results;
these include maps of derived quantities from the spectral data
cubes.}
}
\end{table*}

SDSS-IV data management encompasses the transfer of data among survey
facilities, long-term archiving of data and metadata, documentation,
and distribution to the collaboration and the public. We build on the
data distribution systems developed for SDSS-I through SDSS-III.

The central data system for SDSS-IV is the Science Archive Server
(SAS) hosted by the University of Utah Center for High Performance
Computing. The SAS serves as a data repository with all survey
targeting data, raw data, and reduced data on disk, and has associated
computing to perform reductions and other critical operations. It has
a current capacity of around 1 petabyte, in order to accommodate the
variety of necessary imaging data sets and spectroscopic reduction
versions produced during the survey. A Science Archive Mirror (SAM) at
a separate location contains a copy of all the archived data; the SAM
is housed by the National Energy Research Scientific Computing Center
(NERSC) at the Lawrence Berkeley National Laboratory during the
lifetime of the survey. In addition, the archived data are backed up
on long term tape storage at the High Performance Storage System
(HPSS) at NERSC.  The SAS system also contains the project wiki, used
for documentation and internal communication, and a {\tt subversion}
server used for software version control. These systems are also
backed up at the SAM.

Survey targeting data, plate design data, and other data associated
with the observational planning are stored on the SAS and information
is distributed from there to the University of Washington plate
drilling facility and to APO and LCO as necessary for conducting
operations. Data and metadata from the plate drilling quality
assurance process are backed up to the SAS. At APO, the plate-plugging
metadata, observing logs, telescope telemetry, and the raw data are
transferred each day from the previous night's observing to the SAS
(\citealt{weaver15a}) and backed up on the SAM and HPSS. A similar
system is installed at LCO.

The eBOSS, MaNGA, and APOGEE-2 pipelines are run automatically on each
night's data as they arrive. For eBOSS, this process consists of the
full pipeline through the production of 1D calibrated spectra,
redshifts, and other parameters, for each completed plate. For MaNGA,
this process consists of the Data Reduction Pipeline executed for each
completed plate. However, currently the Data Analysis Pipeline is
experiencing more development and is not run automatically; it is
instead run periodically based on accumulated data and progress in DAP
development. For APOGEE-2, the visit spectrum reductions and radial
velocity determinations are performed automatically.  However, because
the combined spectra require multiple visits and because of its
computational expense, the ASPCAP analysis is performed periodically
on large sets of plates, again based on accumulated data and progress
in ASPCAP development.

The primary point of data access for collaboration members is the
SAS. Collaboration members can access data on the SAS through {\tt
ssh} connections. SAS also provides {\tt http}, {\tt rsync}, and
Globus access to the data files. These methods are available also to
the astronomical community for publicly released data both for the SAS
and SAM. We provide a web interface and an application program
interface (API) on SAS to the eBOSS and APOGEE. A similar set of
interfaces is being developed for MaNGA called Marvin, which will
additionally have a Python module for interaction with the API. The
data directory structure and file format documentation is provided as
a ``data model.''\footnote{\link{http://data.sdss.org/datamodel}}

Public data releases incorporate both the SAS data interface and the
Catalog Archive Server (CAS), hosted at Johns Hopkins University. The
CAS contains catalog data from the SDSS imaging and spectroscopic
survey; it does not currently include images or spectra (other than
JPEG and PNG versions, respectively, for visual browsing). The total
database size is approximately 12 Tb, which is dominated by SDSS
imaging catalogs. The CAS provides web browser-based access in
synchronous mode via the SkyServer web application\footnote{\tt
http://skyserver.sdss.org/} and in asynchronous mode with the CASJobs
batch query service.\footnote{\tt http://skyserver.sdss.org/casjobs/}

The SkyServer (\citealt{szalay02a}) supports multiple levels of data
access ranging from simple form-based queries aimed at novice users to
raw SQL queries for expert users. The SkyServer includes interfaces
displaying the SDSS and 2MASS imaging and the locations of SDSS
spectroscopic and imaging catalog entries, as well as an Explore tool
for each object showing the spectra and listing key
parameters.

CASJobs (\citealt{li08b}) gives each user their own server-side
database called MyDB, along with the ability to submit arbitrarily
complex SQL queries in batch mode and redirect the output to their
MyDB.  Users may import their own data to cross-match with the SDSS
data. There is a Groups feature to allow users to share their data
with collaborators. CASJobs also supports a command-line mode of query
submission. For SDSS-IV, SkyServer and CASJobs are integrated into
the SciServer collaborative data-driven science framework\footnote{\tt
http://sciserver.org/} with seamless single sign-on access to several
new services such as Compute, SciDrive, SciScript and SkyQuery. Compute
includes a Jupyter notebook server that has fast server-side access to
CASJobs and other data sets.

The SDSS data distribution system is heavily used. The CASJobs system
has approximately 2000 unique users each year. The SkyServer system
experiences tens of millions of queries each year. The SAS system is
used to download tens of terabytes of data per year by public
users. The SDSS help desk email account fields around 500 inquiries
per year.

We plan to release data on regular intervals. The released data
include targeting data, raw and reduced spectroscopic data including
of calibrations, derived quantities of several varieties, and
value-added catalogs provided by collaboration members. All metadata
and intermediate data are included and documented. Table
\ref{table:drs} shows our nominal data release plans. The data
releases include not just SDSS-IV data but also data from previous
phases of SDSS, and the services host all previous data releases. New
types of analysis or increments of new data may be added based on
availability. Because of funding uncertainty, the timing of the last
two data releases remains unclear; nevertheless, SDSS-IV is committed
to a final public release of all of its data.

\section{Education and Public Engagement}
\label{sec:epo}
The mission statement of education and public engagement for SDSS-IV
is to make the engineering and scientific results of all SDSS surveys
accessible to the public through formal education, citizen science,
news, and social media. SDSS-IV will continue and expand upon the
activities in these areas of its predecessors.  SDSS public outreach
activities are based on real astronomical data accessed through the
same databases as used by professionals.  These activities expand the
user base of SDSS data and thus its scientific reach, both through
training and directly through investigations made possible with these
scientific tools.

These activities include the public distribution of data, the
development of inquiry-led education material suitable for middle
school and above, the distribution of SDSS plates to educational
venues to support engagement with SDSS data in the classroom,
development of new citizen science projects through collaboration with
the Zooniverse\footnote{\url{http://www.zooniverse.org}} (building on
the success of Galaxy Zoo\footnote{\url{http://www.galaxyzoo.org}}),
regular blogging\footnote{\url{http://blog.sdss.org}} and increased
social media engagement, including multi-lingual
activity.\footnote{\url{http://www.facebook.com/SDSSurveys}; \\
\url{https://twitter.com/sdssurveys}} These activities are coordinated 
by co-Chairs of a Committee on Education and Public Engagement, and
are partly funded by the SDSS-IV project and partly the result of
voluntary activities by collaboration members.

The SkyServer contains material, tutorials, and activities designed
for outreach and education.  Based on SkyServer tools, SDSS
Voyages\footnote{\url{http://voyages.sdss.org}} is created for
educators for designing curricula around astronomical data from the
SDSS. The activities on the site range from very short to extended
projects, aimed at middle and high school students. We have begun a
program associated with the SDSS Voyages activities of distributing to
teachers used plug plates, which so far has reached 32 schools.

\section{Management and Collaboration}
\label{sec:management}
\subsection{Project Management}
\label{sec:project}
The governance and management structure of SDSS-IV continues the
highly successful structure developed over its previous phases.
SDSS-IV is ultimately overseen by the Astrophysical Research
Consortium (ARC) and its Board of Governors.  The ARC Board has
established a set of SDSS-IV Principles of
Operations\footnote{See \link{http://www.sdss.org/collaboration/}.}
which provides the governance and management structure of the project.

Institutions join the collaboration via contributions, both technical
and financial, committed to through Memoranda of Understanding
(MOUs). Scientists at these institutions have data rights to all of
the SDSS-IV surveys. ``Full membership'' yields data rights for all
employees at an institution. ``Associate membership,'' which requires
a smaller contribution, yields data rights for a limited number of
scientists. Technical contributions must directly address items in the
survey budget.

The ARC Board has established an Advisory Council (AC) that oversees
the Director and the project. The AC consists of representatives from
the member institutions. It approves each new MOU and has authority
over significant changes in policy, changes in the project scope, and
fundraising activities. 

\begin{figure*}[t!]
\centering
\includegraphics[width=0.98\textwidth, angle=0]{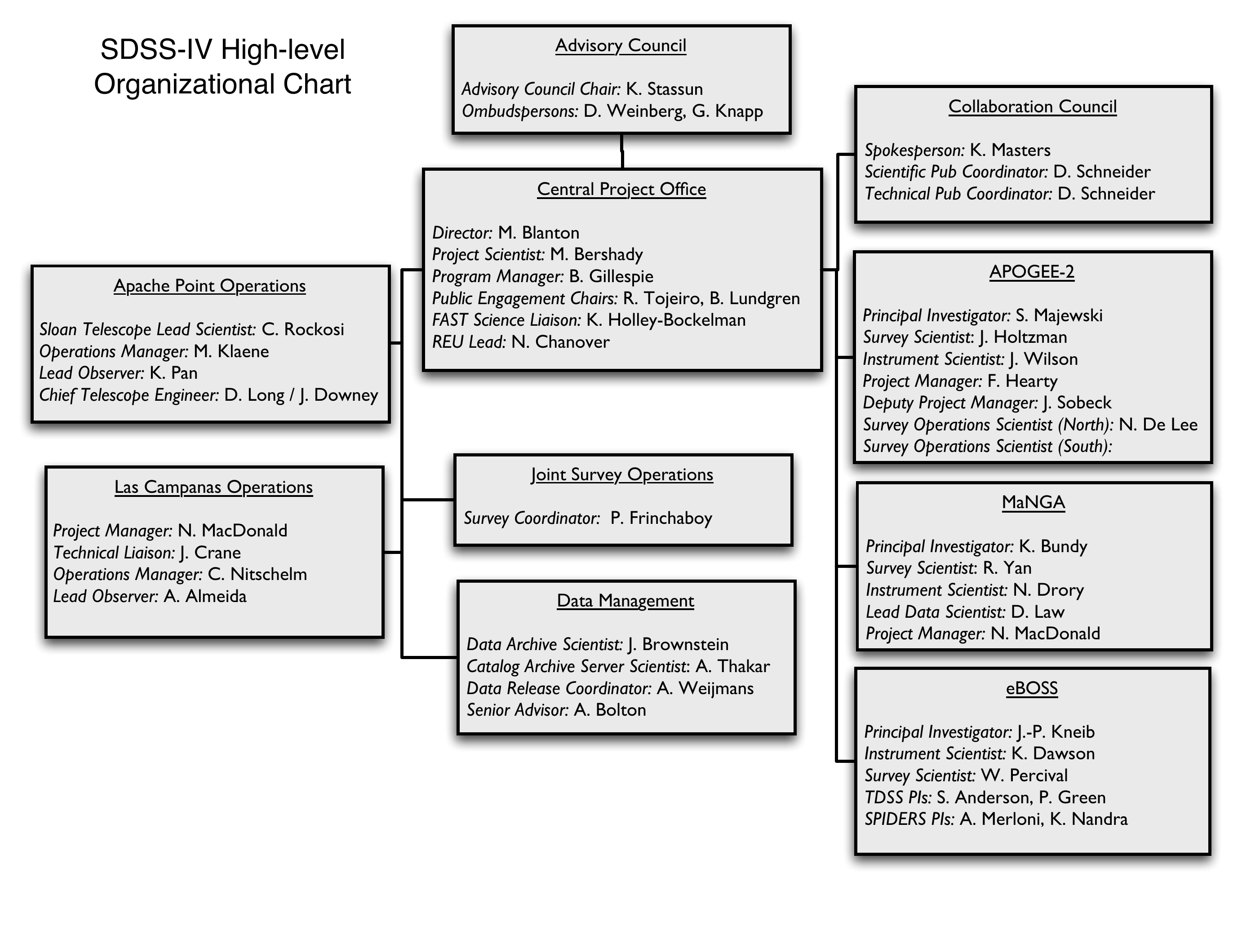}
\caption{ High-level organizational chart for SDSS-IV, as of 
2017 February. Positions have rotated somewhat during the project and will
continue to do so.}
\label{fig:orgchart}
\end{figure*}

Figure \ref{fig:orgchart} shows the high-level organizational chart.
The management structure is designed to unify decision-making and
establish clear lines of authority for the allocation of resources by
the Central Project Office.

The Central Project Office contains the Director, the Project
Scientist, the Program Manager, and the Project Spokesperson. The
Director makes spending, budget, and fundraising decisions, and
resolves decision-making conflicts.  The Project Scientist's role is
to ensure the scientific quality and integrity of the project, through
reviews of the scientific plans and products.  The Program Manager is
the full-time manager of the project, tracking the schedule and
project personnel issues. The co-Chairs of Education and Public
Engagement and the FAST Science Liaison are part of the Central
Project.

The Project Spokesperson is the leader of the Science Collaboration
and represents SDSS-IV to the scientific community. The Science
Collaboration is described more fully in the next subsection.

The leadership teams of each core program in SDSS-IV (APOGEE-2, eBOSS,
MaNGA) have a common structure. Each program has a Principal
Investigator (PI), a Survey Scientist, and an Instrument Scientist.
The PI is responsible for leading each survey, both scientifically and
in terms of its management.  The Survey Scientist is responsible for
the proper execution of the survey. The particular focus of the Survey
Scientist differs from survey to survey, and ranges from overall
scientific strategy to pipeline development.  The Instrument Scientist
is responsible for the development and maintenance of the
instrument. For eBOSS, the instrument is stable and not under
development; in this case the Instrument Scientist takes on many of
the operational tasks. For MaNGA and APOGEE-2, which have major
hardware upgrades and development, the instrument scientists are much
more focused on that development.  For the same reason, MaNGA and
APOGEE-2 have Project Managers to lead the hardware construction.
SPIDERS and TDSS each have PIs but not the other leadership positions.

Several positions exist to support common goals and coordination.  The
Data Management team leads the data management and distribution.  A
Survey Coordinator plans and monitors the survey observational
strategy. The co-Chairs of Education and Public Engagement lead a
committee coordinating the development of educational materials and
public engagement activities.

At APO, the Sloan Telescope Lead Scientist manages the infrastructure
development and maintenance of the telescope and the APO Operations
Manager manages the day-to-day operations, including site
maintenance. At LCO, the LCO Project Manager leads the hardware
development and the LCO Operations Manager manages the day-to-day
survey operations. The LCO site maintenance and telescope maintenance
is handled by the Observatories of the Carnegie Institution for
Science.

Logistical responsibility for handling scientific, technical, and data
release papers rests with the Scientific Publications Coordinator
(SPC), Technical Publications Coordinator (TPC), and Scientific
Spokesperson, respectively. Publications Coordinators ensure
that publications follow standard survey publication processes, and they
maintain a common electronic web-based archive of all scientific,
technical, and data release publications of the SDSS-IV, accessible to
collaboration members.  The TPC coordinates the publication of technical
papers, ensuring that the technical documentation of the project is
disseminated efficiently and promptly.  The SPC is responsible for
tracking SDSS-IV scientific papers through the publication policy
process and assuring that all SDSS-IV papers (scientific, technical,
and data release) reference the appropriate technical papers. The
Scientific Spokesperson has overall responsibility for the
Publications Archive, and coordinates the publication of the data
release papers.

The individuals filling these roles and the teams they lead are
geographically distributed at over twenty institutions. Each team
communicates through email lists, weekly phone meetings, and periodic
in-person meetings. A Management Committee consisting of individuals
in the positions listed here meets weekly to monitor the project
progress.

\subsection{Science Collaboration}
\label{sec:collaboration}
The Science Collaboration is led by the Project Spokesperson, who is
elected for a three-year term by the collaboration. A Collaboration
Council consisting of representatives from the participating
institutions advises the Spokesperson. The Spokesperson and the
Collaboration Council developed the Publication Policy for SDSS-IV.

Following previous SDSS collaborations, the Publication Policy's
guiding principle is that all participants can pursue any project so
long as they notify the entire collaboration of their plans and update
the collaboration as projects progress.  Groups pursuing similar
science projects are encouraged to collaborate, but they are not
required to do so. There is no binding internal refereeing process.
Instead, draft publications using non-public data must be posted to
the whole collaboration for a review period of at least three weeks
prior to submission to any journal or online archive.  Participants
outside of the core analysis team may request co-authorship on a paper
if they played a significant role in producing the data or analysis
tools that enabled it. Scientists who have contributed at least one
year of effort to SDSS-IV infrastructure development or operations can
request ``Architect'' status, which entitles them to request
co-authorship on any science publications for those surveys to which
they contributed.  All SDSS-IV authorship requests are expected to
comply with the professional guidelines of the American Physical
Society.

Each of the SDSS-IV programs has Science Working Groups to coordinate
and promote scientific collaboration within the team.  These working
groups overlap and interact with the SDSS-IV project personnel but are
more focused on science analysis.  The working groups communicate and
collaborate through archived e-mail lists, wiki pages, regular
teleconferences, and in-person meetings.  Importantly, the science
activities of these working groups are not funded by the SDSS-IV
project.

The policies of SDSS-IV allow limited proprietary data rights to
astronomers outside the collaboration under specific conditions that
fall into two categories. First, when an SDSS-IV member leaves for a
non-SDSS institution. The member can ask the Collaboration Council and
Management Committee for ``Continuing External Collaborator'' status
to complete a defined scientific investigation that had been
substantially started before the change in institutions. Second, if
crucial skills to complete science of interest to SDSS-IV members are
not available within the SDSS-IV collaboration, due to either
personnel or time constraints, SDSS-IV members can ask the
collaboration, with approval from the Collaboration Council and
Management Committee, for ``External Collaborator'' status for
non-SDSS members to work on specific aspects of declared projects. The
collaboration evaluates whether the contributions of the non-members
are unique and necessary to produce cutting-edge science from the SDSS
collaboration for a limited number of papers.

The projects, publications, and other activities are tracked in a
central database as part of the SDSS-IV data system. Collaboration
members use a web application to interact with this internal database.
This system lends clarity to the status of approvals and decisions
with regard to internal collaboration activities.

\subsection{Broadening Collaboration Participation}
\label{sec:climate}

The past success of the SDSS collaboration has hinged on tapping into
a diverse talent base. We have worked and continue to work within
SDSS-IV on this issue. Other collaborations may find the SDSS-IV
experience described here informative as they configure their policies
or face similar situations.

The SDSS-IV organization does not directly hire any of the staff, so
all recruitment of staff paid on contracts to institutions from ARC
also must go through each institution's human resources
process. Similarly, in cases of personnel issues, each institution has
its own policies on workplace environment.  The interleaving of
SDSS-IV processes with institutional policies represents an
interesting complication to international, multi-institutional
organizations such as the SDSS.

As discussed in \citet{lundgren15a}, SDSS-IV identified early a
disparity in the gender balance of its leadership structure. In order
to identify the causes of, monitor, and address this issue, we created
a Committee on the Participation of Women in the SDSS (CPWS). The CPWS
initiated regular demographic surveys of the SDSS in order to monitor
the make up of the collaboration and the project over time.  The CPWS
also compiled information on how the project leadership recruitment
proceeded. Near the beginning of SDSS-IV, and in previous phases of
the project, the recruitment for survey positions such as those in
Figure \ref{fig:orgchart} or others such as working group chairs, was
conducted informally and in a relatively federated manner across the
project.

In 2013, SDSS-IV began to implement an early recommendation of the
CPWS to formalize the recruitment process. SDSS-IV policy is that open
project leadership roles are defined and necessary qualifications
discussed prior to searching for candidates.  Roles now usually are
defined with fixed duration to allow rotation and to mitigate the
level of commitment required.  We publicly advertise for candidates
within the collaboration.  Once candidates are identified, the slate
of candidates is reviewed by the Central Project; at this point, if
there is a paucity of female candidates, the reasons for this are
explored and an attempt is made to redress the issue by encouraging
qualified female candidates to apply. The process is tracked by the
Central Project, which needs to approve all appointments.
\citet{lundgren15a} represents an initial attempt to
assess the effectiveness of this process in increasing participation
of women in the survey leadership; the results are as yet unclear for
SDSS-IV.

In the same year, SDSS-IV formed a Committee on the Participation of
Minorities in SDSS (CPMS) to address the underrepresentation of
minorities in the survey.  While the goal of the CPWS was to ensure
gender balance in SDSS leadership, the CPMS was faced with the more
fundamental goal of recruiting and retaining underrepresented minority
talent in the collaboration at all. CPWS identified a lack of
resources, training, and contact with the SDSS collaboration that is a
barrier to full participation of minorities in the survey. In
response, SDSS-IV implemented two immediate and strategic programs to
have the most meaningful impact: the Faculty And Student Team (FAST)
program deliberately focuses on building serious, long-term research
relationships between faculty/student teams and SDSS partners; the
distributed SDSS REU program targets talented minority students at the
undergraduate level, and can be used as a recruitment tool into
graduate school in astronomy.

The FAST program has been independently funded by the Sloan Foundation
for an initial three-year period. It actively recruits and trains
underrepresented minority (URM) talent to participate in SDSS
science. To qualify for FAST, at least one team member is expected to
be a URM and/or to have a track record serving URM scholars.  FAST
scholar teams are matched with established SDSS partners to work on a
research project of mutual interest and receive specialized training,
mentoring, and financial support in order to introduce teams to SDSS
science and to cement their participation within the
collaboration. FAST team faculty become full members of the SDSS
collaboration, with all data rights, access to centralized computing,
and ability to lead projects that this implies. We selected our first
FAST cohort of three teams in 2015 and recruited five FAST teams in 2016.
The distributed SDSS REU program has also been funded by the Sloan
Foundation for one pilot summer in 2016, with six students at four
institutions.

With regard to the climate of the SDSS-IV collaboration, the global
nature of the survey poses unique challenges in developing an
effective and positive work environment.  Project personnel and
science collaboration are distributed at dozens of institutions, in a
number of countries. Opportunities for in-person interaction are
often limited, with most communication happening through email and
phone conversations. There is no central institution recruiting the
leadership and personnel; in addition, a number of project personnel
work on a voluntary basis or for ``in-kind'' credit for their
technical work. Recognizing the potential issues that could arise in
this environment, we requested that an advisory committee from the
American Physical Society conduct a site visit at the 2014
collaboration meeting. There were numerous comments and suggestions
from the visiting committee. In 2015, CPWS crafted these suggestions
into a set of specific recommendations for the project to prioritize
in order to maintain and improve the quality of the climate in the
collaboration.

The CPWS and CPMS have now been combined into a single Committee on
Inclusion in the SDSS (COINS) with the mandate of both original
committees.

In order to address specific issues that may arise within the
collaboration or other problems, the ARC Board has appointed two
Ombudspersons for SDSS-IV that can be consulted to mediate problems
within the collaboration. The position of Ombudsperson is particularly
designed for cases where handling the matter through formal project
channels would lead to a conflict of interest or cases where anonymity
is desired. In addition, SDSS-IV is in the process of developing a
formal Code of Conduct.

\section{Summary}
\label{sec:summary}
We have described SDSS-IV, which began operations in 2014 July, with
plans to continue until mid-2020. The collaboration has over 1,000
participating astronomers from over 50 institutions worldwide. Three
major programs (APOGEE-2, MaNGA, and eBOSS) and two subprograms (TDSS
and SPIDERS) will address a number of key scientific topics using
dual-hemisphere wide-field spectroscopic facilities. The major
elements of this science program are as follows.
\begin{itemize}
\item Milky Way formation history and evolution, using chemical and
  dynamical mapping of all of its stellar components with APOGEE-2.
\item Stellar astrophysics, using APOGEE-2 infrared spectra alone and
  in combination with asteroseismology, using TDSS's optical
  observations of variable stars, and using MaNGA's bright-time
  optical stellar library.
\item Formation history and evolution of the diverse array of galaxy
  types, using chemical and dynamical mapping of stars and gas with
  MaNGA integral field spectroscopy, using the distant galaxy
  populations in the eBOSS LRG and ELG programs, and the cluster
  galaxies in SPIDERS.
\item Quasar properties and evolution using the massive sample of
  quasars in eBOSS, reaching nearly down to Seyfert galaxy
  luminosities out to $z\sim 2$, complemented with quasars selected
  via variability (TDSS) and X-ray emission (SPIDERS).
\item The most powerful cosmological constraints to date from
  large-scale structure, precisely investigating the Hubble diagram
  and the growth of structure in the redshift range $1<z<2$ for the
  first time, using the largest volume cosmological large-scale
  structure survey to date from eBOSS.
\end{itemize}

The science program is coupled to a robust education and public
engagement program. All of the raw and reduced data will be released
on a well-defined schedule using innovative public interfaces.

\acknowledgments

We thank an anonymous referee for numerous comments that improved the
clarity and utility of this paper. 

Funding for the Sloan Digital Sky Survey IV has been provided by the
Alfred P. Sloan Foundation, the U.S. Department of Energy Office of
Science, and the Participating Institutions. SDSS-IV acknowledges
support and resources from the Center for High-Performance Computing
at the University of Utah. The SDSS web site is www.sdss.org.

SDSS-IV is managed by the Astrophysical Research Consortium for the
Participating Institutions of the SDSS Collaboration including the
Brazilian Participation Group, the Carnegie Institution for Science,
Carnegie Mellon University, the Chilean Participation Group, the
French Participation Group, Harvard-Smithsonian Center for
Astrophysics, Instituto de Astrof\'isica de Canarias, The Johns
Hopkins University, Kavli Institute for the Physics and Mathematics of
the Universe (IPMU) / University of Tokyo, Lawrence Berkeley National
Laboratory, Leibniz Institut f\"ur Astrophysik Potsdam (AIP),
Max-Planck-Institut f\"ur Astronomie (MPIA Heidelberg),
Max-Planck-Institut f\"ur Astrophysik (MPA Garching),
Max-Planck-Institut f\"ur Extraterrestrische Physik (MPE), National
Astronomical Observatories of China, New Mexico State University, New
York University, University of Notre Dame, Observat\'ario Nacional /
MCTI, The Ohio State University, Pennsylvania State University,
Shanghai Astronomical Observatory, United Kingdom Participation Group,
Universidad Nacional Aut\'onoma de M\'exico, University of Arizona,
University of Colorado Boulder, University of Oxford, University of
Portsmouth, University of Utah, University of Virginia, University of
Washington, University of Wisconsin, Vanderbilt University, and Yale
University.

\facility{Sloan}
\software{Astropy}

\bibliographystyle{apj}
\bibliography{sdss4} 

\end{document}